\global\def\draftcontrol{0}

   \def\versionno{cascading dfp}

\catcode`\@=11

\expandafter\ifx\csname draftcontrol\endcsname\relax\global\def\draftcontrol{0}
\fi

{\count255=\time\divide\count255 by 60
\xdef\hourmin{\number\count255}
\multiply\count255 by-60\advance\count255 by\time
\xdef\hourmin{\hourmin:\ifnum\count255<10 0\fi\the\count255}}
\def\draftdate{\number\month/\number\day/\number\year\ \ \ \hourmin }

\newcommand\makepapertitle{\par
  \begingroup
    \renewcommand\thefootnote{\@fnsymbol\c@footnote}%
    \def\@makefnmark{\rlap{\@textsuperscript{\normalfont\@thefnmark}}}%
    \long\def\@makefntext##1{\parindent 1em\noindent
            \hb@xt@1.8em{%
                \hss\@textsuperscript{\normalfont\@thefnmark}}##1}%
     \newpage
     \global\@topnum\z@   
     \@makepapertitle
     \thispagestyle{empty}\@thanks
  \endgroup
  \setcounter{footnote}{0}%
  \global\let\thanks\relax
  \global\let\makepapertitle\relax
  \global\let\@makepapertitle\relax
  \global\let\@thanks\@empty
  \global\let\@author\@empty
  \global\let\@date\@empty
  \global\let\@title\@empty
  \global\let\title\relax
  \global\let\author\relax
  \global\let\date\relax
  \global\let\and\relax
  \def\version{\let\version\@version\@gobble}
}
\def\@makepapertitle{%
  \newpage
   \ifnum\draftcontrol=1 {}
   \version\versionno
   \vskip 3em%
   \else
   \hfill\hbox to 3cm {\parbox{4cm}{\@pubnum}\hss}%
   \vskip 3em%
   \fi
   \begin{center}%
   \let \footnote \thanks
     {\LARGE {\@title}}%
     \vskip 1.5em%
     {\normalsize
       \lineskip .5em%
       \begin{tabular}[t]{c}%
         \@author
       \end{tabular}\par}%
     \vskip 1.5em%
     {\@bstract}%
     \end{center}%
     \vskip 1.5em
     \@date%
   \par
}

\gdef\@pubnum{}
\def\pubnum#1{%
  \gdef\@pubnum{#1}}

\gdef\@bstract{}
\def\Abstract#1{%
  \gdef\@bstract{%
   \parbox{\textwidth-0pc}{%
   \centerline{\bf Abstract}\penalty1000%
\kern.2cm%
\noindent
\renewcommand\baselinestretch{1.0}%
{#1}}}
}

\def\ps@paper{\let\@mkboth\@gobbletwo%
     \ifnum\draftcontrol=1
    \def\@oddfoot{\hbox to \textwidth{\tiny \versionno \hfil\tiny\draftdate}%
    \hskip -\textwidth \hbox to \textwidth{\hfil\rm\thepage\hfil}}%
     \else\def\@oddfoot{\hbox to \textwidth{\hfil\rm\thepage\hfil}}
     \fi
     \let\@evenfoot\@oddfoot
}

\def\body{\clearpage
          \pagestyle{paper}
    }

\def\@version#1{\ifnum\draftcontrol=1
\typeout{}\typeout{#1}\typeout{}
\vskip3mm\centerline{\hbox{\fbox{\normalsize{\tt DRAFT -- #1 -- }
                   {\draftdate}}}}\vskip3mm
\fi}
\let\version\@version
\long\def\eqlabel#1{\ifnum\draftcontrol=1
                    \tag@false  
                    \tag*{(\theequation) \hbox to -0.2cm{\hspace{0cm}\small{#1}\hss}}
                    \refstepcounter{equation}
                    \edef\@currentlabel{\theequation}
                    \ltx@label{#1}          
                    \else
                    \label{#1}
                    \fi
                    }
\let\st@bibitem\@bibitem
\let\st@lbibitem\@lbibitem
\ifnum\draftcontrol=1
  \def\@bibitem#1{%
    \st@bibitem{#1}\a@@label{#1}\ignorespaces}
  \def\@lbibitem[#1]#2{%
    \st@lbibitem[#1]{#2}\a@@label{#2}\ignorespaces}
  \def\a@@label#1{%
    \gdef\a@lab{\smash{\normalfont\small#1}}
    \ifvmode
      \if@inlabel
        \global\setbox\@labels\hbox{%
          \llap{\a@lab\let\a@lab\relax
                \kern\@totalleftmargin\kern\marginparsep}%
          \box\@labels}%
      \fi
    \fi}
\fi

\documentclass[12pt,letterpaper]{article}

\usepackage{amsmath,amssymb,array,calc,epsfig,rotating,bm,xcolor}
\usepackage[sort]{cite}
\usepackage{graphicx,esint,float}
\usepackage{psfrag,verbatim}
\usepackage[makeroom]{cancel}
\usepackage{xcolor,url}
\usepackage{hyperref}

\ifnum\draftcontrol=1
\fi

\renewcommand\baselinestretch{1.25}
\renewcommand\section{\@startsection {section}{1}{\z@}%
                                   {-3.5ex \@plus -1ex \@minus -.2ex}%
                                   {2.3ex \@plus.2ex}%
                                   {\normalfont\large\bfseries}}
\renewcommand\subsection{\@startsection{subsection}{2}{\z@}%
                                   {-3.25ex\@plus -1ex \@minus -.2ex}%
                                   {1.5ex \@plus .2ex}%
                                   {\normalfont\normalsize\bfseries}}
\renewcommand\subsubsection{\@startsection{subsubsection}{3}{\z@}%
                                   {-3.25ex\@plus -1ex \@minus -.2ex}%
                                   {1.5ex \@plus .2ex}%
                                   {\normalfont\normalsize\it}}
\renewcommand\paragraph{\@startsection{paragraph}{4}{\z@}%
                                   {-3.25ex\@plus -1ex \@minus -.2ex}%
                                   {1.5ex \@plus .2ex}%
                                   {\normalfont\normalsize\bf}}

\numberwithin{equation}{section}

\def\revise#1       {\raisebox{-0em}{\rule{3pt}{1em}}%
                     \marginpar{\raisebox{.5em}{\vrule width3pt\
                     \vrule width0pt height 0pt depth0.5em
                     \hbox to 0cm{\hspace{0cm}{%
                     \parbox[t]{4em}{\raggedright\footnotesize{#1}}}\hss}}}}

\newcommand\nxt[1]  {\\\fnxt#1}
\newcommand{\ie}{{\it i.e.,}\ }
\newcommand{\eg}{{\it e.g.,}\ }

\newcommand{\mathcolorbox}[2]{\colorbox{#1}{$\displaystyle #2$}}

\def\cala         {{\cal A}}

\def\calm         {{\cal M}}
\def\caln         {{\cal N}}
\def\calo         {{\cal O}}

\def\calr         {{\cal R}}
\def\cals         {{\cal S}}

\def\naturals     {{\mathbb N}}

\def\zet          {{\mathbb Z}}
\def\hw          {{\hat{\Omega}}}
\def\hF          {{\hat{F}}}
\def\hfl          {{\hat{fl}}}
\def\hq          {{\hat{q}}}

\def\del          {\partial}

\def\Re   {{\mathfrak{Re}}}
\def\Im   {{\mathfrak{Im}}}

\def\sqr#1#2{{\vcenter{\vbox{\hrule height.#2pt
 \hbox{\vrule width.#2pt height#1pt \kern#1pt
 \vrule width.#2pt}\hrule height.#2pt}}}}

\newcommand{\ww}{\mathfrak{w}}

\def\a{\alpha}
\def\b{\beta}
\def\w{\omega}

\def\hh{\hat{h}}

\def\hs{\hat{s}}
\def\hf{\hat{f}}
\def\hg{\hat{g}}
\def\hK{\hat{K}}
\def\aa1{\phi}
\def\cc1{\psi}
\def\hh{\hat{h}}

\def\om{\Omega}

\def\hf{\hat{f}}
\def\hK{\hat{K}}

\def\csb{{\chi\rm{SB}}}
\def\cs{{\chi\rm{S}}}

\def\hc{\hat{c}}

\def\df{\delta f}
\def\dk{\delta k}
\def\dg{\delta g}
\def\dh{\delta h}
\def\f0{\text{\boldmath$\varphi$}}
\def\h2{\mathfrak{h}}
\def\vol{{\rm vol}}

\catcode`\@=12

\begin{document}


\title{\bf Stability of the cascading gauge theory de 
Sitter DFPs}

\date{October 30, 2022}

\author{
Alex Buchel\\[0.4cm]
\it $ $Department of Physics and Astronomy\\ 
\it University of Western Ontario\\
\it London, Ontario N6A 5B7, Canada\\
\it $ $Perimeter Institute for Theoretical Physics\\
\it Waterloo, Ontario N2J 2W9, Canada
}

\Abstract{We study stability of the Dynamical Fixed Points (DFPs) of the
cascading gauge theory at strong coupling in de Sitter space-time.  We
compute the spectra of the perturbative fluctuations and identify
stable/unstable DFPs, characterized by the ratio of the strong
coupling scale $\Lambda$ of the gauge theory and the Hubble constant
$H$ of the background space-time.  We discover a new phenomenon in the
spectrum of gravitational fluctuations of a non-conformal holographic
model: distinct branches of the fluctuations for $H\gg \Lambda$
coalesce for sufficiently low $\frac{H}{\Lambda}$, leading to the
removal of some excited modes from the spectrum.  We establish that,
at least in a dual supergravity approximation, the cascading gauge theory
does not have a stable DFP for $H\in (H_{crit_1},H_{crit_2})$.  Initial
states of the theory for $H>H_{crit_2}$ evolve to a stable DFP with
unbroken chiral symmetry; while for $H< H_{crit_1}$ the states evolve
to a de Sitter vacuum with spontaneously broken chiral symmetry.
}

\makepapertitle

\body

\version\versionno
\tableofcontents

\section{Introduction and summary}\label{intro}
The cascading gauge theory\footnote{See \cite{Buchel:2021yay} for a recent review.}
\cite{Klebanov:2000hb} is $\caln=1$ supersymmetric four-dimensional 
$SU(N+M)\times SU(N)$ gauge theory. It is non-conformal, and has a strong coupling scale
$\Lambda$. The high-energy physics of the theory is exotic\footnote{Remarkably,
the cascading gauge theory remains (holographically) renormalizable as a four-dimensional
quantum field theory (QFT) \cite{Aharony:2005zr} when formulated on an arbitrary
background space-time manifold $\calm_4$ .}: it undergoes
perpetual sequence of Seiberg \cite{Seiberg:1994pq} dualities, $N\to N+M$, effectively rendering
the rank parameter $N$ energy dependent \cite{Buchel:2000ch},
\begin{equation}
N=N(E)\ \propto\ M^2\ \ln\frac E\Lambda\,,\qquad {\rm as}\qquad \frac{E}{\Lambda}\to\infty\,.
\eqlabel{runn}
\end{equation}
In the renormalization group flow to the infrared (IR),
the rank parameter $N$ decreases as $N\to N-M$,
with each realization of the Seiberg duality.
In Minkowski space-time, $R^{3,1}$, the moduli space of vacua of the theory was thoroughly
analyzed in \cite{Chai:2021tpt} ---  when $N$ is an integer multiple of $M$,
the cascading gauge theory ends up in the IR as the $\caln=1$ $SU(M)$ Yang-Mills theory.
It confines with a spontaneous breaking of the $U(1)_R$ chiral symmetry,
\begin{equation}
U(1)_R\ \to\ \zet_2\,.
\eqlabel{sbp}
\end{equation}

When $M\gg 1$, the cascading gauge theory has a String Theory holographic dual
\cite{Maldacena:1997re,Aharony:1999ti}
realized by a consistent truncation of Type IIB supergravity on warped
deformed conifold with fluxes \cite{Buchel:2010wp}. Owing to the fact that
the cascading gauge theory in the IR shares the staples of QCD at strong coupling,
namely confinement and the chiral symmetry breaking, the precise holographic dual
allows to explore properties of strongly coupled non-conformal gauge theories which are
difficult (and often impossible) to access otherwise:
the thermal phase diagram \cite{Aharony:2007vg,Buchel:2010wp,Buchel:2018bzp};
the hydrodynamic transport \cite{Buchel:2005cv,Buchel:2009bh},
the gauge theory dynamics in curved space
\cite{Buchel:2001iu,Buchel:2011cc,Buchel:2021yay} and in cosmological setting \cite{Buchel:2013dla,Buchel:2019pjb}. 
The late-time properties of the cascading gauge theory in de Sitter space-time is
the subject of this paper.

Before we present the results of the analysis, we would like to clearly distinguish the
concept of a {\bf de Sitter vacuum} \cite{Birrell:1982ix} of a QFT and a
{\bf de Sitter DFP} of a QFT \cite{Buchel:2021ihu}.  It is useful to start
with a conformal field theory (CFT). Typically\footnote{Some of the
counterexamples are the integrable systems,
Fermi-Pasta-Ulam-Tsingou problem \cite{2008PhT....61a..55D}, and the gravitational collapse in
AdS \cite{Balasubramanian:2014cja}.}, an arbitrary initial state of an interactive CFT
in $R^{3,1}$ thermalizes. In a dual holographic picture this dynamics  is encoded in the
gravitational collapse and the black brane formation \cite{Danielsson:1999fa}. 
Following the second law of thermodynamics, as a CFT state equilibrates, its non-equilibrium entropy density $s(t)$ 
monotonically increases, $\dot{s}(t) \ge 0$, and reaches at late times the {\it finite} thermal entropy
density $s_{thermal}$, determined by the late-time thermal equilibrium temperature $T$,
\begin{equation}
\lim_{t\to\infty} s(t)=s_{thermal}\ <\ \infty\,.
\eqlabel{slate}
\end{equation}
The existence of the above limit, equivalently the equilibration of a generic state,
implies that the entropy production rate vanishes at late times, \ie
\begin{equation}
\lim_{t\to\infty} \frac{\dot{s}(t)}{s(t)}= 0\,.
\eqlabel{sdot}
\end{equation}
Consider now the dynamics of this CFT in 
Friedmann-Lemaitre-Robertson-Walker (FLRW) Universe. Since the background geometry
\begin{equation}
ds_{FLRW}^2=-d\tau^2+a(\tau)^2\ d{\bm x}^2  = a^2\ \left(-dt^2+d{\bm x^2}\right)= a^2\ ds_{Minkowski}^2\,,
\eqlabel{flrw}
\end{equation}
where $a(\tau)$ is a cosmological scale factor and $dt\equiv \frac{d\tau}{a(\tau)}$
is the conformal time,
is Weyl equivalent to Minkowski space-time,
there is a precise translation of the CFT dynamics in $R^{3,1}$ and FLRW. For example, the expectation
values of the theory stress-energy tensor are related as 
\begin{equation}
\langle T_{\mu\nu}(\tau,{\bm x})\rangle\bigg|_{FLRW}=\frac{1}{a^4}\cdot \langle T_{\mu\nu}(t,{\bm x})\rangle\bigg|_{Minkowski}+\frac{c}{8\pi^2}
\biggl(R^{\rho\sigma}R_{\rho\mu\sigma\nu}-\frac{1}{12}R^2\cdot g_{\mu\nu}\biggr)\,,
\eqlabel{stresst}
\end{equation}
where $c$ is the central charge of the CFT, $g_{\mu\nu}$ and $R_{\rho\mu\sigma\nu}$
are the metric \eqref{flrw} and the corresponding Riemann tensor.  
When a CFT has a holographic dual, the Weyl equivalence \eqref{flrw} is nothing but a diffeomorphism
transformation of the gravitational dual \cite{Buchel:2017pto}. Furthermore, when
the non-equilibrium entropy is associated with the apparent horizon (AH) of the gravitational dual,
its Minkowski space-time production rate is identical, Weyl invariant, to the corresponding FLRW
comoving entropy production rate with respect to the conformal time \cite{Buchel:2017pto}.  
This implies that the equilibration of a CFT state in Minkowski space-time is mapped to
the evolution of the corresponding state in FLRW, where the comoving entropy approaches
a constant at late-time. This  late-time state is  a {\bf FLRW vacuum}
of the CFT, characterized by the asymptotically vanishing comoving entropy
production rate.  

While it difficult to map dynamics of a massive QFT in Minkowski
and FLRW Universe from the path integral viewpoint,
the problem is tractable if the theory has a
holographic dual. From the dual gravitational perspective,
a gravitational bulk diffeomorphism relating the two boundary backgrounds
\eqref{flrw} acts on a relevant coupling constant $\lambda_\Delta$ of
a dimension $\Delta<4$ operator $\calo_\Delta$ as \cite{Buchel:2017pto}
\begin{equation}
\lambda_\Delta\ \to \hat{\lambda}_\Delta(t)=a(\tau(t))^{4-\Delta}\ \lambda_\Delta\,,
\eqlabel{ld}
\end{equation}
\ie a massive QFT dynamics with a coupling constant $\lambda_\Delta$
in FLRW is equivalent to the quenched dynamics of the same
theory in Minkowski space-time, where the coupling $\hat{\lambda}_\Delta$
evolves according to \eqref{ld}.
When $a(\tau\to  +\infty)\to {\rm const}$, the QFT coupling constant
is quenched as 
$\hat{\lambda}_\Delta(0)\equiv \lambda_\Delta\ \to \hat{\lambda}_\Delta(+\infty)$.
Holographic quenches of just this type
were extensively studied in \cite{Buchel:2012gw,Buchel:2013lla,Buchel:2013gba}:
\begin{itemize}
\item the theory eventually thermalizes at late times;
\item the thermalization process is irreversible --- the entropy density
production rate is always positive. 
\end{itemize}
The last statement implies that the comoving entropy production rate
of the QFT in FLRW is positive as well. When the FLRW scale factor
$a(\tau)$ diverges as $\tau\to +\infty$, the mapped quenched coupling
$\hat{\lambda}_\Delta$ diverges at late time as well \eqref{ld} --- it is not clear
whether or not the theory thermalizes; irrespectively, it can
be rigorously shown\footnote{This was done explicitly
in case-by-case holographic models
\cite{Buchel:2013lla,Buchel:2017pto,Buchel:2017lhu
,Buchel:2019pjb,Casalderrey-Solana:2020vls,Buchel:2022hjz};
we believe though that the general proof is possible.} 
that the comoving entropy production rate is always positive:
if $s(\tau)$ is the physical entropy density, the entropy current is given
by \cite{Buchel:2021ihu}
\begin{equation}
\cals^\mu=s(\tau)\ u^\mu\,,\qquad u^{\mu}\equiv  (1,0,0,0)\,,
\eqlabel{defs}
\end{equation}
leading to the entropy density production rate $\calr$,
\begin{equation}
\calr(\tau)\equiv \nabla\cdot S=\frac{1}{a(\tau)^3}\ \frac{d}{d\tau}\left(a(\tau)^3s(\tau)\right)
=\frac{1}{a(\tau)^3}\ \frac{d}{d\tau} s_{comoving}(\tau)\ \ge\ 0\,.
\eqlabel{defr}
\end{equation}
If $\calr(\tau)$ vanishes at late times, we say the state of the QFT evolves
to a FLRW vacuum; if the rate approaches a constant, we say that the
state of the QFT
evolves\footnote{More precisely, both for a vacuum and a DFP  we
additionally require that one-point correlation functions
of the stress-energy tensor and gauge-invariant local operators are homogeneous
and time-independent.} to a {\bf Dynamical Fixed Point} \cite{Buchel:2021ihu}. 
In case the FLRW Universe is de Sitter, \ie
\begin{equation}
a(\tau)=e^{H \tau}\,,
\eqlabel{desia}
\end{equation}
where $H$ is the Hubble constant,  the existence of the late-time limit
for the entropy density production rate implies that there is a late-time limit
for a physical entropy density,
\begin{equation}
\lim_{\tau\to\infty}\left(\nabla\cdot \cals\right)= 3 H\ \lim_{\tau\to\infty} s(\tau) = 3H\ s_{ent} \,,
\eqlabel{rlim}
\end{equation}
with the latter being called the {\it vacuum entanglement entropy} \cite{Buchel:2017qwd}.

\begin{figure}[t]
\begin{center}
\psfrag{a}[cc][][1][0]{${H_{crit_1}}$}
\psfrag{b}[cc][][1][0]{${H_{crit_3}}$}
\psfrag{c}[cc][][1][0]{${H_{crit_2}}$}
\psfrag{d}[cc][][1][0]{${H_{crit_4}}$}
\psfrag{e}[cc][][1][0]{${H_{crit_5}}$}
\psfrag{0}[cc][][1][0]{$0$}
\psfrag{q}[cc][][1][0]{TypeB}
\psfrag{w}[cc][][1][0]{TypeA$ _b$}
\psfrag{r}[cc][][1][0]{TypeA$ _s$}
\psfrag{f}[cc][][1.5]{${H}$}
\includegraphics[width=5in]{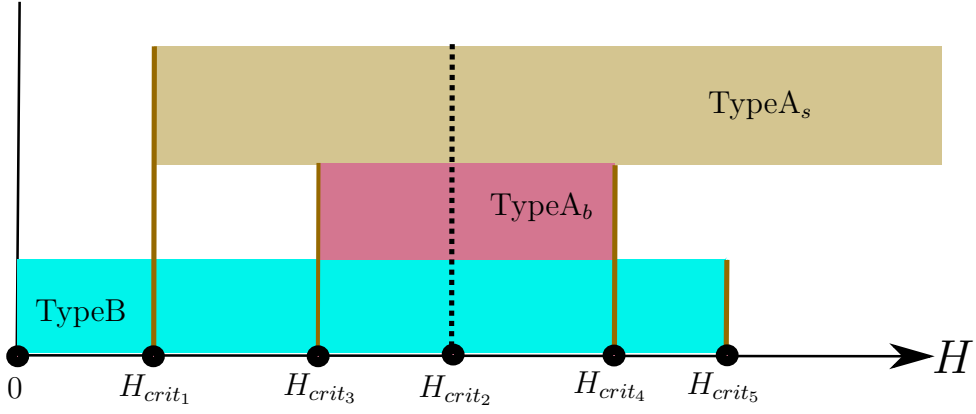}
\end{center}
  \caption{de Sitter vacua (TypeB) and de Sitter DFPs
  (TypeA$ _s$ and TypeA$ _b$) of the cascading gauge theory with a fixed
  strong coupling scale $\Lambda$. The vertical
  brown lines indicate the existence range for different
  phases: TypeA$ _s$ exists for  $H\in (H_{crit_1},+\infty)$;
   TypeA$ _b$ exists for  $H\in (H_{crit_3},H_{crit_4})$;  TypeB
   exists for  $H\in (0,H_{crit_5})$. In the range $H\in (H_{crit_3},H_{crit_2})$,
   indicated by a vertical dashed line, 
   TypeA$ _b$ DFP is the  preferred phase.} \label{phases}
\end{figure}

de Sitter vacua and DFPs of the cascading gauge theory were analyzed in details
in \cite{Buchel:2019pjb}; we identified the following late-time spatially homogeneous
and isotropic phases of the theory:
\nxt TypeA$ _s$ --- the de Sitter DFP with unbroken chiral symmetry,
\begin{equation}
s_{ent}(\Lambda,H)\bigg|_{{\rm TypeA}_s}\ \ne 0\,;
\eqlabel{stypea}
\end{equation}
\nxt TypeA$ _b$ --- the de Sitter DFP with spontaneously broken chiral symmetry,
\begin{equation}
s_{ent}(\Lambda,H)\bigg|_{{\rm TypeA}_b}\ \ne 0\,;
\eqlabel{btypea}
\end{equation}
\nxt TypeB --- the de Sitter vacuum with spontaneously broken chiral
symmetry\footnote{This vacuum is smoothly connected to a supersymmetric
Klebanov-Strassler Minkowski vacuum \cite{Klebanov:2000hb}
in the limit $\frac{H}{\Lambda}\to 0$.},
\begin{equation}
s_{ent}(\Lambda,H)\bigg|_{{\rm TypeB}}\ = 0\,.
\eqlabel{typeb}
\end{equation}
These results are summarized in figure \ref{phases}:
\begin{itemize}
\item all phases can be reliably constructed in the supergravity
approximation within a fixed range of the ratio $\frac H\Lambda$,
specifically,
\begin{equation}
\begin{split}
{\rm TypeA}_s:&\qquad H\in (H_{crit_1,+\infty)}\,,\qquad H_{crit_1}\approx 0.7\Lambda\,,\\
{\rm TypeA}_b:&\qquad H\in (H_{crit_3},H_{crit_4})\,,\qquad H_{crit_3}=0.92(1)\Lambda\,;\ \ \ 
H_{crit_4}\approx 0.93\Lambda\,,\\
{\rm TypeB}:&\qquad H\in (0,H_{crit_5})\,,\qquad H_{crit_5}\approx 0.97\Lambda\,,
\end{split}
\eqlabel{ranges}
\end{equation}
where we used $\approx$ to indicate that the corresponding value of $H_{crit}$
is estimated from the breakdown of the supergravity approximation, see 
\cite{Buchel:2019pjb}. The critical value $H_{crit_3}$ 
can be computed with an arbitrary precision within the supergravity approximation,
hence we used the $=$ sign.
\item Given the ratio $\frac{H}{\Lambda}$, the preferred phase is the one with the
larger vacuum entanglement entropy $s_{ent}$ --- the latter quantity determines
the entropy production rate \eqref{rlim} at late times, and the dual gravitational
evolution always proceeds towards the late-time attractor with the
largest  apparent horizon
comoving area density\footnote{This is nothing but the restatement
of the phase selection principle in approach
to thermal equilibrium in microcanonical ensemble for de Sitter dynamics with
multiple dynamical fixed points.
The latter statement was explicitly verified in the holographic setting in
\cite{Buchel:2021ihu}.}. Thus, whenever a de Sitter DFP exists, \ie
for
\begin{equation}
H\ > H_{crit_1}\,,
\eqlabel{dfprules}
\end{equation}
no state of the cascading gauge theory
would evolve to a vacuum (TypeB).
\item It was established in \cite{Buchel:2019pjb} that
\begin{equation}
\begin{split}
&s_{ent}\bigg|_{{\rm TypeA}_b}\ >\ s_{ent}\bigg|_{{\rm TypeA}_s}\qquad
{\rm for}\qquad H\in (H_{crit_3},H_{crit_2})\,,\\
&s_{ent}\bigg|_{{\rm TypeA}_s}\ >\ s_{ent}\bigg|_{{\rm TypeA}_b}\qquad
{\rm for}\qquad H>H_{crit_2}\,,
\end{split}
\eqlabel{bovers}
\end{equation}
where
\begin{equation}
H_{crit_2}=0.92(5)\Lambda\,,
\eqlabel{hcrit2res}
\end{equation}
computable with arbitrary precision within the supergravity approximation.
Thus, TypeA$ _b$ DFP is the preferred attractor over TypeA$ _s$ DFP whenever
the former exists and for $H< H_{crit_2}$. For  $H>H_{crit_2}$ the
de Sitter dynamical fixed point with unbroken chiral symmetry, \ie TypeA$ _s$,
is the preferred attractor.
\end{itemize}

\begin{figure}[t]
\begin{center}
\psfrag{a}[cc][][1][0]{${H_{crit_1}}$}
\psfrag{b}[cc][][1][0]{${H_{crit_3}}$}
\psfrag{c}[cc][][1][0]{${H_{crit_2}}$}
\psfrag{d}[cc][][1][0]{${H_{crit_4}}$}
\psfrag{e}[cc][][1][0]{${H_{crit_5}}$}
\psfrag{0}[cc][][1][0]{$0$}
\psfrag{q}[cc][][1][0]{TypeB}
\psfrag{w}[cc][][1][0]{TypeA$ _b$}
\psfrag{r}[cc][][1][0]{TypeA$ _s$}
\psfrag{t}[cc][][1][0]{???????}
\psfrag{f}[cc][][1.5]{${H}$}
\includegraphics[width=5in]{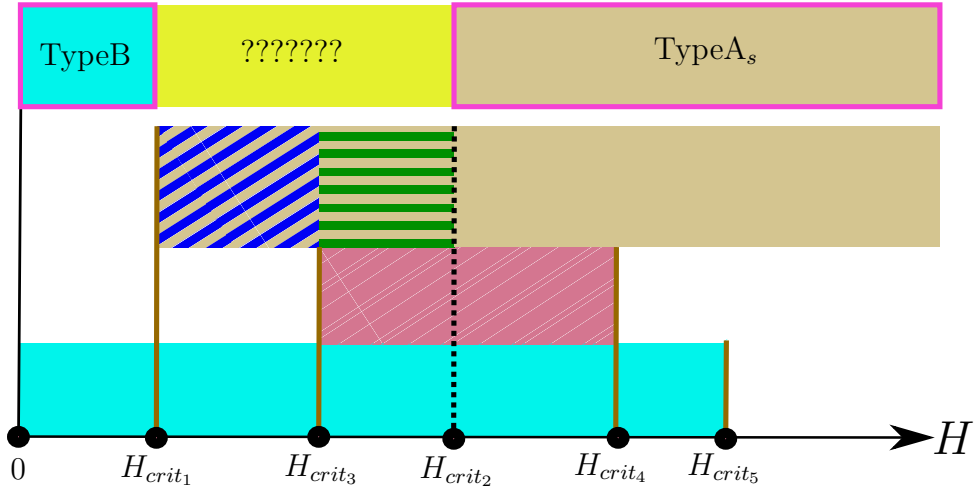}
\end{center}
  \caption{Diagonal blue shaded regions indicate: perturbative instability of TypeA$ _s$ cascading gauge
  theory DFP for $H<H_{crit_3}$, and perturbative instability of TypeA$ _b$ cascading gauge
  theory DFP, whenever it exists. Horizontal green shading for $H_{crit_2}<H<H_{crit_3}$ indicates
  TypeA$ _s$ cascading gauge
  theory DFP which is while perturbatively stable, is unstable to sufficiently
  large amplitude chiral symmetry breaking fluctuations. The cascading gauge theory states in
  de Sitter with $H<H_{crit_1}$ evolve to de Sitter vacuum, TypeB-labeled
  bordered rectangle. The cascading gauge theory states in
  de Sitter with $H>H_{crit_2}$ evolve to  TypeA$ _s$ DFP.
  The late-time dynamics of the cascading gauge theory states
  for $H_{crit_1}<H<H_{crit_2}$, the yellow rectangle, is unknown.} \label{stability}
\end{figure}

In this paper we analyze perturbative stability of TypeA$ _s$ and TypeA$ _b$
DFPs of the cascading gauge theory. Our main results are summarized in figure \ref{stability}.
We find:
\begin{itemize}
\item Precisely at $H=H_{crit_3}$ there is a zero mode of TypeA$ _s$ phase,
associated with the spontaneous breaking of the chiral symmetry \cite{Buchel:2019pjb}.
In the limit $H\to H_{crit_3}+0$ the chiral symmetry breaking order parameters of
TypeA$ _b$ phase, \ie the  expectation values  of the pair of dimension $\Delta=3$ operators
$\calo_3^{\alpha=1,2}$ and the dimension $\Delta=7$ operator $\calo_7$ of the
cascading gauge theory, vanish as $\propto (H-H_{crit_3})^{1/2}$, typical
for a spontaneous symmetry breaking with a mean-field
exponent $\frac 12$ \cite{Buchel:2019pjb}. This zero mode is purely dissipative
away from $H_{crit_3}$, and behaves differently
in the two distinct DFPs. Specifically,
\nxt in the TypeA$ _s$ DFP this mode, we index it with $ _\csb$, has\footnote
{We use reduced frequencies in the paper, $\ww\equiv \frac{\omega}{H}$,}
\begin{equation}
\Im[\ \ww_\csb\ ]\bigg|_{{\rm TypeA}_s}\qquad\ \begin{cases}
&<0\,,\qquad  H>H_{crit_3}\,,\qquad \Longrightarrow\ {\rm stable}\cr
&>0\,,\qquad  H<H_{crit_3}\,,\qquad \Longrightarrow\ {\rm unstable}
\end{cases}\,,
\eqlabel{wa}
\end{equation}
and includes fluctuations of $\calo_3^{\alpha=1,2}$ and $\calo_7$ operators of the
cascading gauge theory;
\nxt in the TypeA$ _b$ DFP, this mode exists only for $H>H_{crit_3}$ (there is no
 TypeA$ _b$ DFP for $H<H_{crit_3}$)
and is unstable,
\begin{equation}
\Im[\ \ww_\csb\ ]\bigg|_{{\rm TypeA}_b}\ 
>0\,,\qquad  H>H_{crit_3}\,.
\eqlabel{wb}
\end{equation}
In the symmetry broken TypeA$ _b$ DFP this mode is much more complicated:
it couples fluctuations of $\calo_3^{\alpha=1,2}$, $\calo_4^{\beta=1,2}$,  $\calo_6$,
 $\calo_7$, and  $\calo_8$ operators of the cascading theory.
 \item Because of \eqref{wa}, TypeA$ _s$ DFP is perturbatively unstable for
 $H< H_{crit_3}$, represented by the diagonal blue shading.
 \item While TypeA$ _s$ DFP is perturbatively stable to chiral symmetry breaking fluctuations
for $H\in (H_{crit_3},H_{crit_2})$ (represented by the horizontal green shading),
it can not be non-perturbatively stable: sufficiently large-amplitude chiral symmetry
breaking fluctuations must force
dynamics\footnote{Identical phenomenon was observed in dynamical simulations in
the model covered in \cite{Buchel:2021ihu}.} towards the preferred TypeA$ _b$ DFP attractor, see 
\eqref{bovers}.
\item TypeA$ _b$ is always perturbatively unstable, represented by the diagonal
blue shading.
\end{itemize}
Given the fluctuation spectra stability analysis, we establish that:
\begin{itemize}
\item all states of the cascading gauge theory in de Sitter with
the Hubble constant $H>H_{crit_2}$ evolve to TypeA$ _s$
DFP (bordered rectangle);
\item all states of the cascading gauge theory in de Sitter with
the Hubble constant $H<H_{crit_1}$ evolve to TypeB late-time attractor ---
the de Sitter vacuum (bordered rectangle).
\item We do not know the late-time dynamics of the cascading gauge theory
in de Sitter for  $H\in (H_{crit_1},H_{crit_2})$ (yellow rectangle) --- in this range both
TypeA$ _s$ and TypeA$ _b$ DFPs are unstable, and the late-time attractor can not
be a de Sitter vacuum (TypeB), which has vanishing entropy density production rate, see
\eqref{typeb}. We expect that in this case the cascading gauge theory states evolve,
spontaneously breaking chiral symmetry, to a naked singularity, similar
to the evolution of the symmetry broken states in the toy model
discussed in \cite{Buchel:2021ihu}
\end{itemize}

The rest of the paper is organized as follows.
In section \ref{dfpreview} we review
the de Sitter vacua and dynamical fixed points
of the cascading gauge theory \cite{Buchel:2019pjb}.
In section \ref{hard} we explain
why perturbative stability analysis of the de Sitter dynamical
fixed points  are difficult. We explain why the general ``master framework''
developed in \cite{Buchel:2022hjz} is not
suitable for the cascading gauge theory, and what straightforward modification
is required. We highlight the difficulty of imposing the
boundary conditions for the gravitational fluctuations, and explain
how to overcome it. In section \ref{stypeas} we study
perturbative stability of TypeA$ _s$ DFP.
We separate fluctuations into sets preserving the chiral symmetry of this DFP,
and the fluctuations that spontaneously break the chiral symmetry.  
We study both sets in the near-conformal regime, \ie when $\ln\frac{H}{\Lambda}\gg 1$
and partial analytic treatment is possible, and follow the fluctuation
spectra to $H\sim \Lambda$. We identify the unstable mode in the
chiral symmetry breaking sector in TypeA$ _s$ DFP when $H<H_{crit_3}$.
The latter mode is marginal at $H=H_{crit_3}$, where the two DFP
TypeA$ _s$ and TypeA$ _b$ are indistinguishable.
We establish that there is no instability in the chiral symmetry preserving
sector of fluctuations in TypeA$ _s$ DFP, at least  for $H > H_{crit_3}$.
In section \ref{stypeab}  we study
perturbative stability of TypeA$ _b$ DFP.
We show that the marginal chiral symmetry breaking mode at $H=H_{crit_3}$
becomes unstable in TypeA$ _b$ DFP, perturbatively in $(H-H_{crit_3})>0$.
We demonstrate that this mode remains unstable at least as $H$ approaches $H_{crit_2}$.
Our numerics indicates that the mode remains unstable even after $H>H_{crit_2}$, but this is
physically irrelevant since in this regime TypeA$ _b$ DFP is not preferred relative
to TypeA$ _s$ DFP, see \eqref{bovers}.
Finally, we conclude with open questions and speculations in
section \ref{conclude}.
Whenever appropriate, we delegate the technical details to appendices
and focus on the physics instead.

Any stability analysis of a gravitational model are necessarily technical. This is particularly the case
for the theory analyzed here --- many equations are too long to be presented even in appendices;
we collected them as a Maple worksheet available at \cite{coleqs}.

\section{de Sitter vacua and DFPs of the cascading theory}
\label{dfpreview}

In this section we summarize the results of \cite{Buchel:2019pjb}.

Consider $SU(2)\times SU(2)\times \zet_2$ invariant states of
the cascading gauge theory on a 4-dimensional manifold
$\calm_4\equiv \del\calm_5$. In the planar limit and at large 't Hooft
coupling, one can consistently truncate the theory to a finite number of
operators \cite{Buchel:2010wp}: a stress-energy tensor $T_{ij}$,
a pair of dimension-3 operators $\calo_3^{\alpha=\{1,2\}}$
(dual to gaugino condensates for each of the gauge group factors),
a pair of dimension-4 operators $\calo_4^{\beta=\{1,2\}}$,
and dimension-6,7,8 operators $\calo_6,\calo_7,\calo_8$. 
Effective gravitational action on a 5-dimensional manifold
$\calm_5$ describing holographic dual of such states was derived
in \cite{Buchel:2010wp}:
\begin{equation}
\begin{split}
&S_5\left[g_{\mu\nu}\leftrightarrow T_{ij},\{\Omega_i,h_i,\Phi\}\leftrightarrow
\{\calo_3^\alpha,\calo_4^\beta,\calo_6,\calo_7,\calo_8\}\right]= \frac{108}{16\pi G_5} 
\int_{\calm_5} {\rm vol}_{\calm_5}\ \Omega_1 \Omega_2^2\om_3^2\ \times\\
&\times \biggl\lbrace 
 R_{10}-\frac 12 \left(\nabla \Phi\right)^2
-\frac 12 e^{-\Phi}\left(\frac{(h_1-h_3)^2}{2\om_1^2\om_2^2\om_3^2}+\frac{1}{\om_3^4}\left(\nabla h_1\right)^2
+\frac{1}{\om_2^4}\left(\nabla h_3\right)^2\right)
\\
&-\frac 12 e^{\Phi}\left(\frac{2}{\om_2^2\om_3^2}\left(\nabla h_2\right)^2
+\frac{1}{\om_1^2\om_2^4}\left(h_2-\frac P9\right)^2
+\frac{1}{\om_1^2\om_3^4} h_2^2\right)
\\
&-\frac {1}{2\Omega_1^2\Omega_2^4\om_3^4}\left(4\Omega_0+ h_2\left(h_3-h_1\right)+\frac 19 P h_1\right)^2
\biggr\rbrace\,,\\
\end{split}
\eqlabel{5action}
\end{equation}
where $\Omega_0$ is a constant\footnote{In the conformal limit of the
cascading gauge theory, $\Omega_0=\frac{L^4}{108}$,
where $L$ is the asymptotic $AdS_5$ radius.},
and $R_{10}$ is given by
\begin{equation}
\begin{split}
R_{10}=R_5&+\left(\frac{1}{2\om_1^2}+\frac{2}{\om_2^2}+\frac{2}{\om_3^2}-\frac{\om_2^2}{4\om_1^2\om_3^2}
-\frac{\om_3^2}{4\om_1^2\om_2^2}-\frac{\om_1^2}{\om_2^2\om_3^2}\right)-2\Box \ln\left(\om_1\om_2^2\om_3^2\right)\\
&-\biggl\{\left(\nabla\ln\om_1\right)^2+2\left(\nabla\ln\om_2\right)^2
+2\left(\nabla\ln\om_3\right)^2+\left(\nabla\ln\left(\om_1\om_2^2\om_3^2\right)\right)^2\biggr\}\,,
\end{split}
\eqlabel{ric5}
\end{equation}
and $R_5$ is the five-dimensional Ricci scalar of the metric on $\calm_5$,
\begin{equation}
ds_{5}^2 =g_{\mu\nu}(y) dy^{\mu}dy^{\nu}\,.
\eqlabel{5met}
\end{equation}
$P$ is the other constant, and is related to the rank-difference of the cascading gauge theory group factors
$M$ as
\begin{equation}
M\equiv \frac{2P}{9\alpha'}\ \in \zet\,,
\eqlabel{defm}
\end{equation}
where $\a'=\ell_s^2$ is the string scale.
Finally, $G_5$ is the five dimensional effective gravitational constant  
\begin{equation}
G_5=\frac{27}{16\pi^3}\ G_{10}\,,
\eqlabel{g5deff}
\end{equation}
where $16 \pi G_{10}=(2\pi)^7g_s^2(\a')^4$ is  the 10-dimensional gravitational constant of 
type IIB supergravity, and $g_s$ is the asymptotic string coupling constant, which we set to 1.

de Sitter vacua and DFPs of the cascading gauge theory are holographically dual
to the solutions 
of the effective action \eqref{5action}, when the boundary metric
is de Sitter with the Hubble constant $H$,
\begin{equation}
ds^2\bigg|_{\calm_4=\del\calm_5}=-d\tau^2 +e^{2 H \tau}{d\bm{x}^2}\,,
\eqlabel{bmetric}
\end{equation}
and all the 7 gauge invariant scalar operators of the theory
$\{\calo_3^{\alpha=1,2},\calo_4^{\beta=1,2},\calo_6,\calo_7,\calo_8\}$
develop  a spatially constant, $\bm{x}$, and time-independent, $\tau$, expectation values.
There are two equivalent ways to represent these cascading gauge theory
states in the dual gravitational bulk:
\nxt Using the Fefferman-Graham (FG) coordinate frame,
\begin{equation}
\begin{split}
&ds_{5}^2 = \frac{1}{h^{1/2}\rho^{2}} \left(-d\tau^2 + e^{2 H \tau}d\boldsymbol{x}^2\right)+
\frac{h^{1/2}}{\rho^{2}} \left(d\rho\right)^2\,,\qquad h=h(\rho)\,,\\
&\Omega_{i=1,2,3}=\Omega_{i=1,2,3}(\rho)\,,\qquad h_{i=1,2,3}=h_{i=1,2,3}(\rho)\,,\qquad \Phi=\Phi(\rho)\,,
\end{split}
\eqlabel{5dfm} 
\end{equation}
with the radial coordinate $\rho$, the $\del\calm_5$ boundary is located at $\rho\to +0$,
\begin{equation}
\rho \in (0,+\infty)\,.
\eqlabel{rangerho}
\end{equation}
Close to the boundary the metric warp factor $h$ takes the form, 
\begin{equation}
h=\frac 18 b +\frac 14 K_0 -\frac 12 b\ln\rho+\calo(\rho\ln\rho)\,,
\eqlabel{hass}
\end{equation}
where $b\equiv P^2$ and
$K_0$ is related to strong coupling scale $\Lambda$ of the cascading gauge theory
as 
\begin{equation}
\Lambda^2=\frac{1}{b}\ e^{-\frac{K_0}{b}}\,.
\eqlabel{deflambda}
\end{equation}
DFPs, TypeA, are such nonsingular gravitational solutions
that
\begin{equation}
{\rm TypeA}:\qquad \lim_{\rho\to \infty} \frac{1}{h^{1/2}\rho^2} = 0\,,
\eqlabel{dfplimit}
\end{equation}
with all the scalars being finite in this limit.
There are two distinct types of the cascading gauge theory DFPs:
TypeA$ _s$ and TypeA$ _b$. The former preserve the $U(1)_R$ (in the large-$N$ supergravity
approximation) chiral symmetry, while the latter spontaneously breaks it to $\zet_2$,
\begin{equation}
\begin{split}
&{\rm TypeA}_s:\qquad \Omega_2\equiv \Omega_3\qquad {\rm and}\qquad h_1\equiv h_3\qquad {\rm and}\qquad h_2\equiv \frac{P}{18}\,,\\
&{\rm TypeA}_b:\qquad \Omega_2\not\equiv \Omega_3\qquad {\rm and}\qquad h_1\not\equiv h_3\qquad {\rm and}\qquad \frac{d}{d\rho} h_2
\not\equiv 0\,.
\end{split}
\eqlabel{detabbyc}
\end{equation}
de Sitter vacua, TypeB, are nonsingular gravitational solutions within the ansatz  \eqref{5dfm},
such that
\begin{equation}
{\rm TypeB}:\qquad \lim_{\rho\to \infty} \Omega_3 = 0\,,
\eqlabel{deblimit}
\end{equation}
with all the other scalars, as well as the $g_{\tau\tau}\equiv -\frac{1}{h^{1/2}\rho^2}$ metric component,
being finite in this limit.
\nxt Using the 
Eddington-Finkelstein (EF) coordinate frame, 
\begin{equation}
\begin{split}
&ds_{5}^2 = 2dt\ \left(dr-a\ dt\right)+\sigma^2 e^{2 Ht}\ d\bm{x}^2\,,\qquad a=a(r)\,,\qquad \sigma=\sigma(r)\,, \\
&\Omega_{i=1,2,3}=\Omega_{i=1,2,3}(r)\,,\qquad h_{i=1,2,3}=h_{i=1,2,3}(r)\,,\qquad \Phi=\Phi(r)\,,
\end{split}
\eqlabel{ef1i}
\end{equation}
with the radial coordinate $r$, the $\del\calm_5$ boundary is located now at $r\to +\infty$,
\begin{equation}
r \in [r_{AH},+\infty)\,,
\eqlabel{ranger}
\end{equation}
and $r_{AH}$ is the location of the apparent horizon in the uplifted 10-dimensional type IIB  supergravity
background, see \cite{Buchel:2019pjb} for detailed discussion,
\begin{equation}
\biggl[\ 3H\cdot \left(\sigma^3 \Omega_1\Omega_2^2\Omega_3^2\right)
+ a\cdot \frac{d}{dr} \left(\sigma^3 \Omega_1\Omega_2^2\Omega_3^2\right)\biggr]\ \bigg|_{r=r_{AH}}=0\,.
\eqlabel{rahf}
\end{equation}
It can be shown that the radial derivative in \eqref{rahf},
provided $\sigma^3 \Omega_1\Omega_2^2\Omega_3^2$ does not vanish ---
which is the case for both TypeA$ _s$ and TypeA$ _b$ DFPs,
is always positive, thus
\begin{equation}
{\rm TypeA}:\qquad a\bigg|_{r=r_{AH}} <0\,,
\eqlabel{nega}
\end{equation}
which further implies that there must be a point $r=r_0>r_{AH}$, such that
\begin{equation}
a\bigg|_{r=r_0}=0\,.
\eqlabel{zeroa}
\end{equation}
In the EF frame description of the cascading gauge theory de Sitter vacua,
\ie TypeB, the apparent horizon is located where $\Omega_3$ vanishes;
this occurs at positive $a$,
\begin{equation}
{\rm TypeB}:\qquad a\bigg|_{r=r_{AH}} >0\,.
\eqlabel{posa}
\end{equation}
EF frame description of the DFPs (or de Sitter vacua) links them directly with
the late-time attractors for the evolution of
the homogeneous and  isotropic states\footnote{The restriction to spatially homogeneous and isotropic
states, rather than {\it any} states, is likely not necessary for the evolution in de Sitter background,
where momentum scale $k$ inhomogeneities are red-shifted as ${k}{e^{-H\tau}}$.} of the boundary gauge theory
\cite{Buchel:2017pto}: specifically, a holographic dual to such an evolution is a gravitational dynamics
of \eqref{5action} with the ansatz
\begin{equation}
\begin{split}
&ds_{5}^2 = 2dt\ \left(dr-A\ dt\right)+\Sigma^2\ d\bm{x}^2\,,\ \qquad A=A(t,r)\,,\qquad \Sigma=\Sigma(t,r)\,,\\
&\Omega_{1,2,3}=\Omega_{1,2,3}(t,r)\,,\qquad h_{1,2,3}=h_{1,2,3}(t,r)\,,\qquad \Phi=\Phi(t,r)\,,
\end{split}
\eqlabel{ef2i}
\end{equation}
leading to\footnote{See \cite{Buchel:2017lhu,Buchel:2021ihu} for examples of implementation of such dynamics.} 
\begin{equation}
\lim_{t\to \infty}\biggl\{A(t,r)\,,\,\frac{\Sigma(t,r)}{e^{H t}}\,,\, 
\Omega_i(t,r)\,,\, h_{i}(t,r)\,,\, \Phi(t,r)\biggr\}=\left\{a(r)\,,\,\sigma(r)\,,\, \Omega_i(r)\,,\, h_i(r)\,,\, \Phi(r)\right\}
\eqlabel{ltlimit1}
\end{equation}
of \eqref{ef1i}.

\begin{figure}[t]
\begin{center}
\psfrag{a}[cc][][1][0]{${H_{crit_1}}$}
\psfrag{q}[cc][][1][0]{TypeB}
\psfrag{w}[cc][][1][0]{TypeA$ _b$}
\psfrag{r}[cc][][1][0]{TypeA$ _s$}
\psfrag{f}[cc][][1.5]{${H}$}
\psfrag{x}{{$\ln \frac{H^2}{\Lambda^2}$}}
\psfrag{y}[bb]{{$4 G_5\ \hat{s}_{ent}$}}
\includegraphics[width=4in]{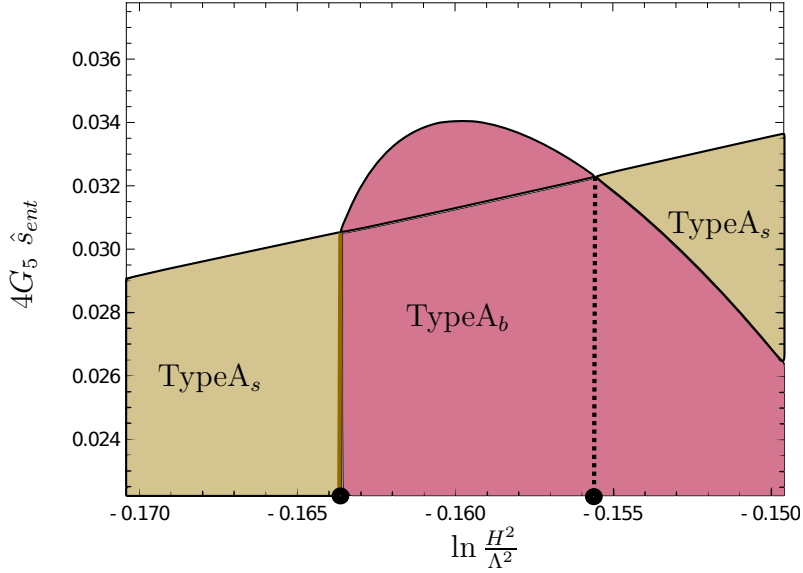}
\end{center}
  \caption{From \cite{Buchel:2019pjb}. Vacuum entanglement entropy densities of
  the chirally symmetric DFP (TypeA$ _s$), and the DFP with spontaneously
  broken chiral symmetry (TypeA$ _b$), as a function of $\ln\frac{H^2}{\Lambda^2}$. 
} \label{figure0}
\end{figure}

Note that besides distinct radial coordinates, $\rho$ (FG) and $r$ (EF),
we used  different bulk times  $\tau$ (FG) and $t$ (EF) in the two frames \eqref{5dfm}
and \eqref{ef1i}. There is a simple coordinate transformation mapping the
{\it full} DFP FG frame geometry, \ie  \eqref{5dfm} with \eqref{dfplimit},
to the $r\in[r_0,+\infty)$ {\it patch}  of the corresponding EF frame geometry
\cite{Buchel:2019pjb},
\begin{equation}
\begin{split}
&r-r_{0}=\frac 1\rho\,,\qquad t=\tau-\int_{0}^\rho\ dz\  \sqrt{h(z)}\,,\\
&a=\frac{1}{2h^{1/2}\rho^2}\,,\qquad \sigma=\frac{1}{\rho h^{1/4}}\ \exp\biggl[
H\ \int_{0}^\rho dz\sqrt{h(z)}
\biggr]\,.
\end{split}
\eqlabel{mapfgef}
\end{equation}
The $r\in[ r_{AH},r_0)$ patch of the DFP EF frame geometry is invisible
in the FG frame. Arguably, EF frame description of a dynamical fixed point
is more important, as its vacuum entanglement entropy density,
relatedly the late-time limit of the entropy density production rate of this DFP
\eqref{rlim}, is identified with the comoving
gravitational entropy density of the
apparent horizon in the corresponding  holographic dual, see eq.(3.9) of
\cite{Buchel:2019pjb},
\begin{equation}
s_{ent}=H^3 b^2\ \hs_{ent}=\frac{1}{4G_5}\ 108\sigma^3\Omega_1\Omega_2^2\Omega_3^2
\bigg|_{r=r_{AH}}\,.
\eqlabel{sent}
\end{equation}
Fig.~\ref{figure0} reproduces the main result of the \cite{Buchel:2019pjb}:
it compares the vacuum entanglement entropy densities of
the chiral symmetry preserving DFP, TypeA$ _s$, and the DFP with spontaneously
broken chiral symmetry, TypeA$ _b$. TypeA$ _b$ exists only for $H>H_{crit_3}$,
represented by a vertical solid brown line,  and is the preferred
late-time attractor for $H<H_{crit_2}$, represented by a dashed black line, see
\eqref{ranges}. Outside the range $H\in (H_{crit_3},H_{crit_2})$, and whenever it
exists, \ie for $H>H_{crit_1}$,  TypeA$ _s$ DFP is the preferred
attractor of the late-time dynamics.

Under the bulk  diffeomorphism \eqref{mapfgef}, the full EF frame background geometry
corresponding to the cascading gauge theory  de Sitter vacua (TypeB) is mapped
to its full corresponding FG frame background geometry. Here,
the vacuum entanglement entropy density vanishes  \cite{Buchel:2019pjb},
\begin{equation}
s_{ent}=H^3 b^2\ \hs_{ent}
\bigg|_{r=r_{AH}}^{\rm TypeB}=0\,,
\eqlabel{sentb}
\end{equation}
\ie at late-times, the entropy density production rate vanishes.
Since the vacuum entanglement entropy density of a dynamical fixed point is always nonzero,
whenever a DFP exists, it is the preferred late-time dynamical attractor,
compare to a de Sitter vacuum at the same Hubble constant.
There are no DFPs of the cascading gauge theory for $H<H_{crit_1}$, see \eqref{ranges}.

\section{Stability analysis framework of the cascading gauge theory de Sitter DFPs}
\label{hard}

Once a de Sitter DFP of a QFT is identified, it is important to analyze its stability
to claim that it is indeed a late-time attractor. A DFP is {\it always}
the preferred late-time state compare to a de Sitter vacuum of a QFT,
however, a DFP can be unstable \cite{Buchel:2021ihu}, in which case
the late-time dynamics is unknown\footnote{It
is definitely not the de Sitter vacuum though!}.

In a holographic setting, it is most natural to analyze stability of a
DFP in the Eddington-Finkelstein coordinate frame of the gravitational dual
\cite{Buchel:2017lhu}. Suppose a holographic QFT in $d$ spatial dimensions has
a de Sitter dynamical fixed point,
\begin{equation}
\begin{split}
&ds_{d+2}^2 = 2dt\ \left(dr-a\ dt\right)+\sigma^2 e^{2 Ht}\ d\bm{x}^2\,,
\end{split}
\eqlabel{efd}
\end{equation}
supported by the bulk scalars $\phi_j=\phi_j(r)$.
We study, homogeneous and isotropic along the
spatial boundary directions, linearized fluctuations $\{F_a$, $F_\sigma$, $F_j\}$ about
the background \eqref{efd},
\begin{equation}
a(r)\to a(r)+F_a(r)e^{-i \omega t}\,,\quad \sigma(r)\to
\sigma(r)+F_\sigma(r)e^{-i \omega t}\,,\quad \phi_j(r)\to \phi_j(r)+F_j(r)e^{-i \omega t}\,,
\eqlabel{flucd}
\end{equation}
Imposing a normalizability of the fluctuations at the asymptotic boundary,
and regularity of the spatial profiles of  $\{F_a$, $F_\sigma$, $F_j\}$
in the background \eqref{efd} as $r\in [r_{AH},\infty)$, we can compute the
spectrum of fluctuations, \ie the set of frequencies $\{\omega\}$.
Since apparent horizon is dissipative, the frequencies will be complex.
Any fluctuation mode with
\begin{equation}
\Im[\omega] > 0
\eqlabel{inst}
\end{equation}
signals an instability of the DFP, represented by \eqref{efd}.

Unfortunately, the above prescription can not be applied to the stability
analysis of the cascading gauge theory DFPs, reviewed in section
\ref{dfpreview}. The stumbling block is the relation between the
EF and the FG frame time coordinates \eqref{mapfgef},
which, given the asymptotic expansion for $h$ \eqref{hass},
makes the EF frame $r\to\infty \Longleftrightarrow \rho\to 0 $
boundary asymptotics intractable\footnote{The presence of
high dimension operators of the cascading gauge theory, such as $\calo_6$,
$\calo_7$ 
and $\calo_8$, requires exquisite control of the asymptotic
boundary data.}. The prescription to circumvent this difficulty was
introduced in \cite{Buchel:2019qcq}.
The cascading gauge theory DFPs are constructed in the FG coordinate frame
\cite{Buchel:2019pjb}.  To compute the vacuum entanglement entropy,
the region of the FG geometry in the vicinity of $r=r_0\Longleftrightarrow \rho=\infty$,
see \eqref{mapfgef}, is mapped into EF coordinate frame, and further extended in this
frame for $r\in [r_{AH},r_0]$. Additional complexities of the EF frame appear
when one studies linearized fluctuations, as in \eqref{flucd}: here, one needs to solve
equations not only for the bulk scalar fluctuations $F_j$, but for the fluctuations
of the metric components as well, \ $\{F_a,F_\sigma\}$.

In \cite{Buchel:2022hjz} we explained how to compute the spectrum
of fluctuations about a DFP directed in the FG coordinate frame,
for any holographic model with an arbitrary ``$d+2$ dimensional Einstein gravity plus
arbitrary bulk scalars''.
The computational framework presented there is highly efficient:
one needs to solve only the fluctuation equations for the bulk scalars,
while the fluctuations of the metric components are determined
algebraically from the latter. Unfortunately, this master equation
framework can not be directly applied to the cascading gauge theory gravitational
dual. Here, the issue is that the holographic models of \cite{Buchel:2022hjz}
must have the standard Einstein-Hilbert term in the gravitational action,
while in the cascading gauge theory gravitational dual the Einstein-Hilbert term is
warped \eqref{5action}:
\begin{equation}
S_{d+2}\propto
\underbrace{\int_{\calm_{d+1}}d^{d+2}\xi \sqrt{-g}\biggl[R+\cdots\biggr]}_{\rm master\ equations}
\ \ {\rm vs.}\ \ S_{5}\propto
\underbrace{\int_{\calm_{5}}\vol_{\calm_5}\ \Omega_1\Omega_2^2\Omega_3^2
\biggl[R_5+\cdots\biggr]}_{\rm cascading\ gauge\ theory}\,.
\eqlabel{mastervscas}
\end{equation}
Of course, we can always Weyl rescale the metric to remove the Einstein-Hilbert
term warp factor, but this would require a new complicated differential relation
between the FG frame radial coordinates, involving fractional powers of $h$. 
This causes the same problems as we faced in the EF coordinate frame:
the boundary $\rho\to 0$ asymptotics become intractable; additionally,
the change of variables dramatically complicates the
master equations for the fluctuations.

Above difficulty is resolved noting that the effective  five-dimensional
gravitational action \eqref{5action} is a Kaluza-Klein reduction of
Type IIB supergravity on warped deformed conifold with fluxes.
Thus, we should be apply the apply the master equations formalism of
\cite{Buchel:2022hjz}, more precisely its obvious variation,
in ten dimensions without any problem.
This is what we do in appendix\footnote{While the discussion
there is attempted to be self-contained,  the reader does need
familiarity with the formalism of \cite{Buchel:2022hjz}.}
\ref{masterfluc}.

We finish this section highlighting the
subtlety developing the near-boundary $\rho\to 0$ asymptotic expansions
of the equations representing the fluctuations.
The equation of motion for a probe massive bulk scalar field dual to an operator of
conformal dimension $\Delta$, on $AdS_5$ background geometry takes the form,
\begin{equation}
\phi=\rho^\Delta\biggl(A_0+\sum_{k=1}^\infty A_k \rho^k\biggr)
+ \rho^{4-\Delta}\biggl(B_0+\sum_{k=1}^\infty B_k \rho^k\biggr)\,.
\eqlabel{asy}
\end{equation}
When $\Delta\in \zet$ or $\Delta\in \zet_{n+\frac 12}$ logarithmic terms appear in this asymptotic
expansion, \ie the series in brackets generalize as
\begin{equation}
\sum_{k=1}^\infty A_k \rho^k\ \to\ \sum_{k=1}^\infty\ \rho^k \sum_{m=0}^{M(k)} A_{k;m} \ln^m\rho \,.
\eqlabel{genser}
\end{equation}
It is important that the number of $\ln\rho$ terms at each fixed order in $k$ is bounded by $M(k)$.
In fact, the metric ansatz \eqref{5dfm} for the cascading gauge theory, along with
the ansatz for the scalars $\Omega_{1,2,3}\propto h^{1/4} f_{a,b,c}^{1/2}$ was proposed
in \cite{Aharony:2005zr}
precisely so that the asymptotic expansions of the metric warp factor $h$, as well as the
scalars $f_{a,b,c}$, have finite number of log-terms at each given order of $\rho^k$.
This is evident in the asymptotic expansions of the background geometry dual to
the cascading gauge theory DFPs, reviewed in appendix \ref{appasymptotics}.
Finite number of log-terms in the asymptotic expansion is a fairly trivial complication.
Rather, we find that the master formalism for the fluctuations, see appendix \ref{masterfluc},
leads to an infinite number of log-terms in their asymptotic expansions at each
finite order of $\rho^k$. In other words, the generalization \eqref{genser}
is yet further generalized:
\begin{equation}
\sum_{k=1}^\infty\ \rho^k \sum_{m=1}^{M(k)} A_{k;m} \ln^m\rho\ \to\ \sum_{k=1}^\infty\ \rho^k\cdot
\cala_k(\ln\rho)\,,
\eqlabel{genser2}
\end{equation}
where $\cala_k(z)$ are now nontrivial functions of $z\equiv\ln \rho$, and in developing
the asymptotic expansions, at each order $\rho^k$, we must solve a coupled system
(if there is more than one bulk scalar)
of differential equations for $\cala_k(z)$. This would be a hopeless task in general.
Lucking for the problem at hand, carefully analyzing the structure of log-term
differential equations we find that their solution is given by (schematically)
\begin{equation}
\cala_k(z)=\frac{1}{(b-4bz +2 K_0)^{n(k)}}\ \sum_{m=0}^{M(k)} A_{k;m}\ z^m\,,
\eqlabel{solveak}
\end{equation}
where $n(k)$ and $M(k)$ are some integers $\sim k$. The denominator factor in \eqref{solveak}
is simply the order $\calo(\rho^0)$ terms of asymptotic expansion of the $h$ factor,
see \eqref{hass}.

\section{Stability analysis of TypeA$ _s$ DFP}
\label{stypeas}

TypeA$ _s$ dynamical fixed point of the cascading gauge theory preserves the chiral symmetry.
There are two decoupled sets of fluctuations about this DFP:
the fluctuations breaking the chiral symmetry ( see section \ref{sec4uns}
with technical details in appendix \ref{ascsb} ), and the fluctuations preserving 
the chiral symmetry  ( see section \ref{sec4s}
with technical details in appendix \ref{ascs} ). 

TypeA$ _s$ DFPs were constructed in various computations
schemes (see appendix C.1 of \cite{Buchel:2019pjb}): either
parameterized by $b$ with $K_0=1$, see \eqref{deflambda},  
\begin{equation}
\ln\frac{H^2}{\Lambda^2}=\frac 1b+\ln b\,,\qquad b\in (0,1]\,,
\eqlabel{bpar1}
\end{equation}
or  with $b=1$ and
\begin{equation}
\ln\frac{H^2}{\Lambda^2}=K_0\,.
\eqlabel{bpar2}
\end{equation}
An excellent agreement was reported in the overlap of the two computational schemes,
\ie for $K_0>1$. The parameterization \eqref{bpar1} is useful to analyze $b\to 0$,
correspondingly $H\gg \Lambda$, near-conformal limit, where perturbative in $b$
treatment is possible. The parameterization \eqref{bpar2} is needed to access TypeA$ _s$
DFP in  $H<\Lambda$ region, not accessible with \eqref{bpar1}.  
We use the same strategy in computing the spectra of fluctuations:
first we perform the computations in the near-conformal limit,
and further extend the results for $H<\Lambda$.

\subsection{Chiral symmetry breaking sector}
\label{sec4uns}

\begin{figure}[t]
\begin{center}
\psfrag{h}[t]{{$1/\ln\frac{H^2}{\Lambda^2}$}}
\psfrag{w}{{$\Gamma_{\csb}^{(3)}$}}
\psfrag{e}[t]{{$\Gamma_{\csb}^{(4)}$}}
\includegraphics[width=3in]{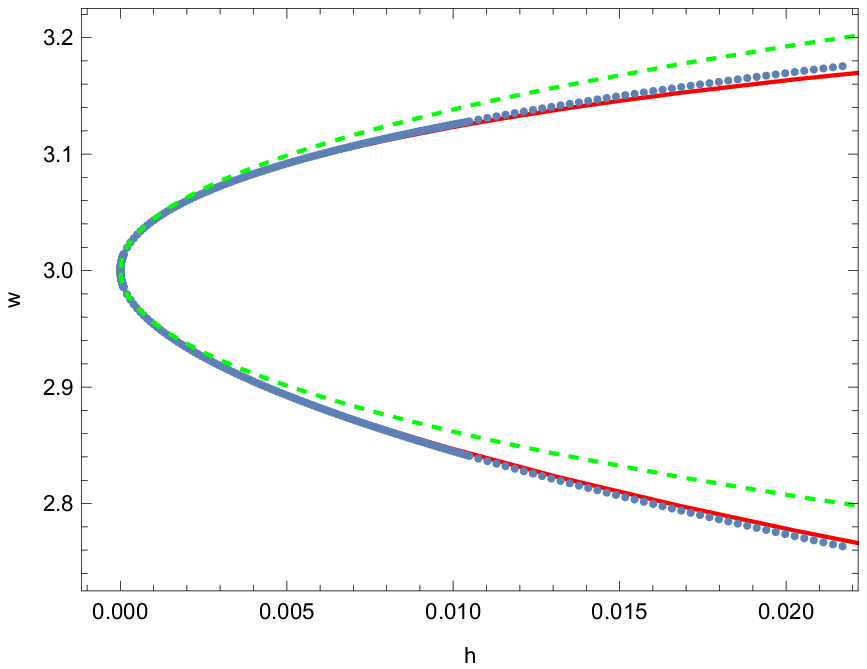}
\includegraphics[width=3in]{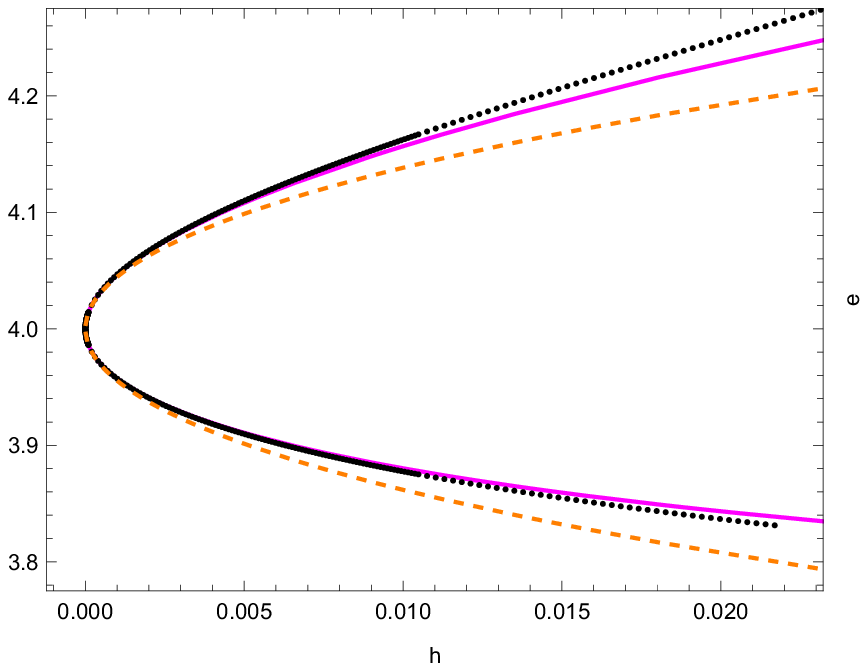}
\end{center}
  \caption{Attenuation $\Gamma_{\csb}^{(n)}\equiv-\Im[\ww_\csb^{(n)}]$ of the
  chiral symmetry breaking fluctuations about cascading gauge theory
  TypeA$ _s$ DFP for $H\gg \Lambda$. Dashed and solid curves
  are correspondingly the leading and the first subleading
  order corrections to the conformal spectra, see \eqref{ancsb34}.
} \label{figure1}
\end{figure}

Chiral symmetry breaking fluctuations about TypeA$ _s$ DFP
activate the cascading gauge theory
operators of conformal dimensions $\Delta=\{3,7\}$.
Thus, in the near conformal limit, \ie for $H\gg \Lambda$,
we expect \cite{Buchel:2022hjz} discrete branches
indexed with $n\in \zet_{\ge 3}$ and
\begin{equation}
\Re[\ww_{\csb}^{(n)}]\ =\ 0\,,\qquad \Im[\ww_\csb^{(n)}]\
\equiv\  -\Gamma_{\csb}^{(n)}\ \ne 0\,.
\eqlabel{csbn}
\end{equation}
We use the subscript $ _\csb$ to indicate that the fluctuations spontaneously
break chiral symmetry of TypeA$ _s$ DFP. We find that the branches with
$3\le n\le 6$ are doubly degenerate in the limit $b\to 0$, while those
with $n\ge 7$ are triple degenerate in the conformal limit.
In fig.~\ref{figure1} dots represent the attenuation $\Gamma_{\csb}^{(n)}$
as a function of $\ln^{-1}\frac{H^2}{\Lambda^2}$ for the lowest $n=3$
mode (the left panel) and the $n=4$ mode (the right panel). The dashed
curves indicate $\calo(\sqrt{b})$ analytic leading order corrections,
see appendix \ref{detailss3}, and the solid lines include next-to-leading
$\calo(b)$ order corrections:
\begin{equation}
\begin{split}
&\Gamma_{\csb}^{(3)}=3\pm \sqrt{2b}-1.57(5)\cdot b\pm \calo(b^{3/2})\,,\\
&\Gamma_{\csb}^{(4)}=4\pm \sqrt{2b}+1.93(4)\cdot b\pm \calo(b^{3/2})\,,
\end{split}
\eqlabel{ancsb34}
\end{equation}
where $b$ is related to $\frac H\Lambda$ as in \eqref{bpar1}.

\begin{figure}[t]
\begin{center}
\psfrag{h}{{$1/\ln\frac{H^2}{\Lambda^2}$}}
\psfrag{w}{{$\Gamma_{\csb}$}}
\includegraphics[width=4in]{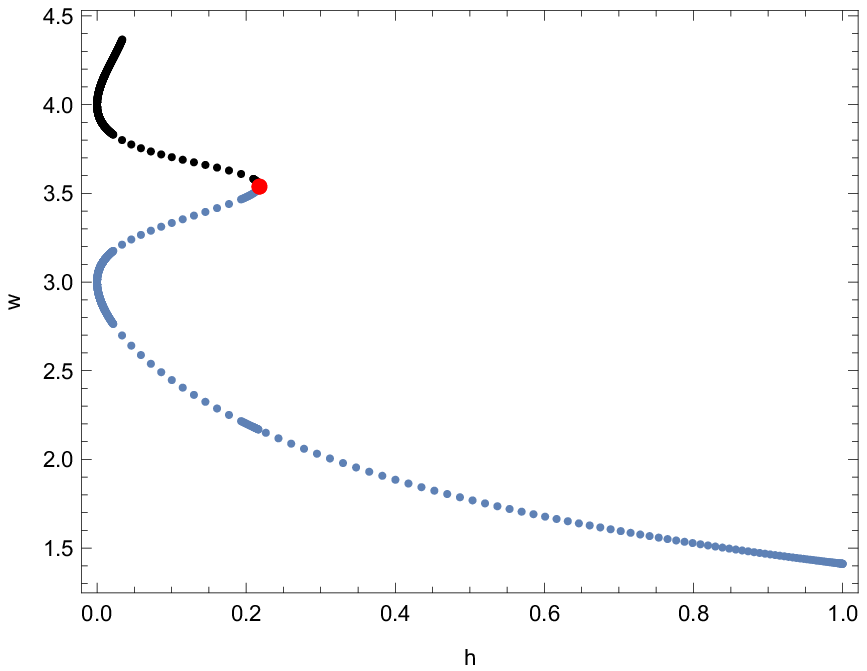}
\end{center}
  \caption{Sub-branches of the distinct in the conformal limit
  branches of the fluctuations
  coalesce as $\frac{H}{\Lambda}$ is lowered. The red dot, see
  \eqref{csbreddot}, highlights
  this phenomenon for $n=4$ and $n=3$ sub-branches.
} \label{figure2}
\end{figure}

As $b$ increases, we discover that the distinct branches of the
fluctuations coalesce, see fig.~\ref{figure2}. Specifically we find that
the lower sub-branch of the $n=4$ branch and the upper sub-branch of
the $n=3$ branch combine at
\begin{equation}
\ln^{-1}\ \frac{H^2}{\Lambda^2}\ =\ 0.217(8)\,,
\eqlabel{csbreddot}
\end{equation}
represented by the red dot, and are removed from the spectrum.
This phenomenon is quite generic,
and is observed for higher $n$ branches as well. 
It can not be universal though: the lower sub-branch of
the $n=3$ branch is the lowest mode in the spectrum, thus, it 
does not have a partner to combine with.

\begin{figure}[t]
\begin{center}
\psfrag{h}[t]{{$1/\ln\frac{H^2}{\Lambda^2}$}}
\psfrag{w}{{$\Gamma_{\csb}^{(n)}$}}
\psfrag{e}[t]{{$\Gamma_{\csb}^{(n)}$}}
\includegraphics[width=3in]{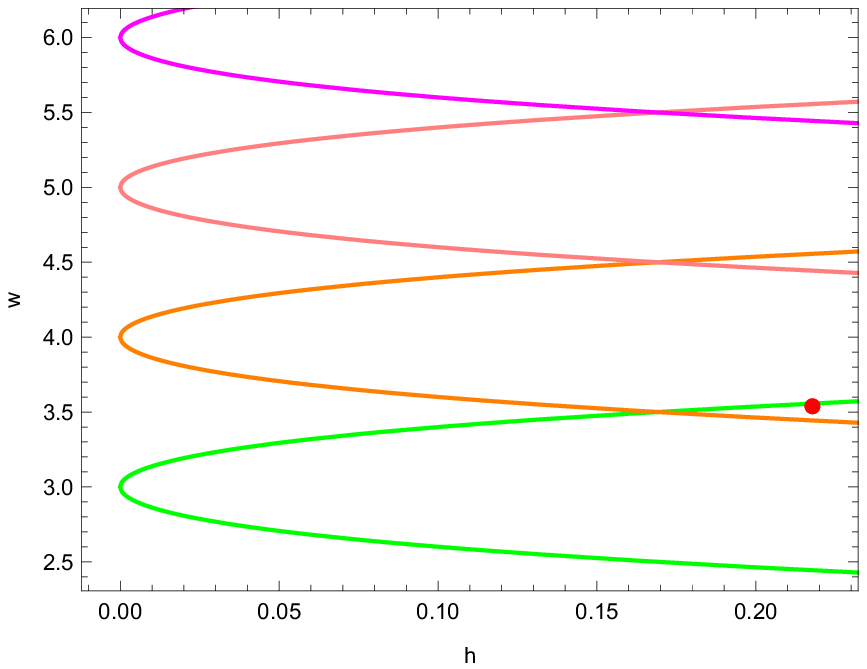}
\includegraphics[width=3in]{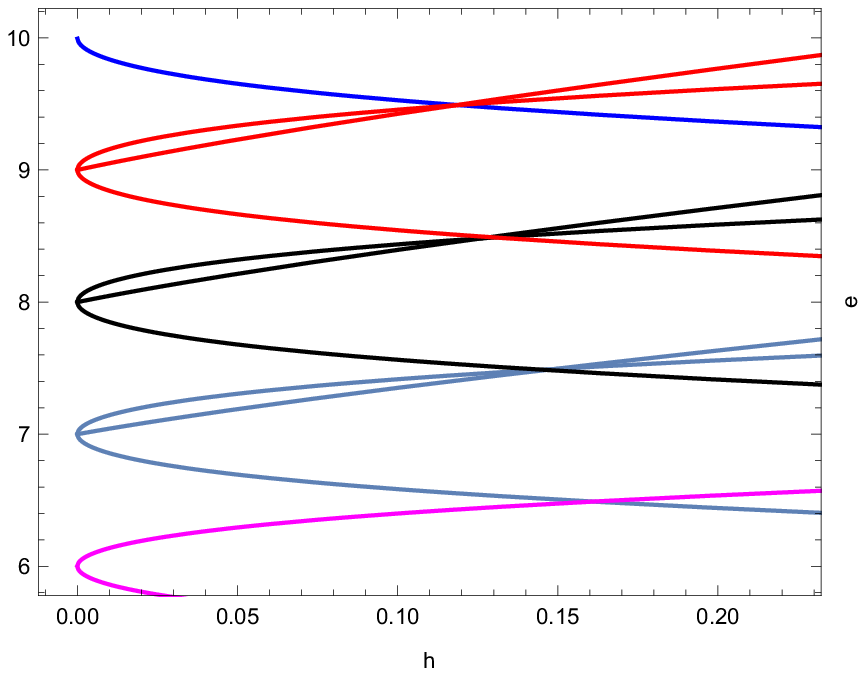}
\end{center}
  \caption{Leading order correction to the conformal
  spectra for the chiral symmetry breaking fluctuations
  at higher $n$. Note that the (perturbative) coalescence of
  various sub-branches
  is quite generic. The red dot (the left panel) is a replot
  of the red dot from fig.~\ref{figure2}.
} \label{figure3}
\end{figure}

It becomes numerically challenging to study
higher $n$ spectral branches at finite $\frac{H}{\Lambda}$.
In particular, we could not stabilize numerics at $n=7$ branch
where the first triple degeneracy occurs. There is no
obstruction to study these branches perturbatively
in the small $b$, the near conformal limit, \eg
see appendices \ref{detailss7} and \ref{detailss7l}. 
In fig.~\ref{figure3} we present leading order correction
to the conformal spectra for $3\le n \le 6$ (the left panel)
and for $6\le n\le 10$ (the right panel).
Note that the non-analytic sub-branches, see appendix
\ref{appnearcsym}, (perturbatively) combine --- as the red dot
(the left panel), replotted from fig.~\ref{figure2}, indicates
the perturbative prediction
 for the coalescence is quite
 reasonable\footnote{Our numerical work, not reported here,
 established joining of $n=5$ and $n=4$, as well as $n=6$ and $n=5$
 sub-branches.}. It appears (the right panel) that
 for $n\ge 7$ the coalescence point involved three sub-branches ---
 this is not the case, as the better resolution of the plots demonstrates.

\begin{figure}[t]
\begin{center}
\psfrag{h}{{$\ln\frac{H^2}{\Lambda^2}$}}
\psfrag{w}{{$\Gamma_{\csb}$}}
\includegraphics[width=4in]{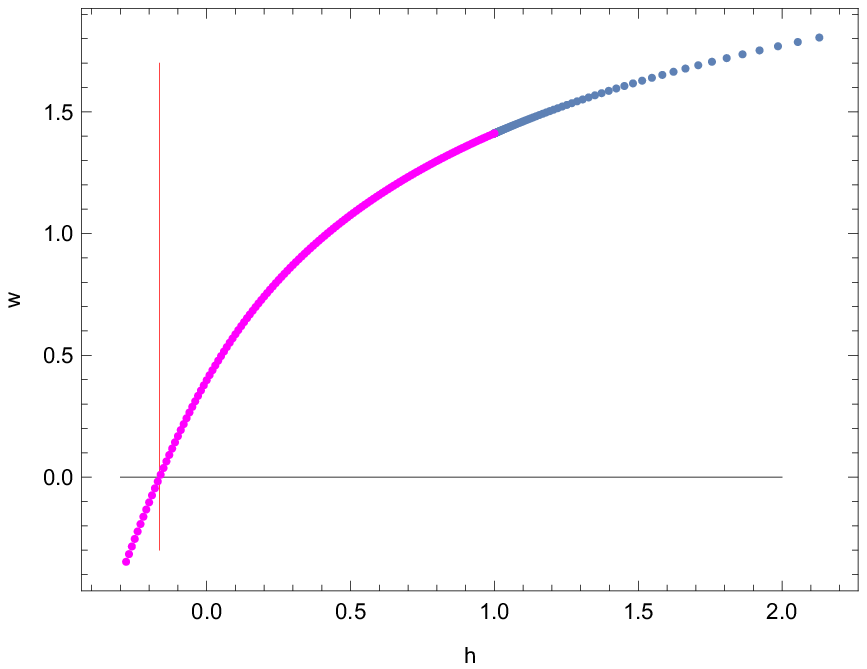}
\end{center}
  \caption{The lower sub-branch of the $n=3$ branch of
  chiral symmetry breaking fluctuations about TypeA$ _s$ DFP
  of the cascading gauge theory 
  becomes unstable for $H<H_{crit_3}$ \eqref{h3l},
  represented by the red vertical line.
} \label{figure4}
\end{figure}

The lower sub-branch of the $n=3$ branch is the lowest lying.
In fig.~\ref{figure2} we followed this branch all the way to $b=1$,
correspondingly to $\ln\frac{H^2}{\Lambda^2}=1$, see \eqref{bpar1}.
In fig.~\ref{figure4} we switch the computational scheme to
that of \eqref{bpar2}, and follow this sub-branch for $H<\Lambda$,
represented by the magenta dots. This mode becomes marginal
at
\begin{equation}
\ln \frac{H_{crit_3}^2}{\Lambda^2}=-0.1636(3)\,,
\eqlabel{h3l}
\end{equation}
represented by the vertical red line, reproducing the
critical Hubble constant $H_{crit_3}$, corresponding to
the origin of the TypeA$ _b$ dynamical fixed point with the
spontaneously broken chiral symmetry, originally
reported in \cite{Buchel:2019pjb}.
Note that for $H<H_{crit_3}$ this mode becomes unstable.
This establishes our first main result:

\bigskip

\noindent\fbox{%
    \parbox{\textwidth}{%
{\color{red} The chirally symmetric TypeA$ _s$ DFP of the cascading gauge theory
is perturbative unstable when $H<H_{crit_3}$, given by \eqref{h3l}. 
 }
}%
}
\bigskip

\subsection{Chiral symmetry preserving sector}
\label{sec4s}

\begin{figure}[ht]
\begin{center}
\psfrag{h}[t]{{$1/\ln\frac{H^2}{\Lambda^2}$}}
\psfrag{w}{{$\Gamma_{\cs}$}}
\psfrag{e}[t]{{$\Gamma_{\cs}$}}
\includegraphics[width=3in]{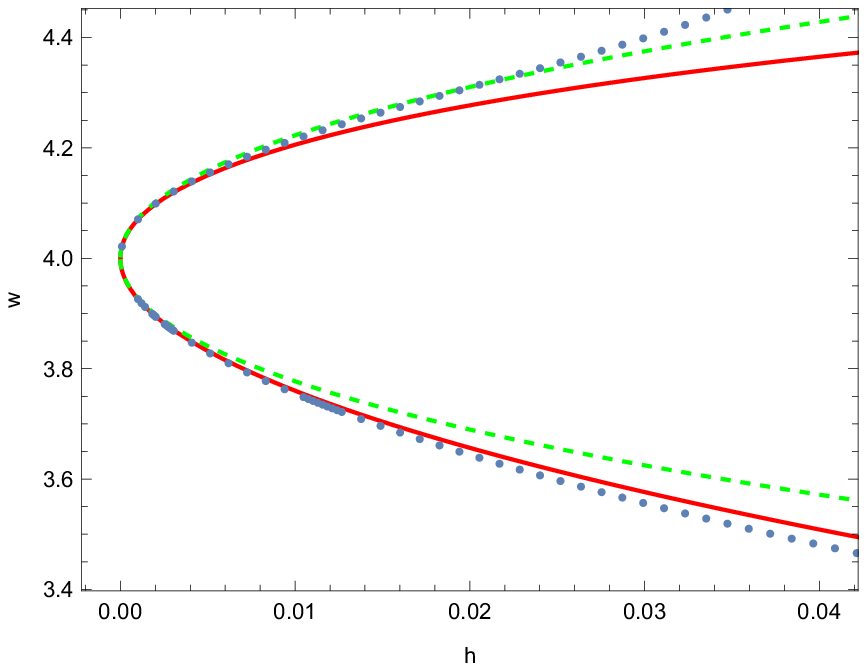}
\includegraphics[width=3in]{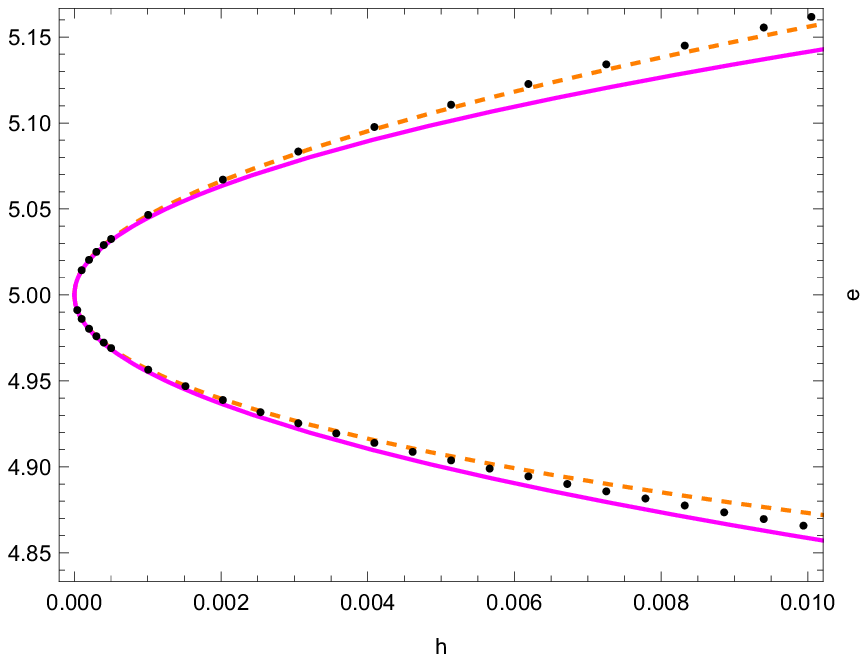}
\end{center}
  \caption{Attenuation $\Gamma_{\csb}^{(n)}\equiv-\Im[\ww_\csb^{(n)}]$ of the
  chirally symmetric fluctuations about cascading gauge theory
  TypeA$ _s$ DFP for $H\gg \Lambda$. Dashed and solid curves
  are correspondingly the leading and the first subleading
  order corrections to the conformal spectra, see \eqref{ancs45}.
} \label{figure1s}
\end{figure}

Chiral symmetry preserving fluctuations about TypeA$ _s$ DFP
activate the cascading gauge theory
operators of conformal dimensions $\Delta=\{4,6,8\}$.
Thus, in the near conformal limit, \ie for $H\gg \Lambda$,
we expect \cite{Buchel:2022hjz} discrete branches
indexed with $n\in \zet_{\ge 4}$ and
\begin{equation}
\Re[\ww_{\cs}^{(n)}]\ =\ 0\,,\qquad \Im[\ww_\csb^{(n)}]\
\equiv\  -\Gamma_{\cs}^{(n)}\ \ne 0\,.
\eqlabel{csn}
\end{equation}
We use the subscript $ _\cs$ to indicate that the fluctuations are 
chirally symmetric.. We find that the branches with
$n=\{4,5\}$ are doubly degenerate in the limit $b\to 0$, while those
with $n\ge 6$ are triple degenerate in the conformal limit.
In fig.~\ref{figure1s} dots represent the attenuation $\Gamma_{\cs}^{(n)}$
as a function of $\ln^{-1}\frac{H^2}{\Lambda^2}$ for the lowest $n=4$
mode (the left panel) and the $n=4$ mode (the right panel). The dashed
curves indicate $\calo(\sqrt{b})$ analytic leading order corrections,
see appendix \ref{sdetailss3}, and the solid lines include next-to-leading
$\calo(b)$ order corrections:
\begin{equation}
\begin{split}
&\Gamma_{\cs}^{(4)}=4\pm \frac{\sqrt{130b}}{5}-1.79(1)\cdot b\pm \calo(b^{3/2})\,,\\
&\Gamma_{\cs}^{(5)}=5\pm \sqrt{2b}+1.45(6)\cdot b\pm \calo(b^{3/2})\,,
\end{split}
\eqlabel{ancs45}
\end{equation}
where $b$ is related to $\frac H\Lambda$ as in \eqref{bpar1}.

\begin{figure}[ht]
\begin{center}
\psfrag{h}{{$1/\ln\frac{H^2}{\Lambda^2}$}}
\psfrag{w}{{$\Gamma_{\cs}$}}
\includegraphics[width=4in]{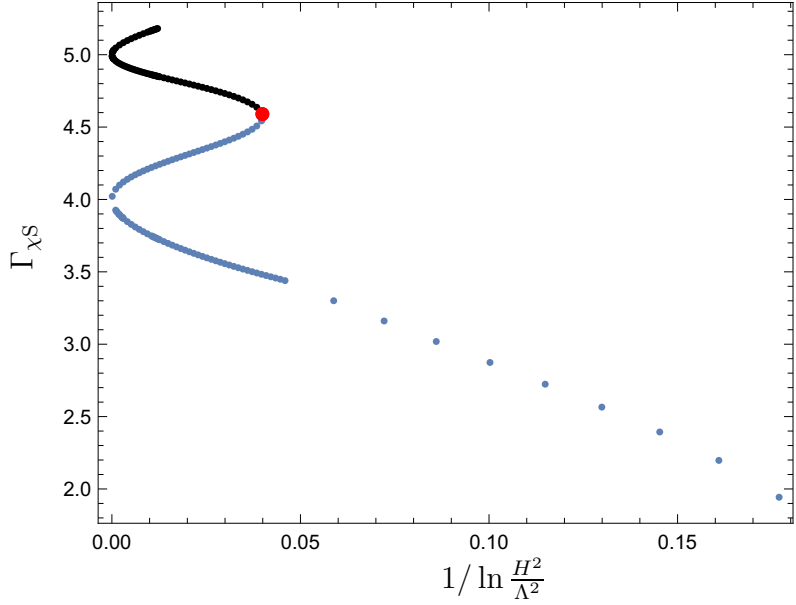}
\end{center}
  \caption{Sub-branches of the distinct in the conformal limit
  branches of the fluctuations
  coalesce as $\frac{H}{\Lambda}$ is lowered. The red dot, see
  \eqref{csreddot}, highlights
  this phenomenon for $n=5$ and $n=4$ sub-branches.
} \label{figure2s}
\end{figure}

As in section \ref{sec4uns},
as $b$ increases, the distinct branches of the
fluctuations coalesce, see fig.~\ref{figure2s}. Specifically we find that
the lower sub-branch of the $n=5$ branch and the upper sub-branch of
the $n=4$ branch combine at
\begin{equation}
\ln^{-1}\ \frac{H^2}{\Lambda^2}\ =\ 0.039(9)\,,
\eqlabel{csreddot}
\end{equation}
represented by the red dot, and are removed from the spectrum.

\begin{figure}[ht]
\begin{center}
\psfrag{h}[t]{{$1/\ln\frac{H^2}{\Lambda^2}$}}
\psfrag{w}{{$\Gamma_{\cs}$}}
\psfrag{e}[t]{{$\Gamma_{\cs}$}}
\includegraphics[width=3in]{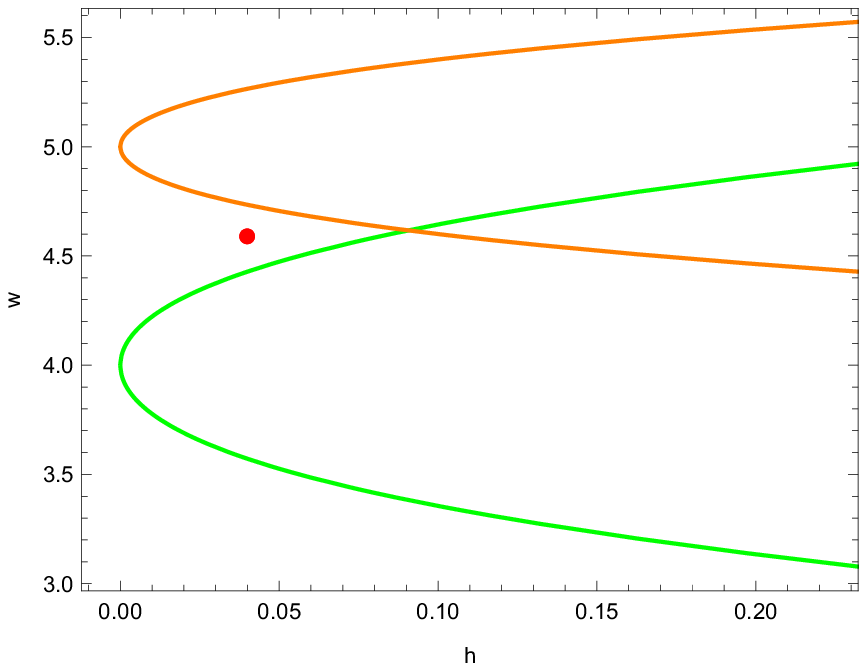}
\includegraphics[width=3in]{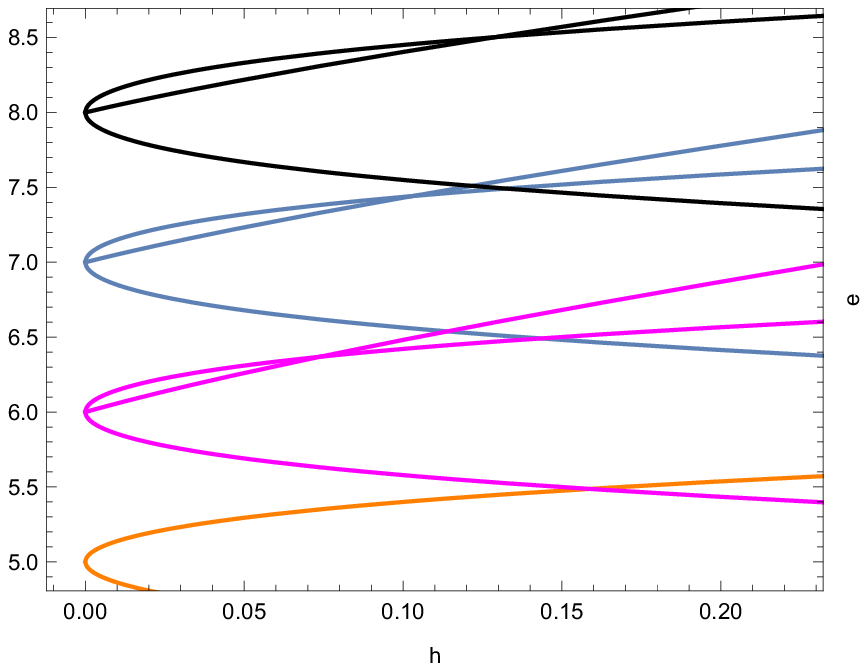}
\end{center}
  \caption{Leading order correction to the conformal
  spectra for the chiral symmetry breaking fluctuations
  at higher $n$. Note  that the (perturbative) coalescence of
  various sub-branches
  is quite generic. The red dot (the left panel) is a replot
  of the red dot from fig.~\ref{figure2s}.
} \label{figure3s}
\end{figure}

In fig.~\ref{figure3s} we present leading order correction
to the conformal spectra for $n=\{4,5\}$ (the left panel)
and for $5\le n\le 8$ (the right panel).
Note that the non-analytic sub-branches, see appendix
\ref{ascs}, (perturbatively) combine ---  the red dot
(the left panel) is replotted from fig.~\ref{figure2s}.

\begin{figure}[ht]
\begin{center}
\psfrag{h}[t]{{$\ln\frac{H^2}{\Lambda^2}$}}
\psfrag{w}{{$\Gamma_{\cs}$}}
\psfrag{e}[t]{{$\Gamma_{\cs}$}}
\includegraphics[width=3in]{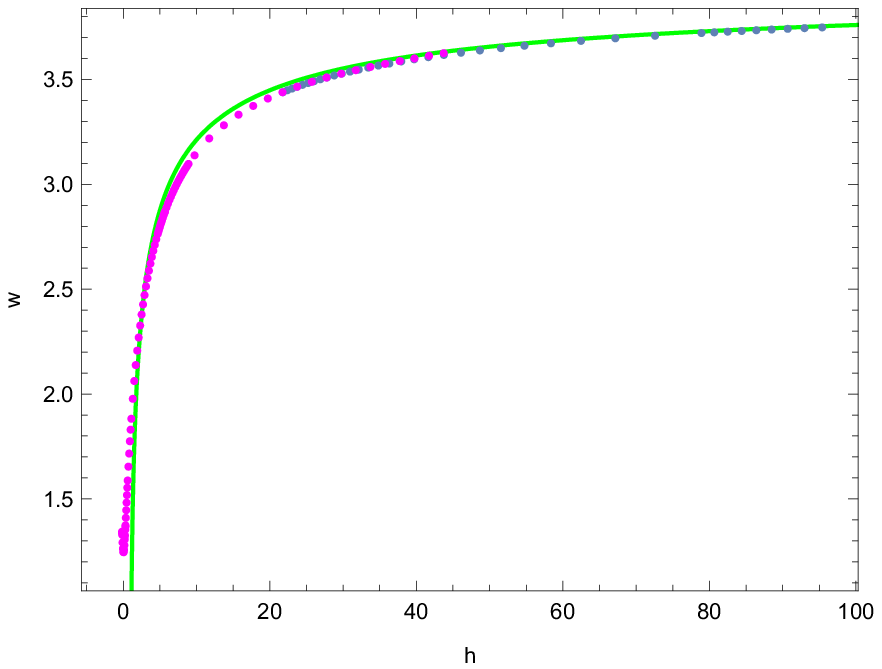}
\includegraphics[width=3in]{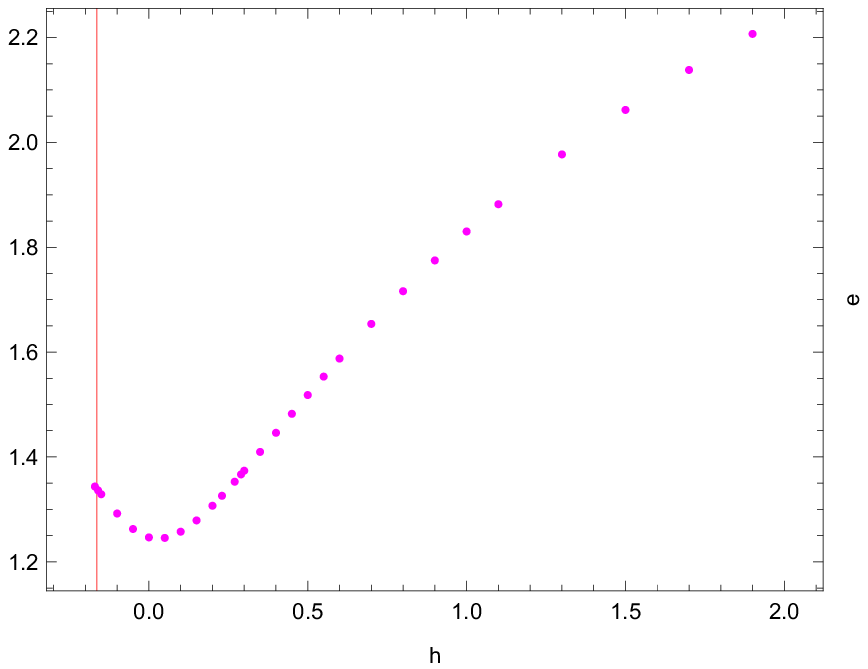}
\end{center}
  \caption{The lower sub-branch of the $n=4$ branch of
  chirally symmetric  fluctuations about TypeA$ _s$ DFP
  of the cascading gauge theory 
  remains perturbatively stable for $H>H_{crit_3}$ \eqref{h3l},
  represented by the red vertical line.
} \label{figure4s}
\end{figure}

In fig.~\ref{figure4s} we show that the lowest lying mode
in the chiral symmetry preserving sector of fluctuations
about TypeA$ _s$ DFP remains perturbatively stable, at least for
$H>H_{crit_3}$, represented by the vertical red line (the right panel).
The solid green curve is the perturbative approximation to the
mode, see $\Gamma_\cs^{(4)}$ in \eqref{ancs45}.
The blue dots are obtained in the computation scheme
\eqref{bpar1}, and the magenta dots are obtained in the computation scheme
\eqref{bpar2}.

This establishes our second main result:

\bigskip

\noindent\fbox{%
    \parbox{\textwidth}{%
{\color{red} The chirally symmetric TypeA$ _s$ DFP of the cascading gauge theory
is perturbative stable when $H>H_{crit_3}$, given by \eqref{h3l}. 
 }
}%
}
\bigskip

\section{Stability analysis of TypeA$ _b$ DFP}
\label{stypeab}

\begin{figure}[t]
\begin{center}
\psfrag{h}{{$\ln\frac{H^2}{\Lambda^2}$}}
\psfrag{a}{{$A$}}
\psfrag{w}{{$\Gamma_{\csb}$}}
\psfrag{d}[t]{{$\Gamma_{\csb}$}}
\includegraphics[width=3in]{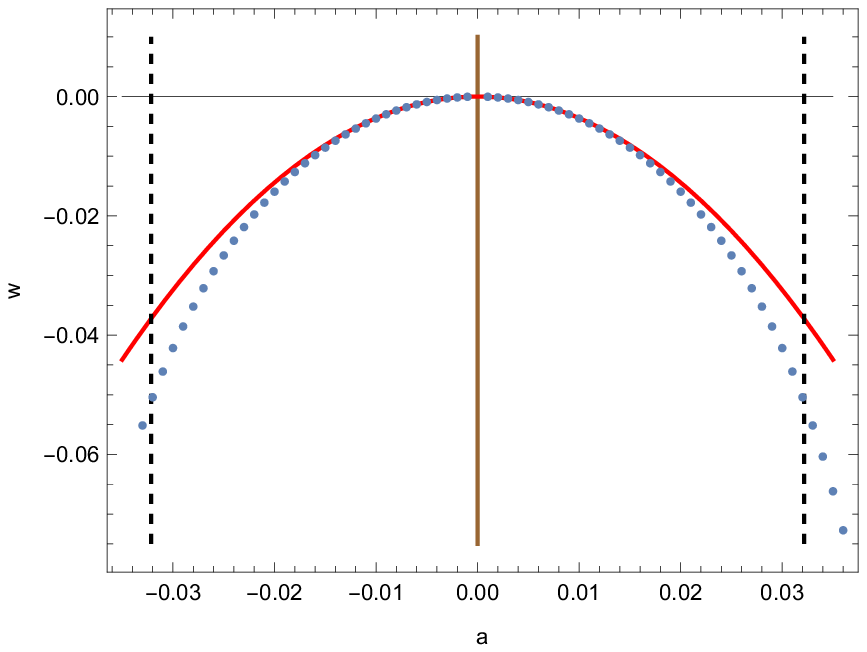}
\includegraphics[width=3in]{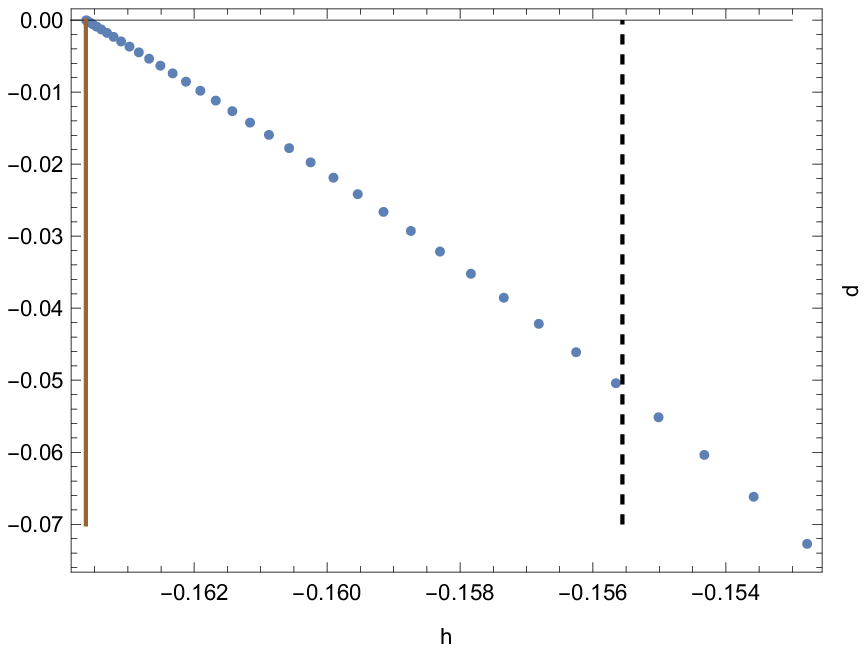}
\end{center}
  \caption{The marginal at $H=H_{crit_3}$ chiral symmetry
  breaking mode becomes unstable in TypeA$ _b$ DFP.
  In the left panel we parameterize this mode
  with the order parameter $A$ of the chiral symmetry breaking
  of TypeA$ _b$ DFP,
  see \eqref{defa}. In the right panel we show the attenuation $\Gamma_{\csb}$
  of this mode as a function of $\ln\frac{H^2}{\Lambda^2}$.
  The solid red curve represent perturbative approximation,
  close to criticality, see \eqref{wpert}. The vertical brown lines
  represent $H=H_{crit_3}$, and the vertical dashed black lines
  represent $H=H_{crit_2}$.
} \label{figure1b}
\end{figure}

TypeA$ _b$ dynamical fixed point of the cascading gauge theory with
spontaneous broken chiral symmetry \cite{Buchel:2019pjb}
exists only for $H> H_{crit_3}$, given by
\eqref{h3l}. Exactly at $H=H_{crit_3}$ TypeA$ _s$ and TypeA$ _b$ DFPs are
indistinguishable. Additionally, at this critical value of the Hubble
constant, the DFP has a marginal chiral symmetry breaking mode ---
this is the lower sub-branch of the $n=3$ fluctuations about TypeA$ _s$ DFP,
see fig.~\ref{figure4}. In fig.~\ref{figure1b} we present the
attenuation of this mode, as a fluctuation about TypeA$ _b$ DFP.
Note that the mode is {\it always } unstable. In the left panel
we present $\Gamma_{\csb}$ with  TypeA$ _b$ DFP
parameterized using the chiral symmetry breaking
order parameter $A$ of this DFP, see \eqref{defa}. This is useful,
as it provides a ready comparison with the perturbative results
of appendix \ref{appdd}, see \eqref{wpert}, represented by a solid red curve.
The translation between the order parameter $A$ and the physical label
\eqref{bpar2} of TypeA$ _b$ DFP is shown in fig.~\ref{K0a};
the latter is further used to
generate the plot in the right panel of  fig.~\ref{figure1b}.
The vertical solid brown lines correspond to $H=H_{crit_3}$,
and the vertical dashed black lines correspond $H=H_{crit_2}$ ---
recall that for $H>H_{crit_2}$, chirally symmetric TypeA$ _s$
DFP is the preferred dynamical attractor compare to
the symmetry broken TypeA$ _b$ DFP, see fig.~\ref{figure0}.

Our final  main result is:

\bigskip

\noindent\fbox{%
    \parbox{\textwidth}{%
{\color{red} TypeA$ _b$ DFP of the cascading gauge theory
is perturbative unstable. 
 }
}%
}
\bigskip

\section{Future directions and speculations}
\label{conclude}

In this paper we presented comprehensive stability analysis of the
de Sitter dynamical fixed points of the cascading gauge theory.
The late-time attractor of the theory is determined by the
ratio of the de Sitter Hubble constant $H$ and the strong coupling
scale $\Lambda$ of the theory. We presented strong evidence that for
$H>H_{crit_2}$, an arbitrary initial state of the gauge theory
would evolve to a chirally symmetric DFP, TypeA$ _s$.
On the other hand, an arbitrary state of the theory with $H<H_{crit_1}$
is expected to evolve to a de Sitter vacuum, with vanishing
comoving entropy density production rate asymptotically.  
Since $H_{crit_1}<H_{crit_2}$, what is the late-time
dynamics of the cascading gauge theory state in de Sitter with the
Hubble constant in the range $H\in (H_{crit_1},H_{crit_2})$
is unknown. In our view, this is the biggest open question.

Note that all the dynamical fixed points of the cascading gauge theory
identified in \cite{Buchel:2019pjb}
have unbroken $SU(2)\times SU(2)$ global symmetry.
The reason for this limitation is simple: we do not know
the dual holographic description of the cascading gauge theory outside
of this $SU(2)\times SU(2)$ symmetric sector \eqref{5action}. 
It is possible that for $H<\Lambda$, the above global symmetry is
spontaneously broken as well, and the new DFPs are stable.
We can only imagine how horrendously complicated it would be to
analyze such DFPs!

Another interesting question is the role confinement plays
in producing de Sitter DFPs of non-conformal field theories
in the first place. In other
non-conformal holographic models, as discussed in
\cite{Buchel:2017lhu} and  \cite{Buchel:2021ihu},
the late time attractor of the de Sitter evolution of these
models is {\it always} a dynamical fixed point, \ie
the state with the non-vanishing comoving entropy production
rate\footnote{We expect that DFPs discussed there will become unreliable
then $H\ll m$, where $m$ is the mass-scale of the models. }.
In other words, there is no analogue of the cascading gauge theory
TypeB de Sitter vacuum. 
Intuitively, the holographic description of
confinement in de Sitter is fairly robust, at least when $H\ll \Lambda$,
and thus it is natural to expect that with a large hierarchy of scales
between the confinement scale and the Hubble constant, as it is in our Universe,
there is no dynamical fixed point at late times.
Can we find a holographic
model where a dynamical fixed point exists in the limit $\frac Hm\to 0$?
Can the idea that a dynamical fixed point requires
deconfinement of the gauge theory be made precise, or shown to be false?

Clearly it is interesting to explore other holographic models
in de Sitter and analyze the corresponding DFPs.

\section*{Acknowledgments}
Research at Perimeter Institute is supported in part by the Government
of Canada through the Department of Innovation, Science and Economic
Development Canada and by the Province of Ontario through the Ministry
of Colleges and Universities. This work is further supported by a
Discovery Grant from the Natural Sciences and Engineering Research
Council of Canada.

\appendix
\section{Symmetry broken DFP --- TypeA$ _b$}
\label{fgframe}

In this appendix we discuss the linearized fluctuations about Type$A _b$
dynamical fixed point of the cascading gauge theory.
In section \ref{masterfluc}
we apply a straightforward generalization of the {\it master equation}
formalism \cite{Buchel:2022hjz} to derive the equations of motion
for the fluctuations. The final  equations are too long/complicated to be collected
in the paper --- they are available as a Maple worksheet in \cite{coleqs}.
In section \ref{appasymptotics} we discuss the boundary conditions,
both for the background geometry and for the fluctuations.

\subsection{Equations of motion}
\label{masterfluc}

As explained in section \ref{hard}, we consider the cascading gauge theory DFPs and
linearized fluctuations about them in ten-dimensional Type IIB supergravity.
The detailed discussion of the uplift can be found in \cite{Buchel:2019pjb}.

For the Fefferman-Graham metric ansatz (with spatially homogeneous
and isotropic background metric of the cascading gauge theory  $\propto d{\bm x}^2$)
we take
\begin{equation}
ds_{10}^2=-\hc_1^2\ d\tau^2+ \hc_2^2\ d{\bm x}^2+ \hc_3^2\ d\rho^2+\underbrace{\hw_1^2\ g_5^2
+ \hw_2^2\ \left(g_3^2+g_4^2\right)+ \hw_3^2\ \left(g_1^2+g_2^2\right)}_{\rm new\ compare\ to\
master\ formalism}\,,
\eqlabel{10dmetrict}
\end{equation}
with 
\begin{equation}
\begin{split}
&\hc_1=\frac{\sqrt{\hat{G}_{tt}}}{\rho\ \hh^{1/4}}\,,\qquad \hc_2=\frac{\sqrt{\hat{G}_{{\bm x} {\bm x}}}}{\rho\ \hh^{1/4}}\,,
\qquad \hc_3=\frac{\hh^{1/4}}{\rho}\,,\\
&\hw_1=\frac 13\ {\hf}_c^{1/2}\left(\hh\right)^{1/4}\,,\qquad
\hw_2=\frac {1}{\sqrt 6}\ {\hf}_a^{1/2}\left(\hh\right)^{1/4}\,,\qquad
\hw_3=\frac {1}{\sqrt 6}\ {\hf}_b^{1/2}\left(\hh\right)^{1/4}\,,
\end{split}
\eqlabel{fcw}
\end{equation}
where we highlighted the part of the metric new compare to the general ansatz of
\cite{Buchel:2022hjz}.
Additionally we set
\begin{equation}
\hh_1=\frac 1P\left(\frac{\hK_1}{12}-36\Omega_0\right)\,,\ \ \hh_2=\frac{P}{18}\hK_2\,,\ \ 
\hh_3=\frac 1P\left(\frac{\hK_3}{12}-36\Omega_0\right)\,,\ \ \hg=e^{\hat\Phi}\,,
\end{equation}
where we use $\ \hat{}\ $ to indicate that the corresponding functions depend on $\rho$ and $\tau$.
In \eqref{10dmetrict}, $g_i$ ( for $i=1,\cdots, 5$ ) are the usual one-forms defined  on the
warped-squashed $T^{1,1}$ \cite{Minasian:1999tt}.

Following \cite{Buchel:2022hjz}, we introduced linearized fluctuation, $\delta\cdots$,
on top of the background solution specified by \cite{Buchel:2019pjb}
\begin{equation}
\bigg\{
f_{a,b,c}\,,\ h\,,\ K_{1,2,3}\,,\ g
\biggr\}\,,
\eqlabel{bacdfp}
\end{equation}
specifically,
\begin{equation}
\begin{split}
&\sqrt{\hat{G}_{tt}}=1+\delta\hat{g}_{11}\,,\qquad \sqrt{\hat{G}_{{\bm x} {\bm x}}}=e^{H\tau}
\left(1+\delta\hat{g}_{22}\right)\,,\qquad
\hf_{a,b,c}=f_{a,b,c}(\rho)+\delta\hf_{a,b,c}\,,\\
&\hh=h(\rho)+\delta\hh\,,\qquad \hK_{1,2,3}=K_{1,2,3}(\rho)+\delta K_{1,2,3}\,,\qquad \hg=g(\rho)+\delta \hg\,.
\end{split}
\eqlabel{fsplit}
\end{equation}
 Notice that the $h$ factor enters both in the
definition of the background DFP metric, \eg see $\hat{c}_1$ in \eqref{fcw},
and the five-dimensional bulk scalars $\Omega_i$ \eqref{fcw}.
This is necessary to produce equations of motion without the fractional powers of
$h$ \cite{Aharony:2005zr}.

Assuming the harmonic time-dependence for the fluctuations, \ie 
\begin{equation}
\begin{split}
&\delta \hg_{11}=e^{-i \w\tau} H_1(\rho)\,,\qquad \delta \hg_{22}=e^{-i \w\tau} H_2(\rho)\,,\qquad
\delta\hf_{a,b,c}=e^{-i \w\tau} H_{a,b,c}(\rho)\,,\\
&\delta\hh=e^{-i \w\tau} H_h(\rho)\,,\qquad  \delta \hK_{1,2,3}=e^{-i \w \tau} H_{K_{1,2,3}}(\rho)\,,\qquad
\delta\hg=e^{-i\w\tau} H_g(\rho)\,,
\end{split}
\eqlabel{fl1}
\end{equation}
we derive 11 equations\footnote{The 11's equation is the Einstein
equation with coordinate indices $ _{\rho\tau}$.} for 10 fluctuations: 
\begin{equation}
\biggl\{ H_{a,b,c}\,,\ H_{1,2}\,,\ H_h\,,\ H_g\,,\ H_{K_{1,2,3}}\,.
\biggr\}
\eqlabel{raweq}
\end{equation}
These equations are collected in \cite{coleqs}.
It is convenient to further introduce 
\begin{equation}
\w=-i H s\,,
\eqlabel{defw}
\end{equation}
effectively measuring all energy scales in Hubble units.
To declutter the formulas we set from now on 
the Hubble constant to unity $H=1$.

The master formalism of \cite{Buchel:2022hjz} allowed to eliminate algebraically ---
solve their corresponding equations ---
the fluctuations $H_1$ and $H_2$ as 
\begin{equation}
{\rm master}:\qquad H_1=H_2=\frac{(d-2)H_h}{4(d-1)h}\bigg|_{d=3\,,\, {\rm in\ our\ case}}=
\frac{H_h}{8h}\,.
\eqlabel{elold}
\end{equation}
Such substitution will not work for the cascading gauge theory gravitational
dual: from the 5d perspective because the Einstein-Hilbert term in the
Kaluza-Klein reduced effective action 
is warped \eqref{mastervscas}, or from the 10d perspective because of extra contributions
in the metric \eqref{10dmetrict}. Instead, we find that a substitution
\begin{equation}
H_1=H_2=-\frac{H_h}{2h}-\frac{H_a}{2f_a}-\frac{H_b}{2f_b}-\frac{H_c}{4f_c}
\eqlabel{mastergauge}
\end{equation}
solves the equations for $H_{1,2}$. 
Furthermore, much like in \cite{Buchel:2022hjz}, the equation for $H_h$
is of the first-order, and can be solved algebraically in terms
of the other fluctuations and their first-order (radial coordinate) derivatives.

As explained in \cite{Buchel:2022hjz},
while the boundary conditions for the
fluctuations in the EF coordinate frame are natural,
they are less obvious in the FG coordinate frame:
\nxt first, near the boundary we require that the fluctuations
are normalizable;
\nxt second, the EF frame bulk regularity condition  is replaced
in the FG frame with the requirement
that all fluctuations behave as
\begin{equation}
H_{\cdots}\ \sim \rho^{s/2}\times {\rm finite}\qquad {\rm as}\qquad \rho\to \infty\,.
\eqlabel{singr0}
\end{equation}
It is thus convenient to extract this singularity from the radial profiles
of the fluctuations, 
\begin{equation}
\begin{split}
&H_{a,b,c}=(1+\rho)^{s/2}\ fl_{a,b,c}(\rho)\,,\ H_{K_{1,2,3}}=(1+\rho)^{s/2}\ fl_{K_{1,2,3}}(\rho)\,,\\
&H_g=(1+\rho)^{s/2}\ fl_g(\rho)\,,\qquad H_h=(1+\rho)^{s/2}\ fl_h(\rho)\,,
\end{split}
\eqlabel{fl2}
\end{equation}
where we modified $\rho\to (1+\rho)$ to avoid introduction of the spurious singularity
near the boundary, \ie as $\rho\to 0$.
As for $H_h$, the expression for $fl_h$ is algebraic, 
\begin{equation}
fl_h=fl_h\biggl[fl_{a,b,c}, \frac{d}{d\rho}fl_{a,b,c}\ ;\  fl_{K_{1,2,3}}, \frac{d}{d\rho}fl_{K_{1,2,3}}\ ;\
 fl_{g}, \frac{d}{d\rho}fl_{g}\ ;\ s\biggr]\,.
\eqlabel{flh}
\end{equation}
We collected the equations for
\begin{equation}
\biggl\{\ fl_{a,b,c}\,,\, fl_{K_{1,2,3}}\,,\, fl_g\,,\, fl_h\ \biggr\}\,,
\eqlabel{fldfp}
\end{equation}
along with
the algebraic expression \eqref{flh}  in \cite{coleqs}.
These final equations, solved subject to normalizability of the fluctuations
at the boundary, and their regularity in the bulk $\rho\in [0,\infty)$,
would determine the spectrum $\{\omega\}$. 
We would like to stress that the regularity condition as $\rho\to \infty$
is much more stronger than the requirement that the modes $fl_{\cdots}$ are finite
in this limit; rather, the regularity mandates \cite{Buchel:2022hjz}
that we have a standard Maclaurin
series expansion for the profiles $fl_{\cdots}$ in variable $y\equiv \frac {1}{\rho}$,
\eg the terms $\rho^{-17/2}$  or $\rho^{-7}\ln\rho$ are not allowed. 
This is necessary so that the fluctuations can be properly
transformed to the EF coordinate frame in the vicinity of $y\propto (r-r_0)\to 0$
(see \eqref{mapfgef}), and further extended in the EF coordinate frame
all the way to the apparent horizon, $r\in [r_{AH},r_0]$.

\subsection{Asymptotics}
\label{appasymptotics}

In this section we discuss the asymptotics of the DFP background
functions \eqref{bacdfp}, and the fluctuations \eqref{fldfp}. 
Keep in find that the equations of motion for $\{f_{a,b}$, $h$, $K_{1,2,3}$, $g\}$
are of the second-order in $\rho$, the equation for $f_c$ is of the first-order;
and the equations of motion for $\{fl_{a,b,c}$, $fl_{K_{1,2,3}}$, $fl_g\}$
are of the second-order. The first-order equation for $fl_h$
is not independent, see \eqref{flh}. All the equations
are nonlinear and coupled. To find a solution to a DFP background,
one needs a single label, corresponding to $\frac{H}{\Lambda}$
\eqref{deflambda}, and $7\times 2 +1\times 1=15$ parameters (from counting
the total order of the background equations of motion).  
Likewise, to solve the fluctuation equations,
assuming that the radial profile functions $fl_{\cdots}$ are real
and $\Re[\ww]=0$, 
one needs
$7\times 2+ 0\times 1 =14$ parameters; one of these parameters
must be $s$ \eqref{defw}.
It is possible to
have fluctuations about a DFP which are not purely imaginary
\cite{Buchel:2017lhu}. Their analysis, using the equations of motion derived
in this paper, are straightforward, but it
will not be performed here: TypeA$ _b$ DFP is found to be unstable to a mode with
 $\Re[\ww]=0$ already;
instabilities of TypeA$ _s$ DFP are anticipated by the marginal modes,
that can be identified independently as in section 5.1 of
\cite{Buchel:2019pjb}. Besides the marginal mode responsible for a branching
of TypeA$ _b$ DFP away from TypeA$ _s$ DFP discussed in section \ref{sec4uns},
none exists. Thus, we do not expect any of TypeA$ _s$ fluctuations with
$\Re[\ww]\ne 0$, if exist, would become 
unstable, at least for $H>H_{crit_3}$.

\subsubsection{Background}

The general UV (as $\rho\to 0$) asymptotic solution of the background equations of motion
describing the phase of the cascading
 gauge theory with spontaneously broken chiral symmetry takes the form
\begin{equation}
\begin{split}
&f_c=1+f_{a,1,0}\ \rho+\sum_{n=2}^\infty\sum_k f_{c,n,k}\ {\rho^n}\ln^k\rho\,,
\end{split}
\eqlabel{ksfc}
\end{equation}
\begin{equation}
\begin{split}
&f_a=1+f_{a,1,0}\ \rho+\sum_{n=2}^\infty\sum_k f_{a,n,k}\ {\rho^n}\ln^k\rho\,,
\end{split}
\eqlabel{ksfa}
\end{equation}
\begin{equation}
\begin{split}
&f_b=1+f_{a,1,0}\ \rho+\sum_{n=2}^\infty\sum_k f_{b,n,k}\ {\rho^n}\ln^k\rho\,,
\end{split}
\eqlabel{ksfb}
\end{equation}
\begin{equation}
\begin{split}
&h=\frac18 b +\frac14 K_0-\frac12 b \ln\rho+\biggl(b \ln\rho-\frac12 K_0\biggr) f_{a,1,0}\ \rho
+\sum_{n=2}^\infty\sum_k h_{n,k}\ {\rho^n}\ln^k\rho\,,
\end{split}
\eqlabel{ksh}
\end{equation}
\begin{equation}
\begin{split}
&K_1=K_0-2 b \ln\rho+b f_{a,1,0}\  \rho+\sum_{n=2}^\infty\sum_k k_{1,n,k}\ {\rho^n}\ln^k\rho\,,
\end{split}
\eqlabel{ksK1}
\end{equation}
\begin{equation}
\begin{split}
&K_2=1+\left(k_{2,3,0}+\frac34   f_{a,1,0} b \ln\rho+3 f_{a,3,0} \ln\rho \right) \rho^3
+\sum_{n=4}^\infty\sum_k k_{2,n,k}\ {\rho^n}\ln^k\rho\,,
\end{split}
\eqlabel{ksK2}
\end{equation}
\begin{equation}
\begin{split}
&K_3=K_0-2 b \ln\rho+bf_{a,1,0}\  \rho+\sum_{n=2}^\infty\sum_k k_{3,n,k}\ {\rho^n}\ln^k\rho\,,
\end{split}
\eqlabel{ksK3}
\end{equation}
\begin{equation}
\begin{split}
&g=1-\frac12 b\ \rho^2+\sum_{n=3}^\infty\sum_k g_{n,k}\ {\rho^n}\ln^k\rho\,.
\end{split}
\eqlabel{ksg}
\end{equation}
It is characterized by 9 parameters:
\begin{equation}
\{K_0\,,\   f_{a,1,0}\,,\ \underbrace{f_{a,3,0}\,,\ k_{2,3,0}}_{\calo_3^\a}\,,\
\underbrace{g_{4,0}\,,\ f_{c,4,0}}_{\calo_4^\b}\,,\ \underbrace{f_{a,6,0}}_{\calo_6}\,,\
\underbrace{f_{a,7,0}}_{\calo_7}\,,\ \underbrace{f_{a,8,0}}_{\calo_8}\}\,,
\eqlabel{uvparks}
\end{equation}
where we indicated the dual cascading gauge theory operators which expectation values
these parameters characterize. 
$K_0$ is related to strong coupling scale $\Lambda$ of the cascading gauge theory
as \eqref{deflambda}.
Finally, $f_{a,1,0}$ corresponds to a diffeomorphism parameter that ensures 
the range of the radial coordinate as in \eqref{rangerho}.

To study the infrared asymptotics, \ie as $y\equiv \frac 1\rho\to 0$, we redefine  
\begin{equation}
h^h\equiv y^{-2}\ h\,,\qquad f^{h}_{a,b,c}\equiv y\ f_{a,b,c}\,.
\eqlabel{defhors4}
\end{equation}
The IR asymptotic expansions 
\begin{equation}
\begin{split}
&f_{a,b,c}^h=\sum_{n=0} f_{a,b,c,n}^h y^n\,,\qquad h^h=\frac {1}{4  }+\sum_{n=1} h^h_{n} y^n\,,\\
&K_{1,2,3}=\sum_{n=0} K_{1,2,3,n}^h y^n\,,\qquad g=\sum_{n=0} g_{n}^h y^n\,,
\end{split}
\eqlabel{phase1ir1}
\end{equation}
are characterized by 7 parameters:
\begin{equation}
\{f_{a,0}^h\,,\ f_{b,0}^h\,,\ f_{c,0}^h\,,\ K_{1,0}^h\,,\ K_{2,0}^h\,,\ K_{3,0}^h\,,\ g_{0}^h\}\,.
\eqlabel{irph1par}
\end{equation}

Notice that in total we have, \eqref{uvparks} and \eqref{irph1par}, $9+7=16=1+15$
parameters, as expected.

\subsubsection{Fluctuations}

The general UV (as $\rho\to 0$) asymptotic solution of the fluctuation equations of motion
is much more complicated:
\begin{equation}
\begin{split}
&fl_{a,b,c}=\sum_{n=2}^\infty\ \rho^n\ F_{a,b,c;n}(z)\,,\qquad fl_{K_{1,3}}=\sum_{n=2}^\infty\ \rho^n\ F_{K_{1,3};n}(z)\\
&fl_{K_{2}}=\sum_{n=3}^\infty\ \rho^n\ F_{K_{2};n}(z)\,,\qquad fl_{g}=\sum_{n=4}^\infty\ \rho^n\ F_{g;n}(z)\,,
\qquad fl_{h}=\sum_{n=2}^\infty\ \rho^n\ F_{h;n}(z)\,,
\end{split}
\eqlabel{fluv}
\end{equation}
where $z\equiv \ln\rho$. At each fixed order $n$ we have a coupled system of 7 second-order ODEs
for
\begin{equation}
\biggl\{\
F_{a,b,c;n}(z)\,,\, F_{K_{1,2,3};n}(z)\,,\, F_{g;n}(z)
\ \biggr\}\,,
\eqlabel{listfluc}
\end{equation}
along with the first-order constraint involving $F_{h;n}(z)$. 
The complexity of these equations grows with $n$.
Since the cascading gauge theory has a gravitational scalar dual to a dimension $\Delta=8$
operator, at the very least the series expansions must be developed to order $n=8$ inclusive.

We present here the simplest set of the equations, \ie for $n=2$:
\begin{equation}
\begin{split}
&0=F_{a;2}''-\frac{b}{b-4 b z+2 K_0} \left(
2 F_{a;2}'+F_{c;2}'+2 F_{b;2}'
\right)
-\frac{4}{b-4 b z+2 K_0} \left(F_{K_1;2}'+3 F_{K_3;2}'\right)
\\&+\frac{1}{(4 b z - 2 K_0 - b)^3} 
\biggl(
(6144 z^2 - 768 z + 256) b^2 + (-6144 z + 384) K_0 b + 1536 K_0^2\biggr) F_{h;2}
\\&+\frac{1}{2 (4 b z - 2 K_0 - b)^2}\biggl(
\left((-1296 z^2 + 264 z - 97) b^2 + (1296 z - 132) K_0 b - 324 K_0^2\right) F_{a;2}\\
&+ \left((-1264 z^2 - 8 z - 31) b^2 + (1264 z + 4) K_0 b - 316 K_0^2\right) F_{b;2}
+ ((-512 z^2 - 24) b^2 \\
&+ 512 b z K_0 - 128 K_0^2) F_{c;2} + ((-320 z - 32) b + 160 K_0) F_{K_1;2}
+ ((128 z - 32) b^2 \\&
- 64 b K_0) F_{K_2;2} - 192 F_{K_3;2} \left(\left(z + \frac13\right) b - \frac{K_0}{2}\right)
\biggr)\,,
\end{split}
\eqlabel{n21}
\end{equation}
\begin{equation}
\begin{split}
&0=F_{b;2}''-\frac{b}{b-4 b z+2 K_0} \left(2 F_{b;2}'+F_{c;2}'+2 F_{a;2}'\right)
-\frac{4}{b-4 b z+2 K_0} \left(F_{K_3;2}'+3 F_{K_1;2}'\right)
\\&+\frac{1}{(4 b z - 2 K_0 - b)^3} \biggl(
(6144 z^2 - 768 z + 256) b^2 + (-6144 z + 384) K_0 b + 1536 K_0^2\biggr) F_{h;2}
\\&+\frac{1}{2 (4 b z - 2 K_0 - b)^2}\biggl(
\left(
(-1296 z^2 + 264 z - 97) b^2 + (1296 z - 132) K_0 b - 324 K_0^2\right) F_{b;2}
\\&+ ((-1264 z^2 - 8 z - 31) b^2 + (1264 z + 4) K_0 b - 316 K_0^2) F_{a;2} + ((-512 z^2 - 24) b^2 \\
&+ 512 b z K_0 - 128 K_0^2) F_{c;2} + ((-320 z - 32) b + 160 K_0) F_{K_3;2} - ((128 z - 32) b^2 \\
&- 64 b K_0) F_{K_2;2} - 192 F_{K_1;2} \left(\left(z + \frac13\right) b - \frac{K_0}{2}\right)
\biggr)\,,
\end{split}
\eqlabel{n22}
\end{equation}
\begin{equation}
\begin{split}
&0=F_{c;2}''-\frac{b}{b-4 b z+2 K_0} \left(2 F_{a;2}'+F_{c;2}'+2 F_{b;2}'\right)
-\frac{4}{b-4 b z+2 K_0} \left( F_{K_3;2}'+F_{K_1;2}'\right)
\\&+\frac{1}{(4 b z - 2 K_0 - b)^3}\biggl(
(6144 z^2 - 768 z + 256) b^2 + (-6144 z + 384) K_0 b + 1536 K_0^2\biggr) F_{h;2}
\\&+\frac{1}{(4 b z - 2 K_0 - b)^2}\biggl(
\left((-512 z^2 + 160 z - 36) b^2 + (512 z - 80) K_0 b - 128 K_0^2\right) F_{c;2}
\\&+ ((-512 z^2 - 24) b^2 + 512 b z K_0 - 128 K_0^2) F_{a;2} + ((-512 z^2 - 24) b^2 + 512 b z K_0
\\&- 128 K_0^2) F_{b;2} + ((-160 z - 16) b + 80 K_0) F_{K_1;2} + ((-160 z - 16) b + 80 K_0) F_{K_3;2} \\
&- 64 F_{g;2} b \left(\left(z - \frac14\right) b - \frac{K_0}{2}\right)
\biggr)\,,
\end{split}
\eqlabel{n23}
\end{equation}
\begin{equation}
\begin{split}
&0=F_{K_1;2}''+\frac{4 b}{b-4 b z+2 K_0} F_{K_1;2}'+b \left(2 F_{g;2}'-F_{a;2}'
-\frac12 F_{c;2}'+3 F_{b;2}'\right)
\\&+\frac{320 b (-2 b z+K_0)}{(b-4 b z+2 K_0)^2} F_{h;2}
+\frac{1}{b-4 b z+2 K_0} \biggl(
4 b (5 K_0-10 b z-b) (2 F_{a;2}+F_{c;2})\\&+24 b (K_0-2 b z) F_{b;2}+4 b (4 b z-2 K_0+b) F_{g;2}+
\left(34 b z-17 K_0-\frac{41}{2} b\right) F_{K_1;2}\\&+16 b (K_0-2 b z) F_{K_2;2}
+\left(-18 b z+9 K_0-\frac{31}{2} b\right) F_{K_3;2}
\biggr)\,,
\end{split}
\eqlabel{n24}
\end{equation}
\begin{equation}
\begin{split}
&0=F_{K_2;2}''+\frac{4 b}{b-4 b z+2 K_0} F_{K_2;2}'+9 (F_{b;2}-F_{a;2})
\\&+\frac{1}{b-4 b z+2 K_0} \biggl(
\frac{18}{b} (K_0-2 b z) (F_{K_1;2}-F_{K_3;2})+(52 b z-26 K_0-5 b) F_{K_2;2}
\biggr)\,,
\end{split}
\eqlabel{n25}
\end{equation}
\begin{equation}
\begin{split}
&0=F_{K_3;2}''+\frac{4 b}{b-4 b z+2 K_0} F_{K_3;2}'+b \left(
2 F_{g;2}'+3 F_{a;2}'-F_{b;2}'-\frac12 F_{c;2}'\right)
\\&+\frac{320 b (-2 b z+K_0)}{b-4 b z+2 K_0)^2} F_{h;2}+\frac{1}{b-4 b z+2 K_0} \biggl(
4 b (5 K_0-10 b z-b) (2 F_{b;2}+F_{c;2})\\
&+24 b (K_0-2 b z) F_{a;2}+4 b (4 b z-2 K_0+b) F_{g;2}+\left(34 b z-17 K_0-\frac{41}{2} b\right) F_{K_3;2}\\
&-16 b (K_0-2 b z) F_{K_2;2}+\left(-18 b z+9 K_0-\frac{31}{2} b\right) F_{K_1;2}\biggr)\,,
\end{split}
\eqlabel{n26}
\end{equation}
\begin{equation}
\begin{split}
&0=F_{g;2}''-\frac{4}{b-4 b z+2 K_0+b} \left( F_{K_3;2}'+F_{K_1;2}'\right)
-\frac{4}{b-4 b z+2 K_0} \biggl(F_{g;2} (-4 b z+2 K_0+5 b)\\
&-2 F_{c;2} b+2 F_{K_1;2}+2 F_{K_3;2}\biggr)\,,
\end{split}
\eqlabel{n27}
\end{equation}
\begin{equation}
\begin{split}
&0=F_{h;2}'+\frac{6 b}{b-4 b z+2 K_0} F_{h;2}
+\frac{b-4 b z+2 K_0}{32} \left(2 F_{a;2}'+F_{c;2}'+2 F_{b;2}'\right)
\\&+\frac14 (F_{K_1;2}+F_{K_3;2})+\frac b8 (2 F_{b;2}+2 F_{a;2}+F_{c;2})\,.
\end{split}
\eqlabel{n28}
\end{equation}
It is straightforward to verify that \eqref{n28} is solved using the algebraic expression for
$F_{h,2}$, derived from \eqref{flh}, 
\begin{equation}
\begin{split}
&F_{h;2}=-\frac{(b-4 b z+2 K_0)^2}{640(K_0-2 b z)} \left(
2 F_{a;2}'+2 F_{b;2}'+F_{c;2}'
+\frac4b \left( F_{K_1;2}'+F_{K_3;2}'\right)\right)
\\&-\frac{4}{5b (K_0-2 b z)} \biggl(
\left(
\frac b4 (z+1) -\frac18 K_0\right) (F_{K_1;2}+ F_{K_3;2})
+b \biggl(
\left( \left(z+\frac{1}{16}\right) b-\frac12 K_0\right)\\
&\times(F_{a;2}+F_{b;2})+\biggl(
\left(\frac14 z+\frac{3}{32}\right) b-\frac18 K_0\biggr) F_{c;2}
+\frac12 F_{g;2} \left( \left(z-\frac14\right) b-\frac12 K_0\right)
\biggr)
\biggr)\\&\times \left(\left(z-\frac14\right) b-\frac12 K_0\right)\,.
\end{split}
\end{equation}
Remarkably, above equations can be solved analytically, 
\begin{equation}
\begin{split}
&F_{a,b,c;2}=-\frac{2 A(3 b-4 b z+2K_0)}{b(b-4 b z+2K_0)^2}\,,\qquad
F_{K_{1,3};2}=-\frac{A}{b-4 b z+2 K_0}\,,\\
&F_{K_2;2}=F_{g;2}=0\,,\qquad F_{h;2}=\frac{A(5b-8 b z+4 K_0)}{4b(b-4 b z+2 K_0)}\,,
\end{split}
\eqlabel{solven2}
\end{equation}
where $A$ is an arbitrary constant, characterizing an overall normalization of the
linearized fluctuations. 

In general, we find that the differential equations for \eqref{listfluc}
are solved with the ansatz
\begin{equation}
\begin{split}
&F_{a,b,c;n}(z)=\frac{1}{(b-4bz+2K_0)^n}\ \sum_{m=0}^{M_{a,b,c;n}}\ fl_{a,b,c;n;m}\ z^m\,,\qquad n\ge 2\,,\\
&F_{K_{1,3};n}(z)=\frac{1}{(b-4bz+2K_0)^{n-1}}\ \sum_{m=0}^{M_{K_{1,3};n}}\ fl_{K_{1,3};n;m}\ z^m\,,\qquad n\ge 2\,,
\\
&F_{K_{2};3}(z)=fl_{K_2;3;0}+fl_{K_2;3;1}\ z\,,
\\
&F_{K_{2};n}(z)=\frac{1}{(b-4bz+2K_0)^{n-4}}\ \sum_{m=0}^{M_{K_{2};n}}\ fl_{K_{2};n;m}\ z^m\,,\qquad n\ge 4\,,\\
&F_{g;n}(z)=\frac{1}{(b-4bz+2K_0)^{n-3}}\ \sum_{m=0}^{M_{g;n}}\ fl_{g;n;m}\ z^m\,,\qquad n\ge 4\,,\\
&F_{h;n}(z)=\frac{1}{(b-4bz+2K_0)^{n-1}}\ \sum_{m=0}^{M_{h;n}}\ fl_{h;n;m}\ z^m\,,\qquad n\ge 2\,,\\
\end{split}
\eqlabel{generalsoln2}
\end{equation}
where $fl_{\cdots;n;m}$ are constants, and the orders of $z$-polynomials in the
numerators of $F_{\cdots;n}$, \ie $M_{\cdots;n}$,  are collected in the table below:
\begin{center}
\begin{tabular}{| c| c|c |c|c|c|c|} 
 \hline
 $n$ & $M_{a;n}=M_{b;n}$ & $M_{c;n}$ & $M_{K_1;n}=M_{K_3;n}$ & $M_{K_2;n}$ & $M_{g;n}$ & $M_{h;n}$\\
 \hline
 2 & 1 & 1 & 0 & $-$ & $-$& 1\\ 
 \hline
 3 & 3 & 2 & 3 & 1 & $-$& 2\\ 
 \hline
 4 & 4 & 4 & 4 & 1 & 2& 4\\ 
 \hline
 5 & 6 & 5 & 6 & 3 & 3& 5\\ 
 \hline
 6 & 8 & 8 & 7 & 4 & 5& 7\\ 
 \hline
 7 & 10 & 9 & 10 & 7 & 6& 8\\ 
 \hline
 8 & 12 & 12 & 11 & 8 & 9& 12\\ 
 \hline
\end{tabular}
\end{center}
The set of independent constants, fully determining the remaining coefficients  $fl_{\cdots;n;m}$,
is given by
\begin{equation}
\biggl\{\ A;\,  fl_{a;3;0}\,,\ fl_{K_2;3;0}\,,\ fl_{g;4;0}\,,\ fl_{K_3;6;5}\,,\ fl_{K_2;7;0}\,,\ fl_{g;8;0};\,
s\
\biggr\}\,,
\eqlabel{uvfl}
\end{equation}
where we also included the frequency parameter $s$, see \eqref{defw}.
In the IR, \ie as $y\equiv \frac 1\rho\to 0$, it is convenient to redefine 
some of the fluctuations as 
\begin{equation}
fl_{a,b,c}\equiv y^{-1}\ fl_{a,b,c}^h\,,\qquad fl_h\equiv y^2\ fl_h^h\,.
\eqlabel{redfflh}
\end{equation}
Note that this redefinition mimic the corresponding redefinitions
of the related background scalars $f_{a,b,c}$ and $h$ in \eqref{defhors4}.
The IR asymptotic expansions take form:
\begin{equation}
\begin{split}
&fl_{a,b,c}^h=\sum_{m=0}^\infty fl_{a,b,c;m}^h\ y^m\,,\qquad fl_{K_{1,2,3}}=\sum_{m=0}^\infty fl_{K_{1,2,3};m}^h\ y^m\,,\\
&fl_{g}=\sum_{m=0}^\infty fl_{g;m}^h\ y^m\,,\qquad fl_{h}^h=\sum_{m=0}^\infty fl_{h;m}^h\ y^m\,.
\end{split}
\eqlabel{flir}
\end{equation}
They are uniquely characterized by
\begin{equation}
\biggl\{\ fl_{a;0}^h\,,\ fl_{b;0}^h\,,\ fl_{c;0}^h\,,\ fl_{K_1;0}^h\,,\ fl_{K_2;0}^h\,,\ fl_{K_3;0}^h\,,\ fl_{g;0}^h\
\biggr\}\,.
\eqlabel{irfl}
\end{equation}

Notice that in total we have, \eqref{uvfl} and \eqref{irfl}, $8+7=15=14+1$, \ie
we have the expected number of parameters, $=14$ (corresponding to the
total order of the non-redundant differential equations of motion for the fluctuations),
and a single arbitrary overall normalization amplitude $A$. We are free to fix $A$ as
we wish. We find it convenient to fix $A$ differently for different branches of the
fluctuations.

\section{Chirally symmetric DFP --- TypeA$ _s$, $\csb$ fluctuations}
\label{ascsb}

In this appendix we discuss the linearized fluctuations about TypeA$ _s$ dynamical fixed point
of the cascading gauge theory, spontaneously breaking the $U(1)_R$ chiral symmetry of this
DFP to $\zet_2$. The corresponding background and the fluctuation equations of motion
are the special case, a consistent truncation, of the
general equations discussed in appendix \ref{fgframe}.
Specifically,
\begin{itemize}
\item for the background we find \cite{Buchel:2019pjb}:
\begin{equation}
f_c\equiv f_2\,,\qquad f_a=f_b\equiv f_3\,,\qquad K_1=K_3\equiv K\,,\qquad K_2\equiv 1\,;
\eqlabel{sdfpb}
\end{equation}
\item for the fluctuations, note the rescaling of the $fl_a=-fl_b$ modes, we find:
\begin{equation}
\begin{split}
&fl_a=-fl_b\equiv f_3\cdot F\,,\qquad fl_{K_1}=-fl_{K_3}\equiv\chi_1\,,\qquad fl_{K_2}\equiv \chi_2\,,\\
&fl_c\equiv0\,,\qquad fl_g\equiv 0\,,\qquad fl_h\equiv 0\,.
\end{split}
\eqlabel{sdfpf}
\end{equation}
\end{itemize}
This particular mode, \ie $\{F$, $\chi_1$, $\chi_2\}$, is featured prominently
throughout the paper, so we discuss it in some details. The corresponding
equations of motion are given by:
\begin{equation}
\begin{split}
&0=F''+\frac{1}{16 b f_3^3 f_2 g^2 h^2 \rho (1+\rho) (f_3' \rho-2 f_3)} \biggl(
f_3^2 h f_2 g \rho^2 (1+\rho) (K')^2\\
&+2 h^2 f_2 f_3^4 b \rho^2 (1+\rho) (g')^2
+2 g^2 f_3^4 f_2 b \rho^2 (1+\rho) (h')^2+20 g^2 h^2 f_3^2 f_2 b \rho^2 (1+\rho)
(f_3')^2\\
&+16 f_3^3 h^2 f_2 g^2 b \rho (\rho s-3 \rho-3) f_3'+16 h g^2 f_3^4 f_2 b \rho (1+\rho) h'
-4 g^3 f_3^2 h (1+\rho) b^2\\&+2 g^2 b (-8 h^2 f_3^2 (\rho + 1) f_2^2
+ 24 h^2 ((\rho^2 (\rho + 1) h + 1 + (-\frac23 s + 1) \rho) f_3 + 2 \rho + 2) f_3^3 f_2
\\&- K^2 (\rho + 1))
\biggr) F'
-\frac{K'}{2f_3^2 h g b} \kappa_1'
+\frac{1}{32 (\rho + 1)^2 \rho^2 h^2 g^2 f_2 f_3^3 b (f_3' \rho - 2 f_3)}\\
&\times \biggl(
20 b \rho^3 s g^2 h^2 f_2 f_3^2 (\rho + 1) (f_3')^2
- 64 g (\frac 14\rho^2 f_2 (\rho + 1)^2 (K')^2 + g (((\rho^2 s f_2 (\rho + 1)^2 (s - 3) h
\\&- ((-6 + (s - 8) \rho) s \rho f_2)\frac14 + 9 (\rho + 1)^2) h f_3^2)\frac12
- 6 h f_2 (\rho + 1)^2 f_3 + b g (\rho + 1)^2) b) f_3 \rho h f_3'
\\&+ ((s + 32) \rho + 32) g f_3^2 f_2 (\rho + 1) \rho^2 h (K')^2
- 4 b (-\rho^3 s g^2 f_2 f_3^4 (\rho + 1) (h')^2\frac12 \\
&- 4 \rho^2 s g^2 h f_2 f_3^4 (\rho + 1) h' - \rho^3 s h^2 f_2 f_3^4 (\rho + 1) (g')^2\frac12
\\&+ g^2 (-16 (((s - \frac94) \rho + s - 3) f_2 s (\rho + 1) \rho^2 h
- ((-3 + (s - 5) \rho) s \rho f_2)\frac14 + 9 (\rho + 1)^2) h^2 f_3^4
\\&- 24 ((s - 8) \rho - 8) f_2 (\rho + 1) h^2 f_3^3 + (\rho + 1) h (((s - 32) \rho - 32) b g
+ 4 \rho s h f_2^2) f_3^2 \\&+ \rho s K^2 (\rho + 1)\frac12))
\biggr) F
-\frac{\kappa_1 f_2 (K') \rho^2 s + 8 \kappa_2 g^2 b^2 (\rho + 1)}{4 (\rho + 1) h g f_3^2 \rho^2 f_2 b}
\,,
\end{split}
\eqlabel{sfl1}
\end{equation}
\begin{equation}
\begin{split}
&0=\kappa_1''+\frac{1}{16 \rho (f_3' \rho - 2 f_3) b f_3^3 f_2 g^2 h^2 (\rho + 1)}
\biggl(
-12 b \rho^2 g^2 h^2 f_2 f_3^2 (\rho + 1) (f_3')^2 \\
&- 16 g b f_3^3 f_2 (\rho h (\rho + 1) g'
+ ((\rho^2 + \rho) h' - h (1 + (s + 1) \rho)) g) \rho h f_3'
\\&+ 2 b \rho^2 h^2 f_2 f_3^4 (\rho + 1) (g')^2 + 32 b \rho g h^2 f_2 f_3^4 (\rho + 1) g'
- 4 g (-b \rho^2 g f_2 f_3^4 (\rho + 1) (h')^2\frac12 \\
&- 12 b \rho g h f_2 f_3^4 (\rho + 1) h'
- \rho^2 h f_2 f_3^2 (\rho + 1) (K')^2\frac14 + (-12 f_2 (\rho^2 (\rho + 1) h + 1
\\&+ (-\frac23 s + 1) \rho) h^2 f_3^4 - 24 h^2 f_2 (\rho + 1) f_3^3 + h (\rho + 1) (4 h f_2^2 + b g) f_3^2
+ K^2 (\rho + 1)\frac12) g b)\biggr) \kappa_1'
\\&+2 K' F'+\frac{1}{32 (\rho + 1)^2 \rho^2 h^2 g^2 f_2 f_3^3 b (f_3' \rho - 2 f_3)}
\biggl(-12 b r^3 s g^2 h^2 f_2 f_3^2 (\rho + 1) (f_3')^2
\\&- 16 g b (\rho^2 s h f_2 (\rho + 1) g'
+ g (\rho^2 s f_2 (\rho + 1) h' + 2 (\rho^2 s f_2 (\rho + 1)^2 (s - 3) h
- \rho s (\rho s + 2) f_2\frac14 \\
&+ 9 (\rho + 1)^2) h)) f_3^3 \rho h f_3'
+ 2 b \rho^3 s h^2 f_2 f_3^4 (\rho + 1) (g')^2 + 32 b \rho^2 s g h^2 f_2 f_3^4 (\rho + 1) g'
\\&- 4 g (-b \rho^3 s g f_2 f_3^4 (\rho + 1) (h')^2\frac12 - 12 b \rho^2 s g h f_2 f_3^4 (\rho + 1) h'
- \rho^3 s h f_2 f_3^2 (\rho + 1) (K')^2\frac14 \\
&+ g b (-16 (((s - \frac94) \rho + s
- 3) f_2 s (\rho + 1) \rho^2 h - ((-3 + (s - 5) \rho) s \rho f_2)\frac14 \\&+ 9 (\rho + 1)^2) h^2 f_3^4
- 24 \rho s h^2 f_2 (\rho + 1) f_3^3
+ \rho s h (\rho + 1) (4 h f_2^2 + b g) f_3^2 \\&+ \rho s K^2 (\rho + 1)\frac12))
\biggr) \kappa_1
+\frac{F f_3^2 h f_2 K' \rho^2 s + 2 b K g (\rho + 1) (\kappa_2 + 2 F)}{(\rho + 1) h f_3^2
\rho^2 f_2}\,,
\end{split}
\eqlabel{sfl2}
\end{equation}
\begin{equation}
\begin{split}
&0=\kappa_2''+\frac{1}{16 \rho (f_3' \rho - 2 f_3) b f_3^3 f_2 g^2 h^2 (\rho + 1)}\biggl(
-12 b \rho^2 g^2 h^2 f_2 f_3^2 (\rho + 1) (f_3')^2 \\&+ 16 g b f_3^3 f_2 \rho h (\rho h (\rho + 1) g'
- ((\rho^2 + \rho) h' - h (1 + (s + 1) \rho)) g) f_3' \\&+ 2 b \rho^2 h^2 f_2 f_3^4 (\rho + 1) (g')^2
- 32 b \rho g h^2 f_2 f_3^4 (\rho + 1) g' - 4 g (-b \rho^2 g f_2 f_3^4 (\rho + 1) (h')^2\frac12
\\&- 12 b \rho g h f_2 f_3^4 (\rho + 1) h' - \rho^2 h f_2 f_3^2 (\rho + 1) (K')^2\frac14 + (-12 f_2 (\rho^2
(\rho + 1) h
+ 1 \\&+ (-\frac23 s + 1) \rho) h^2 f_3^4 - 24 h^2 f_2 (\rho + 1) f_3^3 + h (\rho + 1) (4 h f_2^2 + b g) f_3^2
+ K^2 (\rho + 1)\frac12) g b)
\biggr)  \kappa_2'
\\&+\frac{1}{32 (\rho + 1)^2 \rho^2 h^2 g^2 f_2 f_3^3 b (f_3' \rho - 2 f_3)}\biggl(
-12 b \rho^3 s g^2 h^2 f_2 f_3^2 (\rho + 1) (f_3')^2 \\&+ 16 g (\rho^2 s h f_2 (\rho + 1) g' - g (\rho^2 s f_2 (\rho + 1) h'
+ 2 (\rho^2 s f_2 (\rho + 1)^2 (s - 3) h - \rho s (\rho s + 2) f_2\frac14
\\&+ 9 (\rho + 1)^2) h)) b f_3^3 \rho h f_3' + 2 b \rho^3 s h^2 f_2 f_3^4 (\rho + 1) (g')^2 - 32 b \rho^2 s g h^2 f_2 f_3^4 (\rho + 1)
g' \\&- 4 g (-b \rho^3 s g f_2 f_3^4 (\rho + 1) (h')^2\frac12 - 12 b \rho^2 s g h f_2 f_3^4 (\rho + 1) h'
- \rho^3 s h f_2 f_3^2 (\rho + 1) (K')^2\frac14\\& + g b (-16 (((s - \frac94) \rho + s - 3) f_2 s (\rho + 1) \rho^2 h - ((-3 + (s - 5) \rho)
s \rho f_2)\frac14 \\&+ 9 (\rho + 1)^2) h^2 f_3^4 - 24 \rho s h^2 f_2 (\rho + 1) f_3^3
+ \rho s h (\rho + 1) (4 h f_2^2 + b g) f_3^2 \\&+ \rho s K^2 (\rho + 1)\frac12))
\biggr) \kappa_2
-\frac{18 F}{\rho^2 f_2} + \frac{9 \kappa_1 K}{2 b \rho^2 f_2 g h f_3^2}\,.
\end{split}
\eqlabel{sfl3}
\end{equation}
Since in this sector the fluctuations of $fl_h$ are not activated \eqref{sdfpf}, the asymptotics
are much simpler compare to the general case of section \ref{appasymptotics}.
In the UV, \ie as $\rho\to 0$,
\begin{equation}
F=f_{3,0}\rho^3+\sum_{n=4}^\infty\ \rho^n\cdot \sum_{m} f_{n,m}\ \ln^m\rho\,,
\eqlabel{uvsdpf1}
\end{equation}
\begin{equation}
\kappa_1=\left(\frac23 b (f_{3,0}+\kappa_{2;3,0})+2 f_{3,0} b \ln\rho\right)\rho^3+\sum_{n=4}^\infty\ \rho^n\cdot
\sum_{m} \kappa_{1;n,m}\ \ln^m\rho\,,
\eqlabel{uvsdpf2}
\end{equation}
\begin{equation}
\kappa_2=\left(\kappa_{2;3,0}+3f_{3,0}\ln\rho\right)\rho^3+\sum_{n=4}^\infty\ \rho^n\cdot \sum_{m} \kappa_{2;n,m}\ \ln^m\rho\,.
\eqlabel{uvsdpf3}
\end{equation}
In the IR, \ie as $y\equiv \frac 1\rho\to 0$,
\begin{equation}
F=\sum_{n=0}^\infty f_{n}^h\ y^n\,,\qquad \kappa_{1}=\sum_{n=0}^\infty \kappa_{1;n}^h\ y^n
\,,\qquad \kappa_{2}=\sum_{n=0}^\infty \kappa_{2;n}^h\ y^n \,.
\eqlabel{irsdfp}
\end{equation}
The mode asymptotics 
\begin{equation}
\begin{split}
&{\rm UV}:\qquad \{\ s\,,\, f_{3,0}\,,\, \kappa_{2;3,0}\,,\, \kappa_{2;7,0}\ \}\,,\\
&{\rm IR}:\qquad \{\  f_0^h\,,\, \kappa_{1;0}^h\,,\, \kappa_{2,0}^h \}\,,
\end{split}
\eqlabel{collsdfp}
\end{equation}
are completely specified by  $4+3=1+6$ parameters. One of the parameters from the set  $\{f_{3,0}$, $k_{2,3,0}$, $k_{2,7,0}\}$,
plays the role of the overall normalization $A$ in \eqref{uvfl}, and the number of the
remaining ones match the total order of the coupled differential equations for the fluctuations
\eqref{sfl1}-\eqref{sfl3}: $3\times 2=6$. One of the physical parameter, \ie $s$, determines
the frequency of this $\csb$ mode about TypeA$ _s$ DFP, see \eqref{defw}.

In the rest of this appendix we analyze the near-conformal $b\to 0$, equivalently
$H\gg \Lambda$, limit of this mode. Strictly at $b=0$ the cascading gauge theory is
conformal, and the spectra can be computed analytically \cite{Buchel:2022hjz}.
We discover multiple spectral branches of the fluctuations. 
On some branches we are  able to compute analytically the leading $\calo(\sqrt{b})$,
and numerically the first $\calo(b)$ subleading,
corrections to the conformal spectra, sections \ref{detailss3} and \ref{detailss7}.
On the remaining branches we compute numerically  the leading $\calo(b)$ corrections to
the conformal spectra, section \ref{detailss7l}. Perturbative results obtained here provide
a valuable check of the finite $\frac{H}{\Lambda}$ spectra in the near-conformal
limit, see fig.~\ref{figure1}.

\subsection{Near-conformal limit: $b\to 0$}
\label{appnearcsym}

In the near-conformal limit the background of TypeA$ _s$ DFP is represented by
\begin{equation}
\begin{split}
&f_2=(1+\rho)\ \left(1+\sum_{n=1}^\infty b^n\ f_{2;n}(\rho)\right)\,,\qquad f_3=(1+\rho)\
\left(1+\sum_{n=1}^\infty b^n\ f_{3;n}(\rho)\right)\,,\\
&h=\frac{1}{4(1+\rho)^2}\  \left(1+\sum_{n=1}^\infty b^n\ h_{n}(\rho)\right)\,,\ K=1+\sum_{n=1}^\infty b^n\ k_{n}(\rho)\,,
\ g=1+\sum_{n=1}^\infty b^n\ g_{n}(\rho)\,.
\end{split}
\eqlabel{ktschb1}
\end{equation}
Explicit equations for $\{f_{2n},f_{3n},h_n,k_n,g_n\}$ for $n=1,2$
along with the UV/IR asymptotics
are presented in appendix D.1 of \cite{Buchel:2019pjb}.
There is a useful analytical solution for $k_1$:
\begin{equation}
k_1=\frac\rho4+\frac{1}{4+4\rho}-\frac14-4\ln2+\frac{\rho^3-6\rho^2-24\rho-16}
{8 (1+\rho)^{3/2}}\
\ln\frac{\sqrt{1+\rho}-1}{\sqrt{1+\rho}+1}\,.
\eqlabel{k1anal}
\end{equation}
The coupled system of the linearized fluctuations \eqref{sfl1}-\eqref{sfl3} can be
simplified  introducing
\begin{equation}
\kappa_1=\frac b3 (q_3-q_7)\,,\qquad \kappa_2=\frac 12 (q_3+q_7)\,.
\eqlabel{decfl}
\end{equation}
To leading order in $b$, we find from  \eqref{sfl1}-\eqref{sfl3}:
\begin{equation}
\begin{split}
&0=F''+\frac{2 \rho s-\rho-6}{2\rho (\rho+1)}\ F'
-\frac{3(\rho s-2 \rho-2)}{2(\rho+1)^2 \rho^2}\ F\,,\\
&0=q_3''+\frac{2 \rho s-\rho-6}{2\rho (\rho+1)}\ q_3'
-\frac{3(\rho s-2 \rho-2)}{2(\rho+1)^2 \rho^2}\ q_3+3 k_1'\ F'
+\frac{ 3(k_1' \rho^2 s+4)}{2\rho^2 (\rho+1)}\ F\,,\\
&0=q_7''+\frac{2 \rho s-\rho-6}{2\rho (\rho+1)}\ q_7'
-\frac{3(\rho s+14 \rho+14)}{2(\rho+1)^2 \rho^2}\ q_7
-3 k_1'\ F'-\frac{3 (k_1' \rho^2 s+28)}{2(\rho+1) \rho^2}\ F\,.
\end{split}
\eqlabel{q3q7}
\end{equation}
Solving the decoupled equation for $F$, we find (up to an overall normalization $A_F$)
\begin{equation}
F=A_F\ \frac{\rho^3}{(1+\rho)^s}\ _2 F_1\left(\frac 32,3-s; 3; -\rho\right)\,,\qquad s=3,4,\cdots \,.
\eqlabel{bra}
\end{equation}
Given \eqref{bra}, and using \eqref{k1anal}, it is straightforward to see
that it is impossible to solve the equation for $q_3$ in \eqref{q3q7},
so that this mode
is both normalizable as $\rho\to 0$ and analytic as $\rho \to \infty $ ---
this means that the amplitude of $F$ must always vanish in the limit
$b\to 0$. This is precisely what we find, see \eqref{sab} and \eqref{sc}.

With $F\equiv 0$, we find from \eqref{q3q7} the following leading order as $b\to 0$ solutions:
\begin{equation}
q_3=A_3\ \frac{\rho^3}{(1+\rho)^s}\ _2 F_1\left(\frac 32,3-s; 3; -\rho\right)\,,\qquad s=3,4,\cdots \,,
\eqlabel{brb}
\end{equation}
and
\begin{equation}
q_7=A_7\ \frac{\rho^7}{(1+\rho)^s}\ _2 F_1\left(\frac {11}{2},7-s; 11; -\rho\right)\,,\qquad
s=7,8,\cdots \,.
\eqlabel{brc}
\end{equation}

Extending the leading order solutions \eqref{brb}, \eqref{brc} perturbatively in $b$
we identify three branches:
\nxt A pair of non-analytic\footnote{Related phenomenon was observed
earlier in \cite{Buchel:2011cc} and \cite{Buchel:2020nqs}.} in $b$ branches, $(A_b)$ and $(B_b)$,
\begin{equation}
\begin{split}
&s\bigg|_{A,B}= n+\sum_{k=1}^\infty (\pm )^k s_{n;k}\ b^{k/2}\,,
\qquad n\in \naturals \ge 3\,,\qquad q_3\bigg|_{A,B}=\sum_{k=0}^\infty (\pm )^k q_{3;n;k}\ b^{k/2}\,,\\
&F\bigg|_{A,B}=\sum_{k=1}^\infty (\pm )^k F_{n;k}\ b^{k/2}\,,\qquad
q_7\bigg|_{A,B}=\sum_{k=0}^\infty (\pm )^k q_{7;n;k}\ b^{k/2}\,,
\end{split}
\eqlabel{sab}
\end{equation}
with $q_{3;n;0}$ given by \eqref{brb} and $q_{7;n\ge 7;0}$ given by \eqref{brc} with $s=n$.
$q_{7;n;0}\equiv 0$ for $3\le n< 7$.
\nxt An analytic in $b$ branch $(C_b)$,
\begin{equation}
\begin{split}
&s\bigg|_{C}= n+\sum_{k=1}^\infty  s_{n;k}\ b^{k}\,,
\qquad n\in \naturals \ge 7\,,\qquad q_3\bigg|_{C}=\sum_{k=0}^\infty q_{3;n;k}\
b^{k}\,,\\
&F\bigg|_{C}=\sum_{k=1}^\infty  F_{n;k}\ b^{k}\,,\qquad
q_7\bigg|_{C}=\sum_{k=0}^\infty q_{7;n;k}\ b^{k}\,,
\end{split}
\eqlabel{sc}
\end{equation}
where $q_{7;n;0}$ is given by \eqref{brc} and $q_{3;n;0}$ is given by \eqref{brb}
with $s=n$.

\subsubsection{Details of $s=3\pm\calo(\sqrt{b})$ branches:  (A$ _b$) and (B$ _b$)}
\label{detailss3}

From \eqref{brb}, here  
\begin{equation}
s_{3;0}=3\,,\qquad q_{3;3;0}=
\mathcolorbox{red}{1}\cdot\ \frac{\rho^3}{(1+\rho)^s}\ _2 F_1\left(\frac 32,3-s; 3;
-\rho\right)\bigg|_{s=s_{3;0}}= \frac{\rho^3}{(1+\rho)^3}\,,
\eqlabel{zero3}
\end{equation}
where we highlighted the (fixed) overall normalization of the linearized fluctuations;
the latter implies that in the UV, \ie $\rho\to 0$, expansion of $q_{3;3;k\ge 1}$
the order $\calo(\rho^3)$ terms are absent. 
Because the leading order fluctuation spectra \eqref{bra} and \eqref{brb} are degenerate,
the equations for $F_{3;k}$ will necessarily contain zero modes; specifically, if $F_{3;k\ge 1}$
is a solution, so is $(F_{3;k}+\alpha_k\cdot q_{3;3;0})$ for an arbitrary set of constants $\alpha_k$.
As we will see shortly, the zero modes at order $k$ will be completely fixed at order
$k+1$. We find it convenient to set
\begin{equation}
F_{3;k}\equiv \mathcolorbox{orange}{\alpha_k}\cdot \ q_{3;3;0}+\hF_{3;k}\,,
\eqlabel{deffh3}
\end{equation}
with the understanding that in the UV  expansion of $\hF_{3;k}$
the order $\calo(\rho^3)$ terms are absent.

Using \eqref{zero3}, \eqref{deffh3}, the perturbative ansatz \eqref{ktschb1}
and \eqref{sab}, we find from \eqref{sfl1} the leading order equation for
$\hF_{3;1}$,
\begin{equation}
0=\hF_{3;1}''+\frac{5 \rho-6}{2\rho (\rho+1)}\ \hF_{3;1}'
-\frac{3(\rho-2)}{2(\rho+1)^2 \rho^2}\ \hF_{3;1}\,.
\eqlabel{fh13}
\end{equation}
The most general solution to \eqref{fh13} is specified by two integration constants
$C_1$ and $C_2$:
\begin{equation}
\hF_{3;1}=C_1\cdot  \frac{\rho^3}{(1+\rho)^3}+C_2\cdot \frac{\rho}{(1+\rho)^3}
\left(
\rho^2\ \ln\frac{\sqrt{1+\rho}+1}{\sqrt{1+\rho}-1}-2(\rho+2) \sqrt{1+\rho}
\right)\,.
\eqlabel{fh13sol}
\end{equation}
Normalizability of the fluctuations sets $C_2=0$; the boundary condition imposed by
\eqref{deffh3} further sets $C_1=0$, resulting in
\begin{equation}
\hF_{3,1}\equiv 0\,.
\eqlabel{fh13sol1}
\end{equation}
Likewise, we find
\begin{equation}
q_{7;3;0}=0\,.
\eqlabel{sol730}
\end{equation}

The subleading set of equations involving constants
$\alpha_1$, $s_{3;1}$, and functions $\{q_{3;3;1}$, $q_{7;3;1}$, $\hF_{3;2}\}$ reads:
\begin{equation}
\begin{split}
&0=q_{3;3;1}''+\frac{5 \rho-6}{2(1+\rho) \rho} q_{3;3;1}'
-\frac{3(\rho-2)q_{3;3;1}}{2\rho^2 (1+\rho)^2} +\frac{9\alpha_1 \rho^2 (\rho+2)}{2(1+\rho)^4}
k_1'+\frac{3(4 \alpha_1 (\rho+1)+\rho s_{3;1}) \rho}{2(1+\rho)^5}\,,
\end{split}
\eqlabel{order31}
\end{equation}
\begin{equation}
\begin{split}
&0=q_{7;3;1}''+\frac{5 \rho-6}{2(1+\rho) \rho} q_{7;3;1}'
-\frac{3(17 \rho+14)}{2\rho^2 (1+\rho)^2} q_{7;3;1}
-\frac{9\alpha_1 \rho^2 (\rho+2)}{2(1+\rho)^4} k_1'
-\frac{42 \alpha_1 \rho}{(1+\rho)^4}\,,
\end{split}
\eqlabel{order32}
\end{equation}
\begin{equation}
\begin{split}
&0=\hF_{3;2}''+\frac{5 \rho-6}{2(1+\rho) \rho} \hF_{3;2}'
-\frac{3(\rho-2)}{2\rho^2 (1+\rho)^2} \hF_{3;2}
-\frac{\rho^2 (\rho+2)}{(1+\rho)^4} k_1'+\frac{(3 \alpha_1 \rho s_{3;1}-8 (\rho+1)) \rho}
{2(1+\rho)^5}\,.
\end{split}
\eqlabel{order33}
\end{equation}
Above set can be solved numerically --- and we explain how to do it for the
set of equations at the next order --- here instead we show that the most
important constant, \ie $s_{3;1}$ can be computed analytically:
\begin{itemize}
\item Substituting
\begin{equation}
\hF_{3;2}=q_{3;3;0}\ G_{3;2}\,,
\eqlabel{so31}
\end{equation}
and using \eqref{k1anal}, 
we find a general analytic solution for $G_{3;2}'$,
\begin{equation}
\begin{split}
&G_{3;2}'=-\frac{\rho^3}{16(1+\rho)^{5/2}} \ln\frac{\sqrt{1+\rho}+1}{\sqrt{1+\rho}-1}
+\frac{\sqrt{1+\rho}}{\rho^3} C_1
-\frac{(3 \rho^2+12 \rho+8) }{\rho^3 (1+\rho)}\ \alpha_1 s_{3;1}\\
&+\frac{(\rho+2) (3 \rho^4-8 \rho^3+56 \rho^2+128 \rho+64)}{24(1+\rho)^2 \rho^3}\,.
\end{split}
\eqlabel{so32}
\end{equation}
Normalizability of $\hF_{3;2}$ sets
\begin{equation}
C_1=-\frac{16}{3}+8 \alpha_1 s_{3;1}\,.
\eqlabel{so33}
\end{equation}
As $\rho \to \infty$,
\begin{equation}
G_{3;2}'=-\frac{1}{48} \left(-\frac{384}{5}+144 \alpha_1 s_{3;1}\right)
\rho^{-2}-\frac{1}{48} \left(-384 \alpha_1 s_{3;1}+256\right) \rho^{-5/2}+\calo(\rho^{-3})\,,
\eqlabel{so34}
\end{equation}
thus analyticity of $\hF_{3;2}$, and thus $G_{3;2}'$, in this limit requires 
\begin{equation}
\alpha_1=\frac{2}{3s_{3;1}}\qquad \Longrightarrow\qquad  C_1=0\,,
\eqlabel{so35}
\end{equation}
which is also evident from \eqref{so32}.
\item We continue with \eqref{order31}, setting
\begin{equation}
q_{3;3;1}=\frac{1}{s_{3;1}}\ q_{3;3;0}\ J_{3;3;1}\,,
\eqlabel{so36}
\end{equation}
allows to solve analytically for $J_{3;3;1}'$,
\begin{equation}
\begin{split}
&J_{3;3;1}'=\frac{3\rho^3}{16(1+\rho)^{5/2}} \ln\frac{\sqrt{1+\rho}+1}{\sqrt{1+\rho}-1}
-\frac{3 \rho^2+12 \rho+8}{\rho^3 (1+\rho)}\ s_{3;1}^2+\frac{\sqrt{1+\rho}}{\rho^3}\ C_1
\\&-\frac{(\rho+2) (3 \rho^4-8 \rho^3-72 \rho^2-128 \rho-64)}{8(1+\rho)^2 \rho^3}\,.
\end{split}
\eqlabel{so37}
\end{equation}
Normalizability of $q_{3;3;1}$ sets
\begin{equation}
C_1=8 s_{3;1}^2-16\,,
\eqlabel{so38}
\end{equation}
and  analyticity of $q_{3;3;1}$, and thus $J_{3;3;1}'$, in the limit
$\rho \to \infty$ requires
\begin{equation}
s_{3;1}^2=2\qquad \Longrightarrow\qquad  s_{3;1}=\pm \sqrt{2}\,.
\eqlabel{so39}
\end{equation}
\end{itemize}
Note that to determine $s_{3;1}$, there is no need to solve for $q_{7;3;1}$ ---
of course, this solution is needed for the computation of the higher order corrections $s_{3;k\ge 2}$. 

The sub-subleading set of equations involving constants
$\alpha_2$, $s_{3;2}$, and functions $\{q_{3;3;2}$, $\hF_{3;3}\}$ reads (we omit the equation for
$q_{7;3;2}$ as it is not need to compute $s_{3;2}$; it is required for
the computation of $s_{3;k\ge 3}$):
\begin{equation}
\begin{split}
&0=q_{3;3;2}''+\frac{5 \rho-6}{2(1+\rho) \rho} q_{3;3;2}'
-\frac{3(\rho-2)}{2\rho^2 (1+\rho)^2} q_{3;3;2}+\frac{s_{3;1} q_{3;3;1}'}{1+\rho}
-\frac{3s_{3;1} q_{3;3;1}}{2\rho (1+\rho)^2}
+3 k_1' \hF_{3;2}'\\&+\frac{3(3 k_1' \rho^2+4)}{2(1+\rho) \rho^2} \hF_{3;2}
+\frac{3\rho\alpha_2(3\rho k_1'(\rho+2)+4)}{2(1+\rho)^4}
+\frac{3\rho^2 s_{3;2}}{2(1+\rho)^5}-\frac{3\rho^3 (k_1')^2}{8(1+\rho)^3}
+\frac{\rho^3 k_1'}{(1+\rho)^4}
\\&-\frac{3 \rho^2 (\rho+2) f_{3;1}'}{(1+\rho)^4}
-\frac{9\rho^2 (\rho+2) h_1'}{4(1+\rho)^4}+\frac{3\rho}{8(1+\rho)^5} \biggl(
-3 h_1 (\rho^2+16 \rho+16)\\&+4 (1+\rho) (12 k_1-3 f_{2;1}-20 f_{3;1}+1)
\biggr)\,,
\end{split}
\eqlabel{order321}
\end{equation}
\begin{equation}
\begin{split}
&0=\hF_{3;3}''+\frac{5 \rho-6}{2(1+\rho) \rho} \hF_{3;3}'
-\frac{3(\rho-2)}{2\rho^2 (1+\rho)^2} \hF_{3;3}
+\frac{s_{3;1}}{1+\rho} \hF_{3;2}'
-\frac{3s_{3;1} \hF_{3;2}}{2\rho (1+\rho)^2}
-\frac23 k_1' q_{3;3;1}'\\&
-\frac{k_1' \rho^2+4}{(1+\rho) \rho^2} q_{3;3;1}+\frac23 k_1' q_{7;3;1}'
+\frac{k_1' \rho^2-4}{(1+\rho) \rho^2} q_{7;3;1}
+\frac{\rho^2 s_{3;2}}{s_{3;1} (1+\rho)^5}-\frac{19\rho^3 (k_1')^2}{12s_{3;1} (1+\rho)^3}
\\&-\frac{\rho^3 s_{3;1} k_1'}{3(1+\rho)^4}
-\frac{\rho^2 (\rho+2) h_1'}{2s_{3;1} (1+\rho)^4}
+\frac{3\rho^2 s_{3;1} \alpha_2}{2(1+\rho)^5}+\frac{\rho}{12s_{3;1} (1+\rho)^5}
\biggl(
-3 h_1 (\rho+4) (3 \rho+4)\\&+4 (1+\rho) (12 k_1+15 f_{2;1}-36 f_{3;1}-13)\biggr)\,. 
\end{split}
\eqlabel{order322}
\end{equation}
Eqs. \eqref{order321} and \eqref{order322} are solved subject to the asymptotic
expansions,
\nxt in the UV, \ie as $\rho\to 0$,
\begin{equation}
\begin{split}
&q_{3;3;2}=\biggl(\mathcolorbox{red}{0}+6 \alpha_2 \ln\rho\biggr) \rho^3+\biggl(-3 \alpha_2-\frac32 f_{2,1,0;1}-\frac12 s_{3;2}
+(-3 \alpha_1 s_{3;1}-18 \alpha_2) \ln\rho
\biggr) \rho^4
\\&+\calo(\rho^5\ln\rho)\,,\\
&\hF_{3;3}=\mathcolorbox{orange}{0}\ \rho^3+\biggl(
-\frac32 \alpha_1 f_{2,1,0;1}-\frac12 \alpha_1 s_{3;2}-\frac12 s_{3;1} \alpha_2
\biggr) r^4+\calo(\rho^5\ln\rho)\,,
\end{split}
\eqlabel{uvq332}
\end{equation}
it is completely specified by $\{\alpha_2,s_{3;2}\}$; we further highlighted
 arbitrary constants, fixed to zero by the overall normalization \eqref{zero3} and the
 extraction of the zero mode in $F_{3;3}$ \eqref{deffh3};
\nxt  in the IR, \ie as $y\equiv \frac1\rho\to 0$,
\begin{equation}
q_{3;3;2}=q_{3;3;2;0}^h+\calo(y)\,,\qquad \hF_{3;3}=\hF_{3;3;0}^h+\calo(y)\,,
\eqlabel{irq332}
\end{equation}
it is completely specified by
\begin{equation}
\{q_{3;3;2;0}^h\,,\, \hF_{3;3;0}^h\,,\, \alpha_2\,,\, s_{3;2}\}\,.
\eqlabel{setq332}
\end{equation}
In total, the UV and IR expansions are completely determined by the parameters
\eqref{setq332}, which is precisely what is needed to find a unique solution
for a pair of second order ODEs \eqref{order321} and \eqref{order322}.
Solving these equations we find
\begin{equation}
s_{3;2}=-1.5748(9)\,,\ \alpha_2=0.7936(8)\,,\ q_{3;3;2;0}^h=4.174(3)\,,\
\hF_{3;3;0}^h=\mp  0.40398(7)\,.
\eqlabel{q332resu}
\end{equation}

Once the numerical solution for $\{q_{3;3;2},\hF_{3;3}\}$ is found, the second order
ODE for $q_{7;3;2}$ --- necessary to determine $s_{3;k\ge 3}$ --- is solved adjusting
two parameters,
\begin{equation}
\{q_{7;3;2;7,0}\,,\, q_{7;3;2;0}^h\}\,,
\eqlabel{q372}
\end{equation}
that completely determines its UV and IR asymptotics.

\subsubsection{Details of $s=7\pm\calo(\sqrt{b})$ branches: (A$ _b$) and (B$ _b$) }
\label{detailss7}

From \eqref{brb}, here  
\begin{equation}
\begin{split}
s_{7;0}=7\,,\qquad q_{3;7;0}=&
\mathcolorbox{red}{1}\cdot\ \frac{\rho^3}{(1+\rho)^s}\ _2 F_1\left(\frac 32,3-s; 3;
-\rho\right)\bigg|_{s=s_{7;0}}\\=&
\frac{\rho^3(21\rho^4+112\rho^3+240\rho^2+256\rho+128)}{128(1+\rho)^7}\,,
\end{split}
\eqlabel{7zero3}
\end{equation}
where we highlighted the (fixed) overall normalization of the linearized fluctuations;
the latter implies that in the UV, \ie $\rho\to 0$, expansion of $q_{3;7;k\ge 1}$
the order $\calo(\rho^3)$ terms are absent. 
Because the leading order fluctuation spectra  \eqref{bra}, \eqref{brb} and  \eqref{brc} are degenerate
at $s_{7;0}$,
the equations for $F_{7;k}$ and $q_{7;7;k}$ will necessarily contain zero modes; specifically, if $F_{7;k\ge 1}$
and $q_{7;7;k\ge 1}$
are solutions, so are $(F_{7;k}+\alpha_k\cdot q_{3;7;0})$ and $(q_{7;7;k}+\frac{\beta_k}{\beta_0}\cdot q_{7;7;0})$,
\begin{equation}
q_{7;7;0}=\beta_0\cdot \frac{\rho^7}{(1+\rho)^s}\ _2 F_1\left(\frac {11}{2},7-s; 11;
-\rho\right)\bigg|_{s=s_{7;0}}=\beta_0\cdot
\frac{\rho^7}{(1+\rho)^7}\,,
\eqlabel{q770}
\end{equation}
for an arbitrary set of constants $\{\alpha_k,\beta_k\}$.
As in section \ref{detailss3}, the zero modes at order $k$ will be completely fixed at order
$k+1$. We find it convenient to set
\begin{equation}
F_{7;k\ge 1}\equiv \mathcolorbox{orange}{\alpha_k}\cdot  q_{3;7;0}\ +\hF_{7;k}\,,
\qquad q_{7;7;k\ge 1}\equiv \mathcolorbox{green}{\beta_k}\cdot  \frac{1}{\beta_0} q_{7;7;0}\ +\hq_{7;7;k}\,,
\eqlabel{7deffh3}
\end{equation}
with the understanding that in the UV  expansion of $\hF_{7;k}$
the order $\calo(\rho^3)$ terms are absent, and in the UV  expansion of $\hq_{7;7;k}$
the order $\calo(\rho^7)$ terms are absent.

As in section \ref{detailss3}, the equation for $\hF_{7;1}$ is homogeneous, and the boundary
condition implied by \eqref{7deffh3} sets 
\begin{equation}
\hF_{7,1}\equiv 0\,.
\eqlabel{7fh13sol1}
\end{equation}

The subleading set of equations involving constants
$\alpha_1$, $\beta_0$, $s_{7;1}$, and functions $\{q_{3;3;1}$, $q_{3;7;1}$, $\hF_{3;2}\}$ reads:
\begin{equation}
\begin{split}
&0=q_{3;7;1}''+\frac{13 \rho-6}{2(1+\rho) \rho} q_{3;7;1}'-\frac{3(5 \rho-2)}{2\rho^2 (1+\rho)^2} q_{3;7;1}
+\frac{3\rho^2 (\rho+2)  \alpha_1 k_1'}{256(1+\rho)^8} \biggl(147 \rho^4+560 \rho^3\\
&+944 \rho^2+768 \rho+384\biggr)
+\frac{
\rho}{256(1+\rho)^9}\biggl(
\rho (7 \rho^4+48 \rho^3+144 \rho^2+256 \rho+384) s_{7;1}\\&+12 \alpha_1 (1+\rho) 
(21 \rho^4+112 \rho^3+240 \rho^2+256 \rho+128)
\biggr)\,,
\end{split}
\eqlabel{7order31}
\end{equation}
\begin{equation}
\begin{split}
&0=\hq_{7;7;1}''+\frac{13 \rho-6}{2(1+\rho) \rho} \hq_{7;7;1}'-\frac{21(3 \rho+2)}{2\rho^2 (1+\rho)^2} \hq_{7;7;1}
-\frac{3\rho^2 (\rho+2)\alpha_1 k_1'}{256(1+\rho)^8} \biggl(147 \rho^4+560 \rho^3\\
&+944 \rho^2+768 \rho+384\biggr) -\frac{21\rho \alpha_1}{64(1+\rho)^8} \biggl(
21 \rho^4+112 \rho^3+240 \rho^2+256 \rho+128\biggr)
\\&+\frac{11s_{7;1} \beta_0 \rho^6}{2(1+\rho)^9}\,,
\end{split}
\eqlabel{7order32}
\end{equation}
\begin{equation}
\begin{split}
&0=\hF_{7;2}''+\frac{13 \rho-6}{2(1+\rho) \rho} \hF_{7;2}'-\frac{3(5 \rho-2)}{2\rho^2 (1+\rho)^2} \hF_{7;2}
-\frac{(\rho+2)\rho^2 k_1'}{384 (1+\rho)^8} \biggl(
147 \rho^4+560 \rho^3+944 \rho^2\\&+768 \rho+384\biggr)
+\frac{\rho}{256(1+\rho)^9} \biggl(
\alpha_1 s_{7;1} \rho (7 \rho^4+48 \rho^3+144 \rho^2+256 \rho+384)-8 (1+\rho) \\
&\times (21 \rho^4+112 \rho^3+240 \rho^2+256 \rho+128)\biggr)
+\frac{\rho^5 \beta_0 (7 \rho(\rho+2) k_1'-12)}{3(1+\rho)^8}\,.
\end{split}
\eqlabel{7order33}
\end{equation}
We show here that the most
important constant, \ie $s_{7;1}$ can be computed analytically:
\begin{itemize}
\item Substituting
\begin{equation}
\beta_0=\frac{\alpha_1}{s_{7;1}}\ p\,,\qquad \hq_{7;7;1}=\alpha_1\cdot \frac{1}{\beta_0}q_{7;7;0}\cdot B_{7;7;1}\,,
\eqlabel{7so31}
\end{equation}
and using \eqref{k1anal}, 
we find a general analytic solution for $B_{7;7;1}'$,
\begin{equation}
\begin{split}
&B_{7;7;1}'=-\frac{3(315 \rho^4+1400 \rho^3+2552 \rho^2+2304 \rho+1152)}{20480\rho (1+\rho)^{5/2}}
\ln\frac{\sqrt{1+\rho}+1}{\sqrt{1+\rho}-1}
\\&-\frac{p}{63\rho^{11} (1+\rho)}
\biggl(
77 \rho^{10}-220 \rho^9+792 \rho^8-4224 \rho^7+59136 \rho^6+709632 \rho^5\\
&+2365440 \rho^4
+3784704 \rho^3+3244032 \rho^2+1441792 \rho+262144\biggr)
+\frac{(1+\rho)^{9/2}}{\rho^{11}} C_1
\\&+\frac{1}{51200(1+\rho)^2 \rho^{11}} \biggl(
4725 \rho^{13}+17850 \rho^{12}+42480 \rho^{11}+193440 \rho^{10}+599424 \rho^9\\
&+985856 \rho^8
+277504 \rho^7-5945344 \rho^6-24600576 \rho^5-49201152 \rho^4-56229888 \rho^3\\
&-37486592 \rho^2-13631488 \rho-2097152)\,.
\end{split}
\eqlabel{7so32}
\end{equation}
Normalizability of $\hq_{7;7;1}$ sets
\begin{equation}
C_1=\frac{1024}{25}+\frac{262144}{63}\ p\,.
\eqlabel{7so33}
\end{equation}
As $\rho \to \infty$,
\begin{equation}
\begin{split}
&B_{7;7;1}'=\left(\frac{49}{160}-\frac{11}{9} p\right) \rho^{-2}
+\left(\frac{45}{16}+\frac{33}{7} p\right) \rho^{-3}
+\left(\frac{141}{25}-\frac{121}{7} p\right) \rho^{-4}
\\&+\left(\frac{30242}{5775}+\frac{253}{3} p\right) \rho^{-5}
+\left(-\frac{268984}{25025}-1023 p\right) \rho^{-6}\\&+\left(\frac{1024}{25}
+\frac{262144}{63} p\right) \rho^{-13/2}+\calo(\rho^{-7})\,,
\end{split}
\eqlabel{7so34}
\end{equation}
thus analyticity of $\hq_{7;7;1}$, and thus $B_{7;7;1}'$, in this limit requires 
\begin{equation}
p=-\frac{63}{6400}\qquad \Longrightarrow\qquad  C_1=0\,,
\eqlabel{7so35}
\end{equation}
which is also evident from \eqref{7so32}.
\item We continue with \eqref{7order31}, setting
\begin{equation}
\alpha_1=s_{7;1} v\,,\qquad q_{3;7;1}=\alpha_1\cdot q_{3;7;0}\cdot J_{3;7;1}\,,
\eqlabel{7so36}
\end{equation}
allows to solve analytically for $J_{3;7;1}'$,
\begin{equation}
\begin{split}
&J_{3;7;1}'=\frac{9\rho^3}{32(1+\rho)^{5/2}}
\ln\frac{\sqrt{1+\rho}+1}{\sqrt{1+\rho}-1}+\frac{(1+\rho)^{9/2}C_1}
{(21 \rho^4+112 \rho^3+240 \rho^2+256 \rho+128)^2 \rho^3} \\
&-\frac{1}{15v\rho^3 (21 \rho^4+112 \rho^3+240 \rho^2+256 \rho+128)^2 (1+\rho)} \biggl(
245 \rho^{10}+3140 \rho^9\\&+18936 \rho^8+73088 \rho^7+243968 \rho^6+715776 \rho^5
+1505280 \rho^4+2015232 \rho^3\\&+1622016 \rho^2+720896 \rho+131072\biggr)
\\&-\frac{ \rho+2}{80(1+\rho)^2 (21 \rho^4+112 \rho^3+240 \rho^2+256 \rho+128)^2 \rho^3}
\biggl(
19845 \rho^{12}+158760 \rho^{11}\\&
+353640 \rho^{10}-900480 \rho^9-7755456 \rho^8-23990272 \rho^7
-45821952 \rho^6\\&-60620800 \rho^5-57819136 \rho^4-39976960 \rho^3
-19529728 \rho^2-6291456 \rho\\
&-1048576\biggr)\,.
\end{split}
\eqlabel{7so37}
\end{equation}
Normalizability of $q_{3;7;1}$ sets
\begin{equation}
C_1=-\frac{131072}{15}\ \frac{3v-1}{v}\,,
\eqlabel{7so38}
\end{equation}
and  analyticity of $q_{3;7;1}$, and thus $J_{3;7;1}'$, in the limit
$\rho \to \infty$ requires
\begin{equation}
v=\frac 13\qquad \Longrightarrow\qquad  C_1=0\,.
\eqlabel{7so39}
\end{equation}
\item Consider now \eqref{7order33}: introducing
\begin{equation}
\hF_{7;2}=q_{3;7;0}\cdot G_{7;2}\,,
\eqlabel{7so40}
\end{equation}
we solve for $G_{7;2}'$,
\begin{equation}
\begin{split}
&G_{7;2}'=-\frac{\rho^3}{4000(1+\rho)^{5/2} (21 \rho^4+112 \rho^3+240 \rho^2+256 \rho+128)^2} \biggl(
112455 \rho^8\\
&+1189720 \rho^7+5688536 \rho^6+16165632 \rho^5+30098816 \rho^4
+37888000 \rho^3\\&+31744000 \rho^2+16384000 \rho
+4096000\biggr) \ln\frac{\sqrt{1+\rho}+1}{\sqrt{1+\rho}-1}
\\&-\frac{s_{7;1}^2}{45\rho^3 (21 \rho^4+112 \rho^3+240 \rho^2+256 \rho+128)^2 (1+\rho)} \biggl(
245 \rho^{10}+3140 \rho^9\\&+18936 \rho^8+73088 \rho^7+243968 \rho^6
+715776 \rho^5+1505280 \rho^4+2015232 \rho^3\\&+1622016 \rho^2
+720896 \rho+131072\biggr)
+\frac{(1+\rho)^{9/2}C_1}{(21 \rho^4+112 \rho^3+240 \rho^2+256 \rho+128)^2 \rho^3}
\\&+\frac{\rho+2}{90000(1+\rho)^2 (21 \rho^4+112 \rho^3+240 \rho^2+256 \rho+128)^2 \rho^3}
\biggl(
5060475 \rho^{12}\\&+40042800 \rho^{11}+110307120 \rho^{10}+27525120 \rho^9
-600006144 \rho^8-1699570688 \rho^7\\&-1994001408 \rho^6
+24739840 \rho^5+3575996416 \rho^4+5505679360 \rho^3
+4226940928 \rho^2\\&+1704984576 \rho+284164096\biggr)\,.
\end{split}
\eqlabel{7so41}
\end{equation}
Normalizability of $\hF_{7;2}$ sets
\begin{equation}
C_1=-\frac{35520512}{5625}+\frac{131072}{45}\ s_{7;1}^2\,,
\eqlabel{c1last}
\end{equation}
and analyticity of $\hF_{7;2}$, and thus $G_{7;2}'$, in the limit $\rho\to \infty$ requires
\begin{equation}
s_{7;1}^2=\frac{271}{125}\qquad \Longrightarrow\qquad s_{7;1}=\pm \frac{\sqrt{1355}}{25}\,.
\eqlabel{s71last}
\end{equation}
\end{itemize}

\subsubsection{Details of $s=7+\calo({b})$ branch: (C$ _b$)}
\label{detailss7l}

From \eqref{brc}, here  
\begin{equation}
\begin{split}
s_{7;0}=7\,,\qquad q_{7;7;0}=&
\mathcolorbox{red}{1}\cdot\ \frac{\rho^7}{(1+\rho)^s}\
_2 F_1\left(\frac {11}{2},7-s; 11;
-\rho\right)\bigg|_{s=s_{7;0}}=
\frac{\rho^7}{(1+\rho)^7}\,,
\end{split}
\eqlabel{l7zero3}
\end{equation}
where we highlighted the (fixed) overall normalization of the linearized fluctuations;
the latter implies that in the UV, \ie $\rho\to 0$, expansion of $q_{7;7;k\ge 1}$
the order $\calo(\rho^7)$ terms are absent. 
Because the leading order spectra \eqref{bra}, \eqref{brb} and \eqref{brc} are degenerate
at $s_{7;0}$,
the equations for $F_{7;k}$ and $q_{3;7;k}$ will necessarily contain zero modes; specifically, if $F_{7;k\ge 1}$
and $q_{7;7;k\ge 1}$
are solutions, so are $(F_{7;k}+\frac{\beta_k}{\alpha_0}
\cdot q_{3;7;0})$ and $(q_{3;7;k}+\frac{\alpha_k}{\alpha_0}\cdot q_{3;7;0})$,
\begin{equation}
q_{3;7;0}=\alpha_0\cdot _2F_1\left(\frac {3}{2},3-s; 3;
-\rho\right)\bigg|_{s=s_{7;0}}
=\alpha_0\cdot \frac{\rho^3(21\rho^4+112\rho^3+240\rho^2+256\rho+128)}{128(1+\rho)^7}\,,
\eqlabel{lq370}
\end{equation}
for an arbitrary set of constants $\{\alpha_k,\beta_k\}$.
As in section \ref{detailss3}, the zero modes at order $k$ will be completely fixed at order
$k+1$. We find it convenient to set
\begin{equation}
F_{7;k\ge 1}\equiv \mathcolorbox{orange}{\beta_k}\cdot  \frac{1}{\alpha_0}
q_{3;7;0}\ +\hF_{7;k}\,,
\qquad q_{3;7;k\ge 1}\equiv \mathcolorbox{green}{\alpha_k}\cdot
\frac{1}{\alpha_0}q_{3;7;0}\ +\hq_{3;7;k}\,,
\eqlabel{l7deffh3}
\end{equation}
with the understanding that in the UV  expansion of $\hF_{7;k}$ and
$\hq_{3;7;k}$
the order $\calo(\rho^3)$ terms are absent.

The subleading set of equations involving constants
$\alpha_0$, $\beta_1$, $s_{7;1}$, and functions $\{\hq_{3;7;1}$, $q_{7;7;1}$, $\hF_{7;1}\}$ reads:
\begin{equation}
\begin{split}
&0=\hq_{3;7;1}''+\frac{13 \rho-6}{2(1+\rho) \rho} \hq_{3;7;1}'
-\frac{3(5 \rho-2)}{2\rho^2 (1+\rho)^2} \hq_{3;7;1}+3 k_1' \hF_{7;1}'
+\frac{21k_1'\hF_{7;1}}{2(1+\rho)}+\frac{7\rho^6 (\rho+2)g_1'}{2(1+\rho)^8}
\\&+\frac{6\hF_{7;1}}{\rho^2 (1+\rho)} +\frac{12 \rho^5 g_1}{(1+\rho)^8}
-\frac{\alpha_0 \rho}{168(1+\rho)^9} \biggl(
\rho^2 (147 \rho^4+560 \rho^3+944 \rho^2+768 \rho\\
&+384) (1+\rho)^2 (k_1')^2
+2 \rho (\rho+2) (1+\rho) (147 \rho^4+560 \rho^3+944 \rho^2+768 \rho+384) (3 h_1'
\\&
+4 f_{3;1}')
-4 \rho (7 \rho^4+48 \rho^3+144 \rho^2+256 \rho+384) s_{7;1}
-16 (1+\rho) (273 \rho^4+1232 \rho^3\\&+2384 \rho^2+2304 \rho+1152) k_1
+(1617 \rho^6+12320 \rho^5+40352 \rho^4+74496 \rho^3+83328 \rho^2\\
&+55296 \rho+18432) h_1
+4 (1+\rho) ((273 \rho^4+1232 \rho^3+2384 \rho^2+2304 \rho+1152) f_{2;1}
\\&+(1596 \rho^4+7616 \rho^3+15296 \rho^2+15360 \rho+7680) f_{3;1}
-(147 \rho^4+560 \rho^3+944 \rho^2+768 \rho\\&+384))\biggr)
+\frac{\beta_1 \rho}{14(1+\rho)^8} \biggl(
k_1' \rho (\rho+2) (147 \rho^4+560 \rho^3
+944 \rho^2+768 \rho+384)+84 \rho^4\\&+448 \rho^3+960 \rho^2+1024 \rho+512\biggr)\,,
\end{split}
\eqlabel{l7order31}
\end{equation}
\begin{equation}
\begin{split}
&0=q_{7;7;1}''+\frac{13 \rho-6}{2(1+\rho) \rho} q_{7;7;1}'
-\frac{21(3 \rho+2)}{2\rho^2 (1+\rho)^2} q_{7;7;1}
-\frac{7\rho^7 (k_1')^2}{8(1+\rho)^7}
-3 k_1' \hF_{7;1}'\\&-\frac{21(\rho^2 k_1'+4)}{2\rho^2 (1+\rho)} \hF_{7;1}
-\frac{7 \rho^6 (\rho+2)f_{3;1}'}{(1+\rho)^8}
+\frac{\alpha_0 \rho^2 (\rho+2)g_1'}{42(1+\rho)^8}
\biggl(
147 \rho^4+560 \rho^3+944 \rho^2+768 \rho\\&+384\biggr)
-\frac{21\rho^6 (\rho+2)h_1'}{4(1+\rho)^8}
+\frac{11\rho^6 s_{7;1}}{2(1+\rho)^9}+\frac{\rho}{56(1+\rho)^9}
\biggl(
-32 g_1 (1+\rho) (21 \rho^4+112 \rho^3\\&+240 \rho^2+256 \rho+128) \alpha_0
+\rho^4 (28 (1+\rho) (35 f_{2;1}+4 k_1+20 f_{3;1}+7)-7 (77 \rho^2
+16 \rho\\&+16) h_1)\biggr)
-\frac{\beta_1 \rho}{14(1+\rho)^8} \biggl(
k_1' \rho (\rho+2) (147 \rho^4+560 \rho^3
+944 \rho^2+768 \rho+384)+6720 \rho^2\\&+7168 \rho+3584
+588 \rho^4+3136 \rho^3\biggr)\,,
\end{split}
\eqlabel{l7order32}
\end{equation}
\begin{equation}
\begin{split}
&0=\hF_{7;1}''+\frac{13 \rho-6}{2(1+\rho) \rho} \hF_{7;1}'
-\frac{3(5 \rho-2)}{2\rho^2 (1+\rho)^2} \hF_{7;1}-\frac{\rho^2 (\rho+2)k_1'}{63(1+\rho)^8}
\biggl(
(147 \rho^4+560 \rho^3+944 \rho^2\\&+768 \rho+384) \alpha_0-147 \rho^4\biggr)
-\frac{4\rho}{21(1+\rho)^8} \biggl(
\alpha_0 (21 \rho^4+112 \rho^3+240 \rho^2+256 \rho+128)\\&+21 \rho^4\biggr)\,.
\end{split}
\eqlabel{l7order33}
\end{equation}
Eqs. \eqref{l7order31}---\eqref{l7order33} are solved subject to the asymptotic
expansions,
\nxt in the UV, \ie as $\rho\to 0$,
\begin{equation}
\begin{split}
&\hq_{3;7;1}=\biggl(\mathcolorbox{green}{0}+\frac{256}{7} \beta_1 \ln\rho\biggr) \rho^3
-\biggl(\frac{128}{7} \beta_1+\frac{64}{7} \alpha_0 f_{2,1,0;1}+\frac{64}{21} \alpha_0 s_{7;1}
+\frac{1280}{7} \beta_1 \ln\rho\biggr) \rho^4\\&+\calo(\rho^5\ln\rho)\,,
\end{split}
\eqlabel{luvq332}
\end{equation}
\begin{equation}
\begin{split}
&q_{7;7;1}=-\frac{128}{21} \beta_1 \rho^3
+\frac{640}{21} \beta_1 \rho^4+\biggl(
-\frac{1966}{21} \beta_1+\frac{46}{21} \alpha_0\biggr) \rho^5
+\biggl(
\frac{4756}{21} \beta_1-\frac{92}{7} \alpha_0\biggr) \rho^6
+\biggl(
\\&\mathcolorbox{red}{0}+\biggl(
-2-\frac{256}{35} \alpha_0 k_{4,0;1}+\frac{768}{35} \beta_1 k_{4,0;1}
+\frac{457}{350} \alpha_0-\frac{191}{350} \beta_1\biggr)
\ln\rho\\&+\biggl(
\frac{18}{35} \beta_1-\frac{6}{35} \alpha_0\biggr) \ln^2\rho\biggr) \rho^7+\calo(\rho^8\ln^2\rho)\,,
\end{split}
\eqlabel{luvq3322}
\end{equation}
\begin{equation}
\begin{split}
&\hF_{7;1}=\mathcolorbox{orange}{0}\rho^3+\mathcolorbox{orange}{0}\rho^4
-\frac{88}{63} \alpha_0 \rho^5+\frac{176}{21} \alpha_0 \rho^6+\calo(\rho^7\ln\rho)\,,
\end{split}
\eqlabel{luvq3323}
\end{equation}
it is completely specified by $\{\beta_1,\alpha_0,s_{7;1}\}$; we further highlighted
arbitrary constants, fixed to zero by the overall normalization \eqref{l7zero3}, and the
 extraction of the zero modes  in $F_{7;1}$  and $q_{3;7;1}$ \eqref{l7deffh3};
\nxt  in the IR, \ie as $y\equiv \frac1\rho\to 0$,
\begin{equation}
\hq_{3;7;1}=\hq_{3;7;1;0}^h+\calo(y)\,,\qquad
q_{7;7;1}=q_{7;7;1;0}^h+\calo(y)\,,\qquad 
\hF_{7;1}=\hF_{7;1;0}^h+\calo(y)\,,
\eqlabel{lirq332}
\end{equation}
it is completely specified by
\begin{equation}
\{\hq_{3;7;1;0}^h\,,\, q_{7;7;1;0}^h\,,\,  \hF_{7;1;0}^h\,,\, \beta_1\,,\, \alpha_0\,,\,  s_{7;1}\}\,.
\eqlabel{lsetq332}
\end{equation}
In total, the UV and IR expansions are completely determined by the parameters
\eqref{lsetq332}, which is precisely what is needed to find a unique solution
for three second order ODEs \eqref{l7order31}-\eqref{l7order33}.
Solving these equations we find
\begin{equation}
\begin{split}
&s_{7;1}=4.3945(5)\,,\qquad \beta_1=-3.5047(8)\,,\qquad \alpha_0=4.2000(0)\,,\\
&\hq_{3;7;1;0}^h=-4.4179(7)\,,\qquad q_{7;7;1;0}^h=-1850.3(4)\,,\qquad \hF_{7;1;0}^h=-1.2222(2)\,.
\end{split}
\eqlabel{lq332resu}
\end{equation}
Note that equation \eqref{l7order33} for $\hF_{7;1}$ is decoupled, and involves $k_1$ for which
the analytic expression is available \eqref{k1anal}; solving this equation
using the techniques of section \ref{detailss3}, we find
\begin{equation}
\begin{split}
&\hF_{7;1}=\frac{\rho^3(21\rho^4+112\rho^3+240\rho^2+256\rho+128)}{21(1+\rho)^7}\ G_{7;1}\,,\qquad
\alpha_0=\frac{21}{5}\,,\\
&G_{7;1}'=-\frac{7\rho^3}{10(1+\rho)^{5/2} (21 \rho^4+112 \rho^3+240 \rho^2+256 \rho+128)^2} \biggl(
126 \rho^8+1519 \rho^7+7903 \rho^6\\&+23520 \rho^5+44784 \rho^4+56832 \rho^3+47616 \rho^2+24576 \rho
+6144\biggr)
\ln\frac{\sqrt{1+\rho}+1}{\sqrt{1+\rho}-1}
\\&+\frac{7(\rho+2)\rho}{5(1+\rho)^2 (21 \rho^4+112 \rho^3+240 \rho^2+256 \rho+128)^2} \biggl(
126 \rho^8+1183 \rho^7+4375 \rho^6+4976 \rho^5\\
&-12296 \rho^4-49280 \rho^3-68992 \rho^2
-45056 \rho-11264\biggr)\,, 
\end{split}
\eqlabel{f71an}
\end{equation}
where the analytic expression for $\alpha_0$ is in perfect agreement with the numerical result
\eqref{lq332resu}.

\subsubsection{Select values of $s_{3\le n\le 10;1}$}
\label{selscb}

Extending the computations of sections \ref{detailss3} and \ref{detailss7l},
we collect in the table below leading corrections to the conformal spectra on
branches $(A_b)$, $(B_b)$ and $(C_b)$ for $3\le n\le 10$,
\begin{center}
\begin{tabular}{| c| c| c|} 
 \hline
 $n$ & $s_{n;1}^{(A)\&(B)}$ & $s_{n;1}^{(C)}$ \\
 \hline
 3 & $\pm \sqrt{2}$ & $-$ \\
 \hline
 4 & $\pm \sqrt{2}$ & $-$ \\
 \hline
 5 & $\pm \sqrt{2}$ & $-$ \\
 \hline
 6 & $\pm \sqrt{2}$ & $-$ \\
 \hline
 7 & $\pm \frac{\sqrt{1355}}{25}$ & $4.39(5)$ \\
 \hline
 8 & $\pm \frac{\sqrt{73163}}{175}$ & $4.95(1)$ \\
 \hline
9 & $\pm \frac{\sqrt{1276274}}{700}$ & $5.32(3)$ \\
 \hline
10 & $\pm \frac{\sqrt{604049698}}{14700}$ & did not compute \\
 \hline
\end{tabular}
\end{center}
These results are used to highlight the features of the spectra
presented in fig.~\ref{figure3}. Notice that the leading correction
to the conformal spectra on  branches $(A_b)$, $(B_b)$ is unchanged
for $n\le 6$, 
\begin{equation}
-\Im[\ww]\bigg|_{(A_b)\&(B_b)}=n\pm \sqrt{2b}+\calo(b)\,.
\eqlabel{lcorrections}
\end{equation}

\section{Chirally symmetric DFP --- TypeA$ _s$, chirally symmetric fluctuations}
\label{ascs}

In this appendix we discuss the linearized fluctuations about TypeA$ _s$ dynamical fixed point
of the cascading gauge theory, preserving the $U(1)_R$ chiral symmetry of this
DFP. The corresponding background and the fluctuation equations of motion
are the special case, a consistent truncation, of the
general equations discussed in appendix \ref{fgframe}.
Specifically,
\begin{itemize}
\item for the background we find \cite{Buchel:2019pjb}:
\begin{equation}
f_c\equiv f_2\,,\qquad f_a=f_b\equiv f_3\,,\qquad K_1=K_3\equiv K\,,\qquad K_2\equiv 1\,;
\eqlabel{sdfps}
\end{equation}
\item for the fluctuations: we  keep $\{fl_g\,,\, fl_h\}$ modes, and further restrict 
\begin{equation}
\begin{split}
&fl_a=fl_b\equiv fl_3\,,\qquad fl_{K_1}=fl_{K_3}\equiv fl_K\,,\qquad fl_{K_2}\equiv 0\,,\qquad
fl_c\equiv fl_2\,.
\end{split}
\eqlabel{sdfpfs}
\end{equation}
\end{itemize}
Given \eqref{sdfps} and \eqref{sdfpfs}, the corresponding equations
for the fluctuations and the boundary conditions
can be deduced from those of the symmetry broken DFP discussed in appendix
\ref{fgframe}. 

In the rest of this appendix we analyze the near-conformal $b\to 0$, equivalently
$H\gg \Lambda$, limit of the chiral symmetry preserving fluctuations
in TypeA$ _s$ DFP. Strictly at $b=0$ the cascading gauge theory is
conformal, and the spectra can be computed analytically \cite{Buchel:2022hjz}.
We discover multiple spectral branches of the fluctuations. 
On some branches we are  able to compute analytically the leading $\calo(\sqrt{b})$,
and numerically the first $\calo(b)$ subleading,
corrections to the conformal spectra, sections \ref{sdetailss3}, \ref{detailss6} and
\ref{detailss8}.
On the remaining branches we compute numerically  the leading $\calo(b)$ corrections to
the conformal spectra, sections \ref{sdetailss7l} and \ref{detailssc8}.
Perturbative results obtained here provide
a valuable check of the finite $\frac{H}{\Lambda}$ spectra in the near-conformal
limit, see fig.~\ref{figure1s}.

\subsection{Near-conformal limit: $b\to 0$}

Introducing
\begin{equation}
fl_2=(1+\rho)\ (fl_f+4 fl_w)\,,\qquad fl_3=(1+\rho) (fl_f-fl_w)\,,
\eqlabel{decfls}
\end{equation}
to leading order in $b$, we find\footnote{Note (see the equation for $fl_f$) that the leading  $s=4$ mode
is more subtle; it will be discussed in details in section \ref{sdetailss3}.}
\begin{equation}
\begin{split}
&0=fl_K''+\frac{(2 s-1) \rho-6}{2(1+\rho) \rho}\ fl_K'
-\frac{3s}{2\rho (1+\rho)^2}\ fl_K\,,\\
&0=fl_g''+\frac{(2 s-1) \rho-6}{2(1+\rho) \rho}\ fl_g'
-\frac{3s}{2\rho (1+\rho)^2}\ fl_g
+2 k_1' fl_K'+\frac{k_1' s fl_K}{1+\rho}\,,\\
&0=fl_w'+\frac{(2 s-1)\rho -6}{2\rho (1+\rho)}\
fl_w'-\frac{3(\rho s+8 \rho+8)}{2(1+\rho)^2 \rho^2}\ fl_w
-\frac25 k_1' fl_K'-\frac{k_1' s fl_K}{5(1+\rho)}\,,\\
&0=fl_f''+\frac{(2 s-1)\rho -6}{2\rho (1+\rho)}\ fl_f'
-\frac{3 \rho s+64 \rho+64}{2\rho^2 (1+\rho)^2}\ fl_f
+\frac{8k_1'fl_K'}{5\rho^2 (s-4) (1+s)} \biggl(
\rho^2 s^2-3 \rho^2 s\\&-9 \rho^2-120 \rho-120\biggr)
+\frac{4fl_K}{5(1+\rho) \rho^4 (s-4) (1+s)}
\biggl(\rho k_1' (\rho^3 s (s+6)(s-4)
+10 \rho s(\rho s\\&-15 \rho-12) +480(\rho+2)(\rho+1))
+80 \rho^2 s^2-240 \rho^2 s-240 \rho^2+1920 \rho
+1920\biggr) \,.
\end{split}
\eqlabel{sq3q7}
\end{equation}
Solving the decoupled equation for $fl_K$ , we find (up to an overall normalization $A_K$)
\begin{equation}
fl_K=A_K\ \frac{\rho^4}{(1+\rho)^s}\ _2 F_1\left(\frac 52,4-s; 5; -\rho\right)\,,\qquad s=4,5,\cdots\,. 
\eqlabel{sbra}
\end{equation}
Given \eqref{sbra}, and using \eqref{k1anal}, it is straightforward to see
that it is impossible to solve the equation for $fl_g$ in \eqref{sq3q7},
so that this mode
is both normalizable as $\rho\to 0$ and analytic as $\rho \to \infty $ ---
this means that the amplitude of $fl_K$ must always vanish in the limit
$b\to 0$. This is precisely what we find, see \eqref{ssab} and \eqref{ssc}. 

With $fl_K\equiv 0$, we find from \eqref{sq3q7} the following
leading order as $b\to 0$ solutions:
\begin{equation}
fl_g=A_g\ \frac{\rho^4}{(1+\rho)^s}\ _2 F_1\left(\frac 52,4-s; 5; -\rho\right)\,,\qquad s=4,5,\cdots\,, 
\eqlabel{sbrb}
\end{equation}
\begin{equation}
fl_w=A_w\ \frac{\rho^6}{(1+\rho)^s}\ _2 F_1\left(\frac {9}{2},6-s; 9; -\rho\right)\,,\qquad
s=6,7,\cdots \,,
\eqlabel{sbrc}
\end{equation}
\begin{equation}
fl_f=A_f\ \frac{\rho^8}{(1+\rho)^s}\ _2 F_1\left(\frac {13}{2},8-s; 13; -\rho\right)\,,\qquad
s=8,9,\cdots \,.
\eqlabel{sbrd}
\end{equation}

Extending the leading order solutions \eqref{sbrb}, \eqref{sbrc} and \eqref{sbrd}
perturbatively in $b$ we identify three branches:
\nxt A pair of non-analytic in $b$ branches, $(A_s)$ and $(B_s)$,
\begin{equation}
\begin{split}
&s\bigg|_{A,B}= n+\sum_{k=1}^\infty (\pm )^k s_{n;k}\ b^{k/2}\,,
\qquad n\in \naturals \ge 4\,,\qquad fl_g\bigg|_{A,B}=\sum_{k=0}^\infty (\pm )^k fl_{g;n;k}\ b^{k/2}\,,\\
&fl_K\bigg|_{A,B}=\sum_{k=1}^\infty (\pm )^k fl_{K;n;k}\ b^{k/2}\,,\qquad
fl_w\bigg|_{A,B}=\sum_{k=0}^\infty (\pm )^k fl_{w;n;k}\ b^{k/2}\,,\\
&fl_f\bigg|_{A,B}=\sum_{k=0}^\infty (\pm )^k fl_{f;n;k}\ b^{k/2}\,,
\end{split}
\eqlabel{ssab}
\end{equation}
with $fl_{g;n;0}$ given by \eqref{sbrb},  $fl_{w;n\ge 6;0}$ given by \eqref{sbrc}, and
$fl_{f;n\ge 8;0}$ given by \eqref{sbrd} with $s=n$.   $fl_{w;n;0}\equiv 0$ for $4\le n<6$,
$fl_{f;n;0}\equiv 0$ for $4< n<8$, and $fl_{f;4;0}\ne 0$, see \eqref{flf40solve}. 
\nxt An analytic in $b$  branch $(C_s)$,
\begin{equation}
\begin{split}
&s\bigg|_{C}= n+\sum_{k=1}^\infty  s_{n;k}\ b^{k}\,,
\qquad n\in \naturals \ge 6\,,\qquad fl_g\bigg|_{C}=\sum_{k=0}^\infty fl_{g;n;k}\
b^{k}\,,\\
&fl_K\bigg|_{C}=\sum_{k=1}^\infty  fl_{K,n;k}\ b^{k}\,,\qquad
fl_w\bigg|_{C}=\sum_{k=0}^\infty fl_{w;n;k}\ b^{k}\,,\qquad
fl_f\bigg|_{C}=\sum_{k=0}^\infty fl_{f;n;k}\ b^{k}\,,
\end{split}
\eqlabel{ssc}
\end{equation}
where $fl_{w;n;0}$ is given by \eqref{sbrc}, $fl_{g;n;0}$ is given by \eqref{sbrb},
and $fl_{f;n\ge 8;0}$ given by \eqref{sbrd}
with $s=n$. $fl_{f;n;0}\equiv 0$ for $6\le n<8$.

\subsubsection{Details of $s=4\pm\calo(\sqrt{b})$ branches:  (A$ _s$) and (B$ _s$)}
\label{sdetailss3}

From \eqref{sbrb}, here  
\begin{equation}
s_{4;0}=4\,,\qquad fl_{g;4;0}=
\mathcolorbox{red}{1}\cdot\ \frac{\rho^4}{(1+\rho)^s}\ _2 F_1\left(\frac 52,4-s; 5;
-\rho\right)\bigg|_{s=s_{4;0}}= \frac{\rho^4}{(1+\rho)^4}\,,
\eqlabel{szero3}
\end{equation}
where we highlighted the (fixed) overall normalization of the linearized fluctuations;
the latter implies that in the UV, \ie $\rho\to 0$, expansion of $fl_{g;4;k\ge 1}$
the order $\calo(\rho^4)$ terms are absent. 
Because the leading order fluctuation spectra \eqref{sbra} and \eqref{sbrb} are degenerate,
the equations for $fl_{K;4;k}$ will necessarily contain zero modes; specifically, if $fl_{K;4;k\ge 1}$
is a solution, so is $(fl_{K;4;k}+\alpha_k\cdot fl_{g;4;0})$ for an arbitrary set of constants $\alpha_k$.
As in section \ref{detailss3}, the zero modes at order $k$ will be completely fixed at order
$k+1$. We find it convenient to set
\begin{equation}
fl_{K;4;k}\equiv \mathcolorbox{orange}{\alpha_k}\cdot \ fl_{g;4;0}+\hfl_{K;4;k}\,,
\eqlabel{sdeffh3}
\end{equation}
with the understanding that in the UV  expansion of $\hfl_{K;4;k}$
the order $\calo(\rho^4)$ terms are absent.

As in section \ref{detailss3}, the equation for $\hfl_{K;4;1}$ is homogeneous, and the boundary
condition implied by \eqref{sdeffh3} sets 
\begin{equation}
\hfl_{K;4;1}\equiv 0\,.
\eqlabel{sflk41}
\end{equation}
The leading order equation for $fl_{f;4;0}$ takes form
\begin{equation}
\begin{split}
&0=fl_{f;4;0}''+\frac{7 \rho-6}{2\rho (1+\rho)}\ fl_{f;4;0}'
-\frac{2 (19 \rho+16)}{\rho^2 (1+\rho)^2}\ fl_{f;4;0}+\frac{64(\rho^2+24 \rho+24) \alpha_1}
{5(1+\rho)^5 s_{4;1}}\,,
\end{split}
\eqlabel{sorder31}
\end{equation}
and can be solved analytically,
\begin{equation}
f_{f;4;0}=\frac{128\alpha_1\rho^2}{15s_{4;1}(1+\rho)^3}\,.
\eqlabel{flf40solve}
\end{equation}
The subleading set of equations involving constants
$\alpha_1$, $s_{4;1}$, and functions $\{fl_{g;4;1}$, $\hfl_{K;4;2}$, $fl_{w;4;1}\}$ reads:
\begin{equation}
\begin{split}
&0=fl_{g;4;1}''+\frac{7 \rho-6}{2\rho (1+\rho)}\ fl_{g;4;1}'
-\frac{6}{\rho (1+\rho)^2}\ fl_{g;4;1}+ \frac{4\alpha_1 (\rho+2) \rho^3k_1'}{(1+\rho)^5}
+\frac{5\rho^3 s_{4;1}}{2(1+\rho)^6}\,,
\end{split}
\eqlabel{sorder32}
\end{equation}
\begin{equation}
\begin{split}
&0=\hfl_{K;4;2}''+\frac{7 \rho-6}{2\rho (1+\rho)} \hfl_{K;4;2}'
-\frac{6}{\rho (1+\rho)^2} \hfl_{K;4;2}
+\frac{2k_1'(\rho+2) (32 \alpha_1 (\rho+1)-5 \rho^2 s_{4;1}) \rho}{5(1+\rho)^5 s_{4;1}}\\
&+\frac{1}{30(1+\rho)^6 s_{4;1}} \biggl(
75 \alpha_1 \rho^3 s_{4;1}^2-240 \rho^3 s_{4;1}+2048 \alpha_1 (1+\rho)^2-240 \rho^2 s_{4;1}
\biggr)\,,
\end{split}
\eqlabel{sorder33}
\end{equation}
\begin{equation}
\begin{split}
&0=fl_{w;4;1}''+\frac{7 \rho-6}{2\rho (1+\rho)}\ fl_{w;4;1}'
-\frac{6 (3 \rho+2)}{\rho^2 (1+\rho)^2}\ fl_{w;4;1}
-\frac{4(\rho+2) \rho^3 k_1' \alpha_1}{5(1+\rho)^5}\,.
\end{split}
\eqlabel{sorder34}
\end{equation}
Above set can be solved numerically --- and we explain how to do it for the
set of equations at the next order --- here instead we show that the most
important constant, \ie $s_{4;1}$ can be computed analytically:
\begin{itemize}
\item Substituting
\begin{equation}
\alpha_1=s_{4;1} v\,,\qquad fl_{g;4;1}=s_{4;1}\cdot fl_{g;4;0}\cdot\ G_{g;4;1}\,,
\eqlabel{sso31}
\end{equation}
and using \eqref{k1anal}, 
we find a general analytic solution for $G_{g;4;1}'$,
\begin{equation}
\begin{split}
&G_{g;4;1}'= \frac{3v\rho^3}{16(1+\rho)^{5/2}} \ln\frac{\sqrt{1+\rho}+1}{\sqrt{1+\rho}-1}
+\frac{(1+\rho)^{3/2}}{\rho^5} C_1-\frac{1}{24(1+\rho)^2 \rho^5} \biggl(
(\rho+2)\\
&\times(9 \rho^6-24 \rho^5-152 \rho^4+768 \rho^3+2944 \rho^2
+3072 \rho+1024) v+8 (1+\rho) (5 \rho^4\\
&-40 \rho^3-240 \rho^2-320 \rho-128)
\biggr)\,.
\end{split}
\eqlabel{sso32}
\end{equation}
Analyticity of $fl_{g;4;1}$, and thus $G_{g;4;1}'$, as $\rho\to\infty$ sets
\begin{equation}
C_1=0\,,
\eqlabel{sso33}
\end{equation}
and normalizability of $fl_{g;4;1}$ identifies
\begin{equation}
v=\frac12\,.
\eqlabel{sso35}
\end{equation}
\item We continue with \eqref{sorder33}, setting
\begin{equation}
\hfl_{K;4;2}=fl_{g;4;0}\cdot H_{K;4;2}\,,
\eqlabel{sso36}
\end{equation}
allows to solve analytically for $H_{K;4;2}'$,
\begin{equation}
\begin{split}
&H_{K;4;2}'=-\frac{\rho(15\rho^2-64\rho-64)}{160(1+\rho)^{5/2}} \ln\frac{\sqrt{1+\rho}+1}{\sqrt{1+\rho}-1}
+\frac{(1+\rho)^{3/2}}{\rho^5}\ C_1
-\frac{s_{4;1}^2}{6(1+\rho)\rho^5} 
\biggl(5\rho^4\\&-40\rho^3-240\rho^2-320\rho-128\biggr)+\frac{\rho+2}{240(1+\rho)^2\rho^5}\biggl(
45\rho^6-312\rho^5+840\rho^4\\
&-11008\rho^3-38784\rho^2-39936\rho-13312\biggr)\,.
\end{split}
\eqlabel{sso37}
\end{equation}
 Analyticity of $\hfl_{K;4;2}$, and thus $H_{K;4;2}'$, in the limit
$\rho \to \infty$ requires
\begin{equation}
C_1=0\,,
\eqlabel{sso38}
\end{equation}
while normalizability of $\hfl_{K;4;2}$ sets
\begin{equation}
s_{4;1}^2=\frac{26}{5}\qquad \Longrightarrow\qquad  s_{4;1}=\pm \frac{\sqrt{130}}{5}\,.
\eqlabel{sso39}
\end{equation}
\end{itemize}
Note that to determine $s_{4;1}$, there is no need to solve for $fl_{w;4;1}$ ---
of course, this solution is needed for the computation of higher order corrections $s_{4;k\ge 2}$. 

The sub-subleading set of equations involving constants
$\alpha_2$, $s_{4;2}$, and functions $\{fl_{f;4;1}$, $fl_{g;4;2}$ and $\hfl_{K;4;3}\}$
reads (we omit the equation for
$fl_{w;4;2}$ as it is not need to compute $s_{4;2}$; it is required for
the computation of $s_{4;k\ge 3}$):
\begin{equation}
\begin{split}
&0=fl_{f;4;1}''+\frac{7 \rho-6}{2\rho (1+\rho)}\ fl_{f;4;1}'-\frac{2 (19 \rho+16)}{\rho^2 (1+\rho)^2}\ fl_{f;4;1}
-\frac{8(\rho^2+24 \rho+24)k_1'\hfl_{K;4;2}'}{5\rho^2 s_{4;1}} \\
&+\frac{32(\rho^2+24 \rho+24) (\rho k_1'+2) \hfl_{K;4;2}}{5s_{4;1} \rho^4 (1+\rho)}
-\frac{4(\rho^2+24 \rho+24) (64 \alpha_1 (\rho+1)-5 \rho^2 s_{4;1}) (k_1')^2}
{25s_{4;1}^2 (1+\rho)^4}\\&+\frac{8\alpha_1 \rho^2 (4 \rho^2-3 \rho-12) k_1'}
{5(1+\rho)^5}-\frac{2048 (\rho+2) (\rho^2+24 \rho+24) \alpha_1}{75s_{4;1}^2 (1+\rho)^4 \rho}
\biggl(f_{3;1}'+\frac14 f_{2;1}'+\frac14 h_1'
\biggr)
\\&-\frac{64\alpha_1 (\rho^2+24 \rho+24) s_{4;2}}{5(1+\rho)^5 s_{4;1}^2}
+\frac{64(\rho^2+24 \rho+24) \alpha_2}{5(1+\rho)^5 s_{4;1}}
+\frac{1}{75 (1 + \rho)^5 s_{4;1}^2 \rho^2}\biggl(
-768 (\rho^2 \\&+ 24 \rho + 24) \alpha_1\left(\rho + \frac43\right) (\rho + 4) h_1
- 1024 \alpha_1 (1 + \rho) (\rho^2 + 24 \rho + 24) (f_{2;1}+4f_{3;1}-4k_1)\\
&+ (24576 + 3968 \rho^4 s_{4;1}^2 + (-4288 s_{4;1}^2 + 1024) \rho^3
+ (-4608 s_{4;1}^2 + 25600) \rho^2 + 49152 \rho) \alpha_1 \\&+ 240 \rho^2 s_{4;1} (\rho^2 + 24 \rho + 24)
\biggr)\,,
\end{split}
\eqlabel{sorder321}
\end{equation}
\begin{equation}
\begin{split}
&0=fl_{g;4;2}''+ \frac{7 \rho-6}{2\rho (1+\rho)}\ fl_{g;4;2}'-\frac{6}{\rho (1+\rho)^2}\ fl_{g;4;2}
+2 k_1' \hfl_{K;4;2}'+\frac{4 k_1' \hfl_{K;4;2}}{1+\rho}
\\&+\frac{s_{4;1} fl_{g;4;1}'}{1+\rho}+\frac{4 k_1' \rho^3 (\rho+2) \alpha_2}{(1+\rho)^5}
+\frac{4(\rho+2) (16 \alpha_1 (\rho+1)-5 \rho^2 s_{4;1}) \rho g_1'}{5(1+\rho)^5 s_{4;1}}
+\frac{\alpha_1 \rho^4 s_{4;1} k_1'}{(1+\rho)^5}
\\&+\frac{(\rho+2) \rho^3 (f_{2;1}'+4f_{3;1}')}{(1+\rho)^5}
-\frac{3 s_{4;1} fl_{g;4;1}}{2\rho (1+\rho)^2}
+\frac{1}{30 s_{4;1} (1 + \rho)^6}\biggl(
1024 \alpha_1 (2 \rho + 1)-30 h_1 \rho^4 s_{4;1}
\\&+ (75 s_{4;2} - 240) s_{4;1} \rho^3 + (-240 s_{4;1} + 1024 \alpha_1) \rho^2 \biggr)\,,
\end{split}
\eqlabel{sorder322}
\end{equation}
\begin{equation}
\begin{split}
&0=\hfl_{K;4;3}''+\frac{7 \rho-6}{2\rho (1+\rho)}\ \hfl_{K;4;3}'
-\frac{6}{\rho (1+\rho)^2}\ \hfl_{K;4;3}-\frac34 k_1' fl_{f;4;1}'
-\frac{12 \hfl_{K;4;2}'}{5 (1 + \rho) s_{4;1} \rho^2}
\biggl(
\rho (\rho \\&+ 2) (1 + \rho)^2 (k_1')^2
- \frac83 (1 + \rho)^2 k_1' - \frac{5}{12}{ \rho^2 s_{4;1}^2}\biggr)
+\frac{48\hfl_{K;4;2}}{5 s_{4;1} (1 + \rho)^2 \rho^4} \biggl(
\rho^2 (\rho + 2) (1 + \rho)^2 (k_1')^2 \\
&+ 2 \left(\rho + \frac23\right)
\rho (1 + \rho)^2 k_1' - \frac{5}{32} \rho^3 s_{4;1}^2 - \frac{16}{3} (1+\rho)^2 \biggr) 
+\frac{1}{2 (1 + \rho) \rho^2}
\biggl(
-2 \rho^2 \biggl(
(1 + \rho)\\&\times (fl_{g;4;1}'- 2 fl_{w;4;1}) + \frac32 fl_{f;4;1}
- 4 fl_{w;4;1}\biggr) k_1' + 28 fl_{f;4;1} + 32 fl_{w;4;1}-4 (\rho^2 k_1'\\&+4) fl_{g;4;1}
\biggr)
+\frac{192 \rho (\rho + 2) (1 + \rho)^2 k_1' + 25 \rho^3 s_{4;1}^2 - 512 (1 + \rho)^2}
{10 s_{4;1} (1 + \rho)^6} \alpha_2-\frac{96 \alpha_1s_{4;2}}{5 (1 + \rho)^6 s_{4;1}^2}
\biggl(
\\&\rho (\rho + 2) (1 + \rho)^2 k_1'
- \frac{25}{192} \rho^3 s_{4;1}^2 - \frac83 (1 + \rho)^2\biggr) +\frac{\alpha_1}{75 \rho^2 (1 + \rho)^6 s_{4;1}^2}
\biggl(
-1152 \rho^3 (\rho + 2) \\&\times (1 + \rho)^4
(k_1')^3 - 315 \rho^2 \left(
\rho^4 s_{4;1}^2 + \frac47 \rho^3 s_{4;1}^2 - \frac{1024}{105} (1+\rho)^2\right)
(1 + \rho)^2 (k_1')^2 - 768 (\rho + 2)\\
&\times\biggl(
\rho(\rho+2)(\rho+1) (f_{2;1}' + h_1' + 4 f_{3;1}') + \left(\frac32 \rho^2 + 8 \rho + 8\right) h_1
+ 2( \rho + 1) (f_{2;1} + 4  f_{3;1} \\& - 4 k_1-1) + \frac{3}{8} \rho^2 s_{4;1}^2\biggr)
\rho (1 + \rho)^2 k_1'
+ 75  (\rho + 2) \rho (1 + \rho)\biggl(
\left(\rho^4 s_{4;1}^2 + \frac{2048}{75} (1+\rho)^2\right) f_{2;1}'\\& - 2
\left(\rho^4 s_{4;1}^2 - \frac{1024}{75} (1+\rho)^2\right) h_1'\biggr)
-2 \rho (\rho + 2) (1 + \rho)(75 \rho^5 s_{4;1}^2 g_1' -4096 (1 + \rho)^2
f_{3;1}')  \\&+ (-75 \rho^6 s_{4;1}^2 + 1024 (\rho+4)(3\rho+4)(1+ \rho)^2) h_1
- 600 (1 + \rho) \biggl(
-\frac{512}{75} (1 + \rho)^2 (f_{2;1} \\&+4f_{3;1} -4 k_1-1)+ s_{4;1}^2\left(\rho^4 - \frac{32}{25} \rho^2(\rho+1)
\right)
\biggr)
\biggr) 
+\frac{1}{10 (1 + \rho)^5 s_{4;1}}\biggl(
4 \rho^2 (1 + \rho)^2 (k_1')^2\\
&\times(  3 \rho (2 + \rho) k_1'-8) + (-5 \rho^4 s_{4;1}^2 + 48 \rho(\rho+2)(1+\rho))
k_1' - 128 (1+\rho)\biggr)\,.
\end{split}
\eqlabel{sorder323}
\end{equation}
Eqs. \eqref{sorder321}-\eqref{sorder323} are solved subject to the asymptotic
expansions,
\nxt in the UV, \ie as $\rho\to 0$,
\begin{equation}
\begin{split}
&fl_{f;4;1}=\left(-\frac{128(\alpha_1 s_{4;1}^2+5\alpha_1s_{4;2}
-5\alpha_2 s_{4;1})}{75s_{4;1}^2}-\frac{46592\alpha_1}{225s_{4;1}^2}+\frac{64}{15s_{4;1}}
\right)\rho^2+\cdots\\
&+ \rho^8\biggl(
fl_{f;4;1;8;0}+\alpha_1\left(\frac{17}{200}-\frac{32}{15} k_{4;0;1}\right)\ln\rho-\frac{1}{20}\alpha_1
\ln^2\rho\biggr)
+\calo(\rho^9\ln^2\rho)\,,
\end{split}
\eqlabel{suvq332}
\end{equation}
\begin{equation}
\begin{split}
&fl_{g;4;2}=\biggl(\mathcolorbox{red}{0}+\left(4 \alpha_2+2-\frac{1456\alpha_1}{15s_{4;1}}
\right)\ln\rho\biggr) \rho^4\ln\rho+\calo(\rho^5\ln\rho)\,,
\end{split}
\eqlabel{suvq332a}
\end{equation}
\begin{equation}
\begin{split}
&\hfl_{K;4;3}=\left(-\frac{64(\alpha_1 s_{4;1}^2+5\alpha_1s_{4;2}
-5\alpha_2 s_{4;1})}{75s_{4;1}^2}-\frac{23296\alpha_1}{225s_{4;1}^2}+\frac{32}{15s_{4;1}}
\right)\rho^2\\&+\cdots
+\biggl(\mathcolorbox{orange}{0}\ -4\alpha_1\biggr)\rho^4\ln\rho+\calo(\rho^5\ln\rho)\,,
\end{split}
\eqlabel{suvq33b}
\end{equation}
it is completely specified by $\{\alpha_2,s_{4;2},fl_{f;4;1;8;0}\}$; we further highlighted
 arbitrary constants, fixed to zero by the overall normalization \eqref{szero3} and the
 extraction of the zero mode in $fl_{K;4;3}$ \eqref{sdeffh3};
\nxt  in the IR, \ie as $y\equiv \frac1\rho\to 0$,
\begin{equation}
fl_{f;4;1}=fl_{f;4;1;0}^h+\calo(y)\,,\qquad fl_{g;4;2}=fl_{g;4;2;0}^h+\calo(y)
\,,\qquad \hfl_{K;4;3}=\hfl_{K;4;3;0}^h+\calo(y)\,,
\eqlabel{sirq332}
\end{equation}
it is completely specified by
\begin{equation}
\{fl_{f;4;1;0}^h\,,\, fl_{g;4;2;0}^h\,,\, \hfl_{K;4;3;0}^h\,,\, \alpha_2\,,\, s_{4;2}\}\,.
\eqlabel{ssetq332}
\end{equation}
In total, the UV and IR expansions are completely determined by the parameters
\eqref{ssetq332} and $fl_{f;4;1;8;0}$, which is precisely what is needed to find a unique solution
for three second order ODEs \eqref{sorder321}-\eqref{sorder323}.
Solving these equations we find
\begin{equation}
s_{4;2}=-1.7907(6)\,.
\eqlabel{sq332resu}
\end{equation}

Once the numerical solution for $\{fl_{f;4;1},fl_{g;4;2},\hfl_{K;4;3}\}$ is found, the second order
ODE for $fl_{w;4;2}$ --- necessary to determine $s_{4;k\ge 3}$ --- is solved adjusting
two parameters
\begin{equation}
\{fl_{w;4;2;6,0}\,,\, fl_{w;4;2;0}^h\}\,,
\eqlabel{sq372}
\end{equation}
that completely determine its UV and IR asymptotics.

\subsubsection{Details of $s=6\pm\calo(\sqrt{b})$ branches: (A$ _s$) and (B$ _s$) }
\label{detailss6}

From \eqref{sbrb}, here  
\begin{equation}
\begin{split}
s_{6;0}=6\,,\ fl_{g;6;0}=&
\mathcolorbox{red}{1}\cdot\ \frac{\rho^4}{(1+\rho)^s}\ _2 F_1\left(\frac 52,4-s; 5;
-\rho\right)\bigg|_{s=s_{6;0}}=
\frac{\rho^4(7\rho^2+24\rho+24)}{24(1+\rho)^6}\,,
\end{split}
\eqlabel{s7zero3}
\end{equation}
where we highlighted the (fixed) overall normalization of the linearized fluctuations;
the latter implies that in the UV, \ie $\rho\to 0$, expansion of $fl_{g;6;k\ge 1}$
the order $\calo(\rho^4)$ terms are absent. 
Because the leading order fluctuation spectra  \eqref{sbra}, \eqref{sbrb} and  \eqref{sbrc} are degenerate
at $s_{6;0}$,
the equations for $fl_{K;6;k}$ and $fl_{w;6;k}$ will necessarily contain zero modes;
specifically, if $fl_{K;6;k\ge 1}$
and $fl_{w;6;k\ge 1}$
are solutions, so are $(fl_{K;6;k}+\alpha_k\cdot fl_{g;6;0})$ and
$(fl_{w;6;k}+\frac{\beta_k}{\beta_0} \cdot fl_{w;6;0})$,
\begin{equation}
fl_{w;6;0}=\beta_0\cdot\frac{\rho^6}{(1+\rho)^s}\ _2 F_1\left(\frac {9}{2},6-s; 9;
-\rho\right)\bigg|_{s=s_{6;0}}=
\beta_0\cdot\frac{\rho^6}{(1+\rho)^6}\,,
\eqlabel{sq770}
\end{equation}
for an arbitrary set of constants $\{\alpha_k,\beta_k\}$.
As in section \ref{detailss3}, the zero modes at order $k$ will be completely fixed at order
$k+1$. We find it convenient to set
\begin{equation}
fl_{K;6;k\ge 1}\equiv \mathcolorbox{orange}{\alpha_k}\cdot  fl_{g;6;0}\
+\hfl_{K;6;k}\,,
\qquad fl_{w;6;k\ge 1}\equiv \mathcolorbox{green}{\beta_k}\cdot
\frac{1}{\beta_0}fl_{w;6;0}\ +\hfl_{w;6;k}\,,
\eqlabel{s7deffh3}
\end{equation}
with the understanding that in the UV  expansion of $\hfl_{K;6;k}$
the order $\calo(\rho^4)$ terms are absent, and in the UV  expansion of
$\hfl_{w;6;k}$
the order $\calo(\rho^6)$ terms are absent.

As in section \ref{detailss3}, the equation for $\hfl_{K;6;1}$
is homogeneous, and the boundary
condition implied by \eqref{s7deffh3} sets 
\begin{equation}
\hfl_{K;6,1}\equiv 0\,.
\eqlabel{s7fh13sol1}
\end{equation}

The subleading set of equations involving constants
$\alpha_1$, $\beta_0$, $s_{6;1}$, and functions $\{fl_{g;6;1}$, $\hfl_{K;6;2}$,
$\hfl_{w;6;1}$, $fl_{f;6;1}\}$ reads:
\begin{equation}
\begin{split}
&0=fl_{g;6;1}''+\frac{11\rho-6}{2\rho (1+\rho)}\ fl_{g;6;1}'-\frac{9}{(1+\rho)^2 \rho}\ fl_{g;6;1}
+\frac{\rho^3 s_{6;1} (5 \rho^2+24 \rho+40)}{16(1+\rho)^8}\\
&+\frac{\alpha_1 \rho^3 (\rho+2) (7 \rho^2+16 \rho+16)k_1'}{4(1+\rho)^7} \,,
\end{split}
\eqlabel{s7order31}
\end{equation}
\begin{equation}
\begin{split}
&0=\hfl_{K;6;2}''+\frac{11 \rho-6}{2\rho (1+\rho)}\ \hfl_{K;6;2}'-\frac{9}{(1+\rho)^2 \rho}\ \hfl_{K;6;2}
-\frac{(\rho+2) (7 \rho^2+16 \rho+16) \rho^3 k_1'}{8(1+\rho)^7}\\
&+\frac{(3 \alpha_1 \rho s_{6;1} (5 \rho^2+24 \rho+40)-(16 (7 \rho^2+24 \rho+24)) (1+\rho)) \rho^2}{48(1+\rho)^8}
\\&+\frac{2 \beta_0 \rho^4 (3 \rho k_1' (\rho+2)+8)}{(1+\rho)^7}\,,
\end{split}
\eqlabel{s7order32}
\end{equation}
\begin{equation}
\begin{split}
&0=\hfl_{w;6;1}''+\frac{11 \rho-6}{2\rho (1+\rho)}\ \hfl_{w;6;1}'-\frac{3(7 \rho+4)}{(1+\rho)^2 \rho^2}\ \hfl_{w;6;1}
+\frac{9\rho^5 \beta_0 s_{6;1}}{2(1+\rho)^8}\\&-\frac{\alpha_1 \rho^3 (\rho+2)(7 \rho^2+16 \rho+16) k_1'}{20(1+\rho)^7} \,,
\end{split}
\eqlabel{s7order33}
\end{equation}
\begin{equation}
\begin{split}
&0=fl_{f;6;1}''+\frac{11 \rho-6}{2\rho (1+\rho)}\ fl_{f;6;1}'-\frac{41 \rho+32}{(1+\rho)^2 \rho^2}\ fl_{f;6;1}\\&+
\frac{4(5 \rho^2+8 \rho+8) (7 \rho^2+24 \rho+24) \alpha_1}{7(1+\rho)^7}+\frac{4\alpha_1 \rho^3 (\rho+2) (3 \rho^2+4 \rho+4)k_1'}{5(1+\rho)^7}\,. 
\end{split}
\eqlabel{s7order34}
\end{equation}
We show here that the most
important constant, \ie $s_{6;1}$ can be computed analytically:
\begin{itemize}
\item Substituting
\begin{equation}
\alpha_1={s_{6;1}}\cdot v\,,\qquad fl_{g;6;1}=s_{6;1}\cdot fl_{g;6;0}\cdot J_{g;6;1}\,,
\eqlabel{s7so31}
\end{equation}
and using \eqref{k1anal}, 
we find a general analytic solution for $J_{g;6;1}'$,
\begin{equation}
\begin{split}
&J_{g;6;1}'=\frac{3\rho^3 v}{16(1+\rho)^{5/2}}
\ln\frac{\sqrt{1+\rho}+1}{\sqrt{1+\rho}-1}+\frac{(1+\rho)^{7/2}}{\rho^5 (7 \rho^2+24 \rho+24)^2}\ C_1
\\&-\frac{1}{40\rho^5 (1+\rho)^2 (7 \rho^2+24 \rho+24)^2} \biggl(
(\rho + 2) (735 \rho^{10} + 3080 \rho^9 - 6200 \rho^8 - 63104 \rho^7 \\&- 151104 \rho^6 - 100864 \rho^5 + 233984 \rho^4 + 622592 \rho^3
+ 647168 \rho^2 + 327680 \rho \\&+ 65536) v
+8 (\rho + 1) (75 \rho^8 + 624 \rho^7 + 1968 \rho^6 - 576 \rho^5 - 20160 \rho^4
- 53760 \rho^3 \\&- 64512 \rho^2 - 36864 \rho - 8192)
\biggr)\,.
\end{split}
\eqlabel{s7so32}
\end{equation}
Analyticity of $fl_{g;6;1}$, and thus $J_{g;6;1}'$, in the limit
$\rho\to \infty$ requires 
\begin{equation}
C_1=0\,,
\eqlabel{s7so33}
\end{equation}
while normalizability of $fl_{g;6;1}$ sets 
\begin{equation}
v=\frac 12\,.
\eqlabel{s7so35}
\end{equation}
\item We continue with \eqref{s7order33}, setting
\begin{equation}
\hfl_{w;6;1}=s_{6;1}\cdot \frac{1}{\beta_0}fl_{w;6;0}\cdot H_{w;6;1}\,,
\eqlabel{s7so36}
\end{equation}
allows to solve analytically for $H_{w;6;1}'$,
\begin{equation}
\begin{split}
&H_{w;6;1}'=-\frac{\rho (35 \rho^2+96 \rho+96)}{6400(1+\rho)^{5/2}}
\ln\frac{\sqrt{1+\rho}+1}{\sqrt{1+\rho}-1}+\frac{(1+\rho)^{7/2}}{\rho^9}\ C_1
\\&-\frac{1}{336000(1+\rho)^2 \rho^9} \biggl(
(9600 (1+\rho)) (45 \rho^8-144 \rho^7+672 \rho^6-8064 \rho^5-80640 \rho^4\\&-215040 \rho^3-258048 \rho^2-147456 \rho-32768) \beta_0-7 (\rho+2)
(525 \rho^{10}+40 \rho^9\\&-7000 \rho^8-11008 \rho^7-14208 \rho^6+74752 \rho^5+560128 \rho^4+1245184 \rho^3\\&
+1294336 \rho^2+655360 \rho+131072)
\biggr)\,.
\end{split}
\eqlabel{s7so37}
\end{equation}
Analyticity of $\hfl_{w;6;1}$, and thus $H_{w;6;1}'$, in the limit
$\rho \to \infty$ requires
\begin{equation}
C_1=0\,,
\eqlabel{s7so38}
\end{equation}
while normalizability of $\hfl_{w;6;1}$ sets 
\begin{equation}
\beta_0=-\frac{7}{1200}\,.
\eqlabel{s7so39}
\end{equation}
\item Consider now \eqref{s7order32}: introducing
\begin{equation}
\hfl_{K;6;2}=fl_{g;6;0}\cdot G_{K;6;2}\,,
\eqlabel{s7so40}
\end{equation}
we solve for $G_{K;6;2}'$,
\begin{equation}
\begin{split}
&G_{K;6;2}'=-\frac{3\rho^3 (3185 \rho^4+21504 \rho^3+57504 \rho^2+72000 \rho+36000)}
{2000(1+\rho)^{5/2} (7 \rho^2+24 \rho+24)^2}
\ln\frac{\sqrt{1+\rho}+1}{\sqrt{1+\rho}-1}
\\&+\frac{(1+\rho)^{7/2}}{\rho^5 (7 \rho^2+24 \rho+24)^2}\ C_1
+\frac{1}{5000\rho^5 (1+\rho)^2 (7 \rho^2+24 \rho+24)^2}
\biggl(
(\rho+2) (47775 \rho^{10}\\&+195160 \rho^9+166200 \rho^8-164672 \rho^7
+220128 \rho^6-2597632 \rho^5-19464448 \rho^4\\&-43270144 \rho^3
-44978176 \rho^2-22773760 \rho-4554752)-500 (1+\rho)
(75 \rho^8+624 \rho^7\\
&+1968 \rho^6-576 \rho^5-20160 \rho^4-53760 \rho^3
-64512 \rho^2-36864 \rho-8192) s_{6;1}^2
\biggr)\,.
\end{split}
\eqlabel{s7so41}
\end{equation}
 Analyticity of $\hfl_{K;6;2}$, and thus $G_{K;6;2}'$, in the limit $\rho\to \infty$ requires
\begin{equation}
C_1=0\,,
\eqlabel{sc1last}
\end{equation}
while normalizability of $\hfl_{K;6;2}$ sets
\begin{equation}
s_{6;1}^2=\frac{278}{125}\qquad \Longrightarrow\qquad s_{6;1}=\pm \frac{\sqrt{1390}}{25}\,.
\eqlabel{ss71last}
\end{equation}
\end{itemize}

The remaining equation, \ie \eqref{s7order34}, does not constrain $s_{6;1}$ ---
it is required to determine the higher-order corrections $s_{6,k\ge 2}$.

\subsubsection{Details of $s=6+\calo({b})$ branch: (C$ _s$)}
\label{sdetailss7l}

From \eqref{sbrc}, here  
\begin{equation}
\begin{split}
s_{6;0}=6\,,\qquad fl_{w;6;0}=&
\mathcolorbox{red}{1}\cdot\ \frac{\rho^6}{(1+\rho)^s}\
_2 F_1\left(\frac {9}{2},6-s; 9;
-\rho\right)\bigg|_{s=s_{6;0}}=
\frac{\rho^6}{(1+\rho)^6}\,,
\end{split}
\eqlabel{sl7zero3}
\end{equation}
where we highlighted the (fixed) overall normalization of the linearized fluctuations;
the latter implies that in the UV, \ie $\rho\to 0$, expansion of $fl_{w;6;k\ge 1}$
the order $\calo(\rho^6)$ terms are absent. 
Because the leading order fluctuation spectra of \eqref{sbra}, \eqref{sbrb} and
\eqref{sbrc}
are degenerate
at $s_{6;0}$,
the equations for $fl_{g;6;k}$ and $fl_{K;6;k}$ will necessarily contain zero modes; specifically,
if $fl_{g;6;k\ge 1}$
and $fl_{K;6;k\ge 1}$
are solutions, so are $(fl_{g;6;k}+\frac{\alpha_k}{\alpha_0}
\cdot fl_{g;6;0})$ and $(fl_{K;6;k}+\frac{\beta_k}{\alpha_0}\cdot fl_{g;6;0})$,
\begin{equation}
fl_{g;6;0}=\alpha_0\cdot _2F_1\left(\frac {5}{2},4-s; 5;
-\rho\right)\bigg|_{s=s_{6;0}}
=\alpha_0\cdot \frac{\rho^4(7\rho^2+24\rho+24)}{24(1+\rho)^6}\,,
\eqlabel{slq370}
\end{equation}
for an arbitrary set of constants $\{\alpha_k,\beta_k\}$.
As in section \ref{detailss3}, the zero modes at order $k$ will be completely fixed at order
$k+1$. We find it convenient to set
\begin{equation}
fl_{g;6;k\ge 1}\equiv \mathcolorbox{orange}{\alpha_k}\cdot  \frac{1}{\alpha_0}
fl_{g;6;0}\ +\hfl_{g;6;k}\,,
\qquad fl_{K;6;k\ge 1}\equiv \mathcolorbox{green}{\beta_k}\cdot
\frac{1}{\alpha_0}fl_{g;6;0}\ +\hfl_{K;6;k}\,,
\eqlabel{sl7deffh3}
\end{equation}
with the understanding that in the UV  expansion of $\hfl_{g;6;k}$ and
$\hfl_{K;6;k}$
the order $\calo(\rho^4)$ terms are absent.

The subleading set of equations involving constants
$\alpha_0$, $\beta_1$, $s_{6;1}$, and functions $\{\hfl_{g;6;1}$, $\hfl_{K;6;1}$, $fl_{w;6;1}\}$ reads
(we do not discuss the equation for $fl_{f;6;1}$ --- it is needed to determine
$s_{6;k\ge 2}$, but it does not affect the computation of $s_{6;1}$) :
\begin{equation}
\begin{split}
&0=\hfl_{g;6;1}''+\frac{11 \rho-6}{2\rho (1+\rho)}\ \hfl_{g;6;1}'
-\frac{9}{\rho (1+\rho)^2}\ \hfl_{g;6;1}+2 k_1' \hfl_{K;6;1}'
+\frac{6 k_1' \hfl_{K;6;1}}{1+\rho}\\
&+\frac{\rho^3 k_1' (\rho+2) (7 \rho^2+16 \rho+16) \beta_1}{4(1+\rho)^7}
+\frac{2 \rho^6 (k_1')^2}{(1+\rho)^6}+\frac{8 \rho^4}{(1+\rho)^7}-\frac{ \rho^2\alpha_0}
{48(1+\rho)^8} \biggl(
-3 \rho (\rho+2) \\&\times(1+\rho) (7 \rho^2+16 \rho+16) (f_{2;1}'+4f_{3;1}'
-4 g_1')+9 \rho^2 (7 \rho^2+24 \rho+24) h_1-3 \rho (5 \rho^2+24 \rho\\
&+40) s_{6;1}+16 (1+\rho) (7 \rho^2+24 \rho+24)
\biggr)\,,
\end{split}
\eqlabel{sl7order31}
\end{equation}
\begin{equation}
\begin{split}
&0=\hfl_{K;6;1}''+\frac{11 \rho-6}{2\rho (1+\rho)}\ \hfl_{K;6;1}'
-\frac{9}{\rho (1+\rho)^2}\ \hfl_{K;6;1}+\frac{2 (3 k_1' \rho (\rho+2)+8) \rho^4}{(1+\rho)^7}
\\&-\frac{\rho^2 (3 k_1' \rho (\rho+2) (7 \rho^2+16 \rho+16)
+56 \rho^2+192 \rho+192) \alpha_0}{24(1+\rho)^7}\,,
\end{split}
\eqlabel{sl7order32}
\end{equation}
\begin{equation}
\begin{split}
&0=fl_{w;6;1}''+\frac{11 \rho-6}{2\rho (1+\rho)}\ fl_{w;6;1}'
-\frac{3 (7 \rho+4)}{\rho^2 (1+\rho)^2}\ fl_{w;6;1}-\frac25 k_1' \hfl_{K;6;1}'
-\frac{6k_1' \hfl_{K;6;1}}{5(1+\rho)}\\&+\frac{9\rho^5 s_{6;1}}{2(1+\rho)^8}
-\frac{19\rho^6 (k_1')^2}{20(1+\rho)^6}
+\frac{(7 \rho^2+24 \rho+24) ((k_1')^2 \rho^2 (1+\rho)+4) \rho^2 \alpha_0}
{120(1+\rho)^7}
\\&-\frac{\rho^3 k_1' (\rho+2) (7 \rho^2+16 \rho+16) \beta_1}{20(1+\rho)^7}
-\frac{\rho^4}{20(1+\rho)^8} \biggl(
2 \rho (\rho + 2) (\rho + 1) (33 f_{2;1}'
- 48 f_{3;1}' \\&- 5 h_1') + 5 (7 \rho + 4) (3 \rho - 4) h_1
+ 80 (\rho + 1) \left(k_1 + \frac{19}{4} f_{2;1} - 11 f_{3;1} - \frac{3}{20}\right)
\biggr) \,.
\end{split}
\eqlabel{sl7order33}
\end{equation}
Eqs. \eqref{sl7order31}---\eqref{sl7order33} are solved subject to the asymptotic
expansions,
\nxt in the UV, \ie as $\rho\to 0$,
\begin{equation}
\begin{split}
&\hfl_{g;6;1}=\biggl(\mathcolorbox{orange}{0}+(2\alpha_0+4\beta_1)\ln\rho\biggr) \rho^4+\calo(\rho^5\ln\rho)\,,
\end{split}
\eqlabel{sluvq332}
\end{equation}
\begin{equation}
\begin{split}
&\hfl_{K;6;1}=\mathcolorbox{green}{0}\rho^4+\mathcolorbox{green}{0}\rho^5+
\left(\frac23-\frac{1}{72}\alpha_0\right) \rho^6
+\calo(\rho^7)\,,
\end{split}
\eqlabel{sluvq3322}
\end{equation}
\begin{equation}
\begin{split}
&fl_{w;6;1}=\left( \frac{2}{15} \alpha_0+\frac{4}{15} \beta_1\right) \rho^4
+\left(-\frac23 \alpha_0-\frac43 \beta_1\right) \rho^5
+\biggl(\mathcolorbox{red}{0}+\left(
\frac{7}{120} \alpha_0-\frac{1}{120} \beta_1\right) \ln\rho\biggr) \rho^6 \\
&+\calo(\rho^7\ln\rho)\,,
\end{split}
\eqlabel{sluvq3323}
\end{equation}
it is completely specified by $\{\alpha_0,\beta_1,s_{6;1}\}$; we further highlighted
arbitrary constants, fixed to zero by the overall normalization \eqref{sl7zero3}, and the
 extraction of the zero modes  in $fl_{g;6;1}$  and $fl_{K;6;1}$ \eqref{sl7deffh3};
\nxt  in the IR, \ie as $y\equiv \frac1\rho\to 0$,
\begin{equation}
\hfl_{g;6;1}=\hfl_{g;6;1;0}^h+\calo(y)\,,\qquad
\hfl_{K;6;1}=\hfl_{K;6;1;0}^h+\calo(y)\,,\qquad 
fl_{w;6;1}=fl_{w;6;1;0}^h+\calo(y)\,,
\eqlabel{slirq332}
\end{equation}
it is completely specified by
\begin{equation}
\{\hfl_{g;6;1;0}^h\,,\, \hfl_{K;6;1;0}^h\,,\,  fl_{w;6;1;0}^h\,,\, \alpha_0\,,\, \beta_{1}\,,\,  s_{6;1}\}\,.
\eqlabel{slsetq332}
\end{equation}
In total, the UV and IR expansions are completely determined by the parameters
\eqref{slsetq332}, which is precisely what is needed to find a unique solution
for three second order ODEs \eqref{sl7order31}-\eqref{sl7order33}.
Solving these equations we find
\begin{equation}
\begin{split}
&s_{6;1}=6.0318(6)\,.
\end{split}
\eqlabel{slq332resu}
\end{equation}

\subsubsection{Details of $s=8\pm\calo(\sqrt{b})$ branches: (A$ _s$) and (B$ _s$) }
\label{detailss8}

From \eqref{sbrb}, here  
\begin{equation}
\begin{split}
s_{8;0}=8\,,\ fl_{g;8;0}=&
\mathcolorbox{red}{1}\cdot\ \frac{\rho^4}{(1+\rho)^s}\ _2 F_1\left(\frac 52,4-s; 5;
-\rho\right)\bigg|_{s=s_{8;0}}\\=&
\frac{\rho^4 (33 \rho^4+192 \rho^3+448 \rho^2+512 \rho+256)}{256(1+\rho)^8}\,,
\end{split}
\eqlabel{s87zero3}
\end{equation}
where we highlighted the (fixed) overall normalization of the linearized fluctuations;
the latter implies that in the UV, \ie $\rho\to 0$, expansion of $fl_{g;8;k\ge 1}$
the order $\calo(\rho^4)$ terms are absent. 
Because the leading order fluctuation spectra \eqref{sbra}, \eqref{sbrb}, \eqref{sbrc}
and \eqref{sbrd} are degenerate at $s_{8;0}$,
the equations for  $fl_{K;8;k}$, $fl_{w;8;k}$ and $fl_{f;8;k}$ will necessarily contain zero modes;
specifically, if $fl_{K;8;k\ge 1}$, $fl_{w;8;k\ge 1}$
and $fl_{f;8;k\ge 1}$
are solutions, so are $(fl_{K;8;k}+\alpha_k\cdot fl_{g;8;0})$,
$(fl_{w;8;k}+\frac{\beta_k}{\beta_0} \cdot fl_{w;8;0})$ and
$(fl_{f;8;k}+\frac{\gamma_k}{\gamma_0} \cdot fl_{f;8;0})$,
\begin{equation}
\begin{split}
&fl_{w;8;0}=\beta_0\cdot\frac{\rho^6}{(1+\rho)^s}\ _2 F_1\left(\frac {9}{2},6-s; 9;
-\rho\right)\bigg|_{s=s_{8;0}}=
\beta_0\cdot\frac{\rho^6 (11\rho^2+40\rho+40)}{40(1+\rho)^8}\,,\\
&fl_{f;8;0}=\gamma_0\cdot\frac{\rho^8}{(1+\rho)^s}\ _2 F_1\left(\frac {13}{2},8-s; 13;
-\rho\right)\bigg|_{s=s_{8;0}}=
\gamma_0\cdot\frac{\rho^8}{(1+\rho)^8}\,,
\end{split}
\eqlabel{s8q770}
\end{equation}
for an arbitrary set of constants $\{\alpha_k,\beta_k,\gamma_k\}$.
As in section \ref{detailss3}, the zero modes at order $k$ will be completely fixed at order
$k+1$. We find it convenient to set
\begin{equation}
\begin{split}
&fl_{K;8;k\ge 1}\equiv \mathcolorbox{orange}{\alpha_k}\cdot  fl_{g;8;0}\
+\hfl_{K;8;k}\,,
\qquad fl_{w;8;k\ge 1}\equiv \mathcolorbox{green}{{\beta_k}}\cdot \frac{1}{\beta_0}
fl_{w;8;0}\ +\hfl_{w;8;k}\,,\\
&fl_{f;8;k\ge 1}\equiv \mathcolorbox{pink}{{\gamma_k}}\cdot \frac{1}{\gamma_0}
fl_{f;8;0}\ +\hfl_{f;8;k}\,,
\end{split}
\eqlabel{s87deffh3}
\end{equation}
with the understanding that in the UV  expansion of $\hfl_{K;8;k}$
the order $\calo(\rho^4)$ terms are absent,  in the UV  expansion of $\hfl_{w;8;k}$
the order $\calo(\rho^6)$ terms are absent, and in the UV  expansion of
$\hfl_{f;8;k}$
the order $\calo(\rho^8)$ terms are absent.

As in section \ref{detailss3}, the equation for $\hfl_{K;8;1}$
is homogeneous, and the boundary
condition implied by \eqref{s87deffh3} sets 
\begin{equation}
\hfl_{K;8,1}\equiv 0\,.
\eqlabel{s87fh13sol1}
\end{equation}

The subleading set of equations involving constants
$\alpha_1$, $\beta_0$, $\gamma_0$, $s_{8;1}$, and functions $\{fl_{g;8;1}$, $\hfl_{K;8;2}$,
$\hfl_{w;8;1}$, $\hfl_{f;8;1}\}$ reads:
\begin{equation}
\begin{split}
&0=fl_{g;8;1}''+\frac{3(5 \rho-2)}{2(1+\rho) \rho}\ fl_{g;8;1}'-\frac{12}{\rho (1+\rho)^2}\ fl_{g;8;1}
+\frac{s_{8;1} \rho^3}{512(1+\rho)^{10}} \biggl(45 \rho^4+320 \rho^3+960 \rho^2\\
&+1536 \rho+1280\biggr)
+\frac{k_1' \alpha_1 \rho^3 (\rho+2) (33 \rho^4+144 \rho^3+272 \rho^2+256 \rho+128)}
{32(1+\rho)^9}\,,
\end{split}
\eqlabel{s87order31}
\end{equation}
\begin{equation}
\begin{split}
&0=\hfl_{K;8;2}''+\frac{3(5 \rho-2)}{2\rho (1+\rho)}\ \hfl_{K;8;2}'
-\frac{12}{\rho (1+\rho)^2}\ \hfl_{K;8;2}
-\frac{(\rho+2)\rho^3 k_1'}{64(1+\rho)^9}
\biggl(33 \rho^4+144 \rho^3\\
&+272 \rho^2+256 \rho+128\biggr) +\frac{\alpha_1 s_{8;1} \rho^3
(45 \rho^4+320 \rho^3+960 \rho^2+1536 \rho+1280)}{512(1+\rho)^{10}} \\
&+\frac{\rho^4 (\rho k_1' (\rho+2) (11 \rho^2+30 \rho+30)+22 \rho^2+80 \rho+80) \beta_0}
{5(1+\rho)^9}-\frac{\gamma_0 \rho^6 (3 \rho k_1' (\rho+2)-14)}{(1+\rho)^9}
\\&-\frac{\rho^2 (33 \rho^4+192 \rho^3+448 \rho^2+512 \rho+256)}{32(1+\rho)^9}\,,
\end{split}
\eqlabel{s87order32}
\end{equation}
\begin{equation}
\begin{split}
&0=\hfl_{w;8;1}''+\frac{3(5 \rho-2)}{2\rho (1+\rho)}\ \hfl_{w;8;1}'-\frac{12 (2 \rho+1)}{\rho^2 (1+\rho)^2}\
\hfl_{w;8;1}+\frac{(63 \rho^2+280 \rho+360) \beta_0 \rho^5 s_{8;1}}{80(1+\rho)^{10}}
\\
&-\frac{k_1'\alpha_1 (\rho+2) (33 \rho^4+144 \rho^3+272 \rho^2+256 \rho+128) \rho^3}{160(1+\rho)^9}\,,
\end{split}
\eqlabel{s87order33}
\end{equation}
\begin{equation}
\begin{split}
&0=\hfl_{f;8;1}''+\frac{3(5 \rho-2)}{2\rho (1+\rho)}\ \hfl_{f;8;1}'-\frac{4 (11 \rho+8)}{\rho^2 (1+\rho)^2}\
\hfl_{f;8;1}+\frac{(\rho+2)\alpha_1 \rho^3k_1'}{120(1+\rho)^9} \biggl(
154 \rho^4+607 \rho^3\\&+991 \rho^2+768 \rho+384\biggr)
+\frac{13s_{8;1} \rho^7\gamma_0}{2(1+\rho)^{10}}+\frac{\alpha_1 (37 \rho^2+24 \rho+24)}{144(1+\rho)^9}
\biggl(33 \rho^4+192 \rho^3+448 \rho^2\\&+512 \rho+256\biggr)\,.
\end{split}
\eqlabel{s87order34}
\end{equation}
We show here that the most
important constant, \ie $s_{8;1}$ can be computed analytically:
\begin{itemize}
\item Substituting
\begin{equation}
\alpha_1={s_{8;1}}\cdot v\,,\qquad fl_{g;8;1}=s_{8;1}\cdot fl_{g;8;0}\cdot J_{g;8;1}\,,
\eqlabel{s87so31}
\end{equation}
and using \eqref{k1anal}, 
we find a general analytic solution for $J_{g;8;1}'$,
\begin{equation}
\begin{split}
&J_{g;8;1}'=\frac{3\rho^3 v}{16(1+\rho)^{5/2}}\
\ln\frac{\sqrt{1+\rho}+1}{\sqrt{1+\rho}-1}+\frac{(1+\rho)^{11/2}\ C_1}{
\rho^5 (33 \rho^4+192 \rho^3+448 \rho^2+512 \rho+256)^2}
\\&-\frac{1}{840\rho^5 (1+\rho)^2 (33 \rho^4+192 \rho^3+448 \rho^2+512 \rho+256)^2} \biggl(
(\rho+2) (343035 \rho^{14}\\&+3076920 \rho^{13}+8036280 \rho^{12}-17283840 \rho^{11}-182325120 \rho^{10}
-613628928 \rho^9\\&-1212027904 \rho^8-1507459072 \rho^7-1026850816 \rho^6+74186752 \rho^5
+1002176512 \rho^4\\&+1149239296 \rho^3+700448768 \rho^2+234881024 \rho+33554432) v
+8 (1+\rho)\\& \times
(14175 \rho^{12}
+186200 \rho^{11}+1114000 \rho^{10}+3973568 \rho^9+9149312 \rho^8+11880448 \rho^7
\\&-2265088 \rho^6-45760512 \rho^5-98402304 \rho^4-112459776 \rho^3-74973184 \rho^2\\
&-27262976 \rho-4194304)
\biggr)\,.
\end{split}
\eqlabel{s87so32}
\end{equation}
Analyticity of $fl_{g;8;1}$, and thus $J_{g;8;1}'$, in the limit
$\rho\to \infty$ requires 
\begin{equation}
C_1=0\,,
\eqlabel{s87so33}
\end{equation}
while normalizability of $fl_{g;8;1}$ sets 
\begin{equation}
v=\frac 12\,.
\eqlabel{s87so35}
\end{equation}
\item We continue with \eqref{s87order33}, setting
\begin{equation}
\hfl_{w;8;1}=s_{8;1}\cdot \frac{1}{\beta_0}fl_{w;8;0}\cdot H_{w;8;1}\,,
\eqlabel{s87so36}
\end{equation}
allows to solve analytically for $H_{w;8;1}'$,
\begin{equation}
\begin{split}
&H_{w;8;1}'=-\frac{\rho}{7168(1+\rho)^{5/2} (11 \rho^2+40 \rho+40)^2}
\biggl(7623 \rho^6+69696 \rho^5+270400 \rho^4
\\&+573440 \rho^3+716800 \rho^2+516096 \rho+172032\biggr)
\ln\frac{\sqrt{1+\rho}+1}{\sqrt{1+\rho}-1}
\\&+\frac{(1+\rho)^{11/2}}{(11 \rho^2+40 \rho+40)^2 \rho^9}\ C_1
-\frac{1}{225792(1+\rho)^2 (11 \rho^2+40 \rho+40)^2 \rho^9}
\biggl(
3584 (1+\rho)\\&\times (3969 \rho^{12}+28616 \rho^{11}+69184 \rho^{10}
+45824 \rho^9+27456 \rho^8-439296 \rho^7
\\&-6150144 \rho^6-24600576 \rho^5-49201152 \rho^4-56229888 \rho^3-37486592 \rho^2
\\&-13631488 \rho-2097152) \beta_0-3 (\rho+2) (160083 \rho^{14}+1036728 \rho^{13}
+727608 \rho^{12}\\&-12423936 \rho^{11}-51946368 \rho^{10}-99511296 \rho^9
-109388800 \rho^8-48627712 \rho^7\\&+163610624 \rho^6+624689152 \rho^5+1112276992 \rho^4
+1149239296 \rho^3+700448768 \rho^2\\&+234881024 \rho+33554432)
\biggr)\,.
\end{split}
\eqlabel{s87so37}
\end{equation}
Analyticity of $\hfl_{w;8;1}$, and thus $H_{w;8;1}'$, in the limit
$\rho \to \infty$ requires
\begin{equation}
C_1=0\,,
\eqlabel{s87so38}
\end{equation}
while normalizability of $\hfl_{w;8;1}$ sets 
\begin{equation}
\beta_0=-\frac{3}{112}\,.
\eqlabel{s87so39}
\end{equation}
\item Substituting 
\begin{equation}
\hfl_{f;8;1}=s_{8;1}\cdot \frac{1}{\gamma_0}fl_{f;8;0}\cdot B_{f;8;1}\,,
\eqlabel{sb87so36}
\end{equation}
allows to solve analytically for $B_{f;8;1}'$,
\begin{equation}
\begin{split}
&B_{f;8;1}'=\frac{539 \rho^4+2428 \rho^3+4220 \rho^2+3584 \rho+1792}{35840(1+\rho)^{5/2} \rho}
\ln\frac{\sqrt{1+\rho}+1}{\sqrt{1+\rho}-1}+\frac{(1+\rho)^{11/2}}{\rho^{13}} C_1\\
&-\frac{1}{62092800(1+\rho)^2 \rho^{13}} \biggl(
268800 (1+\rho) (273 \rho^{12}-728 \rho^{11}+2288 \rho^{10}-9152 \rho^9\\
&+54912 \rho^8
-878592 \rho^7-12300288 \rho^6-49201152 \rho^5-98402304 \rho^4-112459776 \rho^3
\\&-74973184 \rho^2-27262976 \rho-4194304) \gamma_0+11 (\rho+2) (169785 \rho^{14}
+312060 \rho^{13}\\&+2529660 \rho^{12}+17283840 \rho^{11}+50552960 \rho^{10}+78883328 \rho^9
+71192064 \rho^8\\&+63537152 \rho^7+279535616 \rho^6+937033728 \rho^5
+1668415488 \rho^4+1723858944 \rho^3\\&+1050673152 \rho^2+352321536 \rho+50331648)
\biggr)\,.
\end{split}
\eqlabel{sb87so37}
\end{equation}
Analyticity of $\hfl_{f;8;1}$, and thus $B_{f;8;1}'$, in the limit
$\rho \to \infty$ requires
\begin{equation}
C_1=0\,,
\eqlabel{sb87so38}
\end{equation}
while normalizability of $\hfl_{f;8;1}$ sets 
\begin{equation}
\gamma_0=\frac{11}{11200}\,.
\eqlabel{sb87so39}
\end{equation}
\item Consider now \eqref{s87order32}: introducing
\begin{equation}
\hfl_{K;8;2}=fl_{g;8;0}\cdot G_{K;8;2}\,,
\eqlabel{s87so40}
\end{equation}
we solve for $G_{K;8;2}'$,
\begin{equation}
\begin{split}
&G_{K;8;2}'=-\frac{3\rho^3}{9800(1+\rho)^{5/2} (33 \rho^4+192 \rho^3+448 \rho^2+512 \rho+256)^2} \biggl(
373527 \rho^8\\&+4265712 \rho^7+21950064 \rho^6+66764544 \rho^5+131942272 \rho^4+174469120 \rho^3
\\&+151818240 \rho^2+80281600 \rho+20070400\biggr)
\ln\frac{\sqrt{1+\rho}+1}{\sqrt{1+\rho}-1} \\&+\frac{(1+\rho)^{11/2}}
{\rho^5 (33 \rho^4+192 \rho^3+448 \rho^2+512 \rho+256)^2}\ C_1
\\&+\frac{1}{514500\rho^5 (1+\rho)^2 (33 \rho^4+192 \rho^3+448 \rho^2+512 \rho+256)^2} \biggl(
(\rho+2) (\\
&117661005 \rho^{14}+1029936600 \rho^{13}+3458168280 \rho^{12}+4151454720 \rho^{11}
\\&-5844802560 \rho^{10}-27053374464 \rho^9-37550953472 \rho^8-30962548736 \rho^7
\\&-76202098688 \rho^6-243316424704 \rho^5-433231888384 \rho^4
-447628705792 \rho^3\\&-272824795136 \rho^2-91486158848 \rho
-13069451264)-2450 (1+\rho) (14175 \rho^{12}\\&+186200 \rho^{11}+1114000 \rho^{10}+3973568 \rho^9
+9149312 \rho^8+11880448 \rho^7-2265088 \rho^6\\&-45760512 \rho^5-98402304 \rho^4
-112459776 \rho^3-74973184 \rho^2-27262976 \rho\\
&-4194304) s_{8;1}^2\biggr)\,.
\end{split}
\eqlabel{s87so41}
\end{equation}
 Analyticity of $\hfl_{K;8;2}$, and thus $G_{K;8;2}$, in the limit $\rho\to \infty$ requires
\begin{equation}
C_1=0\,,
\eqlabel{s8c1last}
\end{equation}
while normalizability of $\hfl_{K;8;2}$ sets
\begin{equation}
s_{8;1}^2=\frac{3116}{1225}\qquad \Longrightarrow\qquad s_{8;1}=\pm \frac{2\sqrt{779}}{35}\,.
\eqlabel{s8s71last}
\end{equation}
\end{itemize}

\subsubsection{Details of $s=8+\calo({b})$ branch: (C$ _s$)}
\label{detailssc8}

From \eqref{sbrc}, here  
\begin{equation}
\begin{split}
s_{8;0}=8\,,\  fl_{w;8;0}=&
\mathcolorbox{red}{1}\cdot\ \frac{\rho^6}{(1+\rho)^s}\
_2 F_1\left(\frac {9}{2},6-s; 9;
-\rho\right)\bigg|_{s=s_{8;0}}=
\frac{\rho^6(11\rho^2+40\rho+40)}{40(1+\rho)^8}\,,
\end{split}
\eqlabel{s8l7zero3}
\end{equation}
where we highlighted the (fixed) overall normalization of the linearized fluctuations;
the latter implies that in the UV, \ie $\rho\to 0$, expansion of $fl_{w;8;k\ge 1}$
the order $\calo(\rho^6)$ terms are absent.
Because the leading order fluctuation spectra \eqref{sbra}, \eqref{sbrb}, \eqref{sbrc}
and \eqref{sbrd} 
are degenerate 
at $s_{8;0}$,
the equations for $fl_{g;8;k}$, $fl_{K;8;k}$ and $fl_{f;8;k}$ will necessarily contain zero modes;
specifically,
if $fl_{g;8;k\ge 1}$, $fl_{K;8;k\ge 1}$ and $fl_{f;8;k\ge 1}$
are solutions, so are $(fl_{g;8;k}+\frac{\alpha_k}{\alpha_0}\cdot fl_{g;8;0})$,
$(fl_{K;8;k}+\frac{\beta_k}{\alpha_0}\cdot fl_{g;8;0})$ and
$(fl_{f;8;k}+\frac{\gamma_k}{\gamma_0}\cdot fl_{f;8;0})$,
\begin{equation}
\begin{split}
fl_{g;8;0}=&\alpha_0\cdot _2F_1\left(\frac {5}{2},4-s; 5;
-\rho\right)\bigg|_{s=s_{8;0}}\\
 =&\alpha_0\cdot \frac{\rho^4(33\rho^4+192\rho^3+448\rho^2+512\rho+256)}{256(1+\rho)^8}
\,,\\
fl_{f;8;0}=&\gamma_0\cdot _2F_1\left(\frac {13}{2},8-s; 13;
-\rho\right)\bigg|_{s=s_{8;0}}=\gamma_0\cdot \frac{\rho^8}{(1+\rho)^8}\,,
\end{split}
\eqlabel{s8lq370}
\end{equation}
for an arbitrary set of constants $\{\alpha_k,\beta_k,\gamma_k\}$.
As in section \ref{detailss3}, the zero modes at order $k$ will be completely fixed at order
$k+1$. We find it convenient to set
\begin{equation}
\begin{split}
&fl_{g;8;k\ge 1}\equiv \mathcolorbox{orange}{\alpha_k}\cdot  \frac{1}{\alpha_0}
fl_{g;8;0}\ +\hfl_{g;8;k}\,,
\qquad fl_{K;8;k\ge 1}\equiv \mathcolorbox{green}{\beta_k}\cdot
\frac{1}{\alpha_0}fl_{g;8;0}\ +\hfl_{K;8;k}\,,\\
&fl_{f;8;k\ge 1}\equiv \mathcolorbox{pink}{\gamma_k}\cdot
\frac{1}{\alpha_0}fl_{f;8;0}\ +\hfl_{f;8;k}\,,
\end{split}
\eqlabel{s8l7deffh3}
\end{equation}
with the understanding that in the UV  expansion of $\hfl_{g;8;k}$ and
$\hfl_{K;8;k}$
the order $\calo(\rho^4)$ terms are absent, and in the UV  expansion of $\hfl_{f;8;k}$ and
the order $\calo(\rho^8)$ terms are absent .

The subleading set of equations involving constants
$\alpha_0$, $\beta_1$, $\gamma_0$, $s_{8;1}$, and functions
$\{\hfl_{g;8;1}$, $\hfl_{K;8;1}$, $fl_{w;8;1}$ , $\hfl_{f;8;1}\}$ reads:
\begin{equation}
\begin{split}
&0=\hfl_{g;8;1}''+\frac{3(5 \rho-2)}{2\rho (1+\rho)} \hfl_{g;8;1}'
-\frac{12}{\rho (1+\rho)^2} \hfl_{g;8;1}
+2 k_1' \hfl_{K;8;1}'+\frac{8 k_1' \hfl_{K;8;1}}{1+\rho}
\\&+\frac{\rho^3 k_1' (\rho+2)
(33 \rho^4+144 \rho^3+272 \rho^2+256 \rho+128) \beta_1}{32(1+\rho)^9}
-\frac{\rho^6 \gamma_0 (3 \rho^2 (k_1')^2 (1+\rho)-28)}{4(1+\rho)^9}
\\&+\frac{\rho^3 \alpha_0 s_{8;1} (45 \rho^4+320 \rho^3+960 \rho^2+1536 \rho+1280)}{512(1+\rho)^{10}}
+\frac{11 \rho^2}{20 (1 + \rho)^{10}}
\biggl(
\frac{5}{352} \alpha_0 \rho (\rho + 2)\\
&\times (1 + \rho)
(33 \rho^4 + 144 \rho^3 + 272 \rho^2 + 256 \rho + 128) (f_{2;1}' + 4 f_{3;1}'
- 4 g_1') + \rho^4 \biggl(\rho^2 \\&+ \frac{40}{11} (\rho + 1)\biggr) (1 + \rho)^2 (k_1')^2
- \frac{75\alpha_0 \rho^2 h_1}{32} \biggl(\rho^4 + \frac{64}{11} \rho^3
+ \frac{448}{33} \rho^2 +\frac{512}{33} \rho + \frac{256}{33}\biggr) 
\\&-\frac{15}{8}(1+\rho)\biggl(
\alpha_0\left(\rho^4+\frac{64}{11}\rho^3+\frac{448}{33}\rho^2+\frac{512}{33}\rho+\frac{256}{33}\right)
-\frac{32}{15}\rho^4-\frac{256}{33}\rho^3-\frac{256}{33}\rho^2\biggr)
\biggr)\,,
\end{split}
\eqlabel{s8l7order31}
\end{equation}
\begin{equation}
\begin{split}
&0=\hfl_{K;8;1}''+\frac{3(5 \rho-2)}{2\rho (1+\rho)} \hfl_{K;8;1}'
-\frac{12}{\rho (1+\rho)^2} \hfl_{K;8;1}
-\frac{\rho^2 \alpha_0}{64(1+\rho)^9}
\biggl(\rho k_1' (\rho+2) (33 \rho^4\\&+144 \rho^3+272 \rho^2+256 \rho+128)
+384 \rho^3+66 \rho^4+896 \rho^2+1024 \rho+512\biggr)
\\&-\frac{3 \rho k_1' (\rho+2)-14) \gamma_0 \rho^6}{(1+\rho)^9}
+\frac{(\rho k_1' (\rho+2) (11 \rho^2+30 \rho+30)+22 \rho^2+80 \rho+80) \rho^4}
{5(1+\rho)^9}\,,
\end{split}
\eqlabel{s8l7order32}
\end{equation}
\begin{equation}
\begin{split}
&0=fl_{w;8;1}''+\frac{3(5 \rho-2)}{2\rho (1+\rho)} fl_{w;8;1}'
-\frac{12 (2 \rho+1)}{\rho^2 (1+\rho)^2} fl_{w;8;1}
-\frac25 k_1' \hfl_{K;8;1}'-\frac{8k_1' \hfl_{K;8;1}}{5(1+\rho)}
\\&-\frac{\rho^3 k_1' (\rho+2) (33 \rho^4+144 \rho^3+272 \rho^2+256 \rho+128)
\beta_1}{160(1+\rho)^9}
+\frac{\rho^5 (63 \rho^2+280 \rho+360) s_{8;1}}{80(1+\rho)^{10}}
\\&+\frac{
(33 \rho^4+192 \rho^3+448 \rho^2+512 \rho+256) (\rho^2 (k_1')^2 (1+\rho)+4)
\rho^2 \alpha_0}{1280(1+\rho)^9}
-\frac{\gamma_0 \rho^6}{20(1+\rho)^9}
\biggl(
\\&\rho^2 (k_1')^2
(1+\rho)-32 \rho (\rho+2) (f_{3;1}'-f_{2;1}')+80 (f_{2;1}-f_{3;1})+12
\biggr)
-\frac{\rho^4}{800 (1 + \rho)^{10}} \biggl(
\\&8 \rho (\rho + 2) (1 + \rho) (11 \rho^2 + 30 \rho + 30)
(11 f_{2;1}' - 16 f_{3;1}') + 209 \biggl(\rho^2 + \frac{40}{11} (\rho +1)\biggr)
\biggl(
-\frac{10}{19} \rho \\&\times (\rho + 2)(1 + \rho) h_1' + \rho^2 (1 + \rho)^2 (k_1')^2
+ \biggl(\frac{215}{19} \rho^2 - \frac{80}{19} (\rho +1)\biggr) h_1
+ \frac{80(1+\rho)}{19} \biggl(k_1 \\&+ \frac{19}{4} f_{2;1} 
- 11 f_{3;1} - \frac{3}{20}\biggr) 
\biggr)
\biggr)\,,
\end{split}
\eqlabel{s8l7order33}
\end{equation}
\begin{equation}
\begin{split}
&0=\hfl_{f;8;1}''+\frac{3(5 \rho-2)}{2\rho (1+\rho)} \hfl_{f;8;1}'
-\frac{4 (11 \rho+8)}{\rho^2 (1+\rho)^2} \hfl_{f;8;1}
+\frac{2(31 \rho^2-120 \rho-120) k_1' \hfl_{K;8;1}'}{45\rho^2}
\\&+\frac{16 (k_1' \rho (28 \rho^3-5 \rho^2+30 \rho+60)+185 \rho^2+120 \rho+120)
\hfl_{K;8;1}}{45(1+\rho) \rho^4}
+\frac{\beta_1}{720(1+\rho)^9} \biggl(
6 \rho^3 \\&\times(\rho+2) (154 \rho^4+607 \rho^3+991 \rho^2+768 \rho+384) k_1'
+5 (37 \rho^2+24 \rho+24) (33 \rho^4+192 \rho^3\\&+448 \rho^2+512 \rho+256)
\biggr)
+\frac{\alpha_0}{11520 (1 + \rho)^9} \biggl(
-\rho^2 (1 + \rho) (31 \rho^2 - 120 \rho - 120) (33 \rho^4 \\&+ 192 \rho^3 + 448 \rho^2
+ 512 \rho + 256) (k_1')^2 + 240 \rho^5 (\rho + 2) (11 \rho^2 + 35 \rho + 35) g_1' + 4 (149 \rho^2 \\
&+ 120 (\rho +1)) (33 \rho^4 + 192 \rho^3 + 448 \rho^2 + 512 \rho + 256)
\biggr)
-\frac{27 \rho^4 \gamma_0}{40 (1 + \rho)^{10}} \biggl(
-\frac{2}{243} \rho (\rho + 2)\\&\times (1 + \rho) (157 \rho^2 - 660 \rho - 660) (f_{2;1}'
+ 4 f_{3;1}')
- \frac{830}{243} \rho (\rho + 2) (1 + \rho) \left(\rho^2 - \frac{276}{83} (\rho+1)
\right)\\&\times h_1'
+ \left(\rho^2 - \frac{20}{9} (\rho+1)\right) \rho^2 (1 + \rho)^2 (k_1')^2
+ \biggl(
-\frac{11840}{81} + \frac{455}{27} \rho^4 - \frac{58900}{243} \rho^3 - \frac{94420}{243} \rho^2
\\&- \frac{23680}{81} \rho\biggr) h_1
- \frac{20}{243} (1 + \rho) (1331 \rho^2 + 732 \rho + 732) (f_{2;1} + 4 f_{3;1})
+ \frac{59440}{243} \biggl(\rho^2 \\&+ \frac{444}{743} (\rho +1)\biggr) (1 + \rho) k_1
+ \frac{2000}{81} + \biggl(\frac{8116}{243}
- \frac{260 s_{8;1}}{27}\biggr) \rho^3 + \frac{14116}{243} \rho^2 + \frac{4000}{81} \rho
\biggr)
\\&+\frac{187 \rho^2}{600 (1 + \rho)^{10}}
\biggl(
-\frac{26}{17} \rho (\rho + 2) \biggl(
\rho^4 + \frac{1900}{429} \rho^3 + \frac{4300}{429} \rho^2
+ \frac{800}{143} (2\rho +1)\biggr) (1 + \rho) f_{2;1}'
\\&+ \frac{328}{51} \biggl(\rho^2 + \frac{120}{41} (\rho + 1)\biggr)
\rho (\rho + 2) (1 + \rho) \biggl(\rho^2 + \frac{30}{11} (\rho
+ 1)\biggr) f_{3;1}' +  \biggl(\rho^2 + \frac{40}{11} (\rho +1)\biggr)
\\&\times\biggl(
\frac{50}{51} \left(\rho^2 + \frac{12}{5} (\rho + 1)\right) \rho (\rho + 2)
(1 + \rho) h_1'
+ \rho^2 \left(\rho^2 - \frac{60}{17} (\rho +1)\right) (1 + \rho)^2 (k_1')^2
\\&+ \frac{25}{17} \left(\rho^2 + \frac{12}{5} (\rho + 1)\right) (\rho + 4)
\left(\rho + \frac43\right) h_1 - \frac{400}{51}(1 + \rho)
\biggl(
-\biggl(\frac{31}{4} \rho^2 + \frac{39}{5} (\rho+1)\biggr) f_{2;1}
\\&+ \left(\frac{13}{2} \rho^2 + \frac{24}{5} (\rho + 1)\right) f_{3;1}
+ \left(\rho^2 + \frac{12}{5} (\rho + 1)\right) k_1 + \frac{287}{100} \rho^2 + 3 (\rho + 1)
\biggr)
\biggr)
\biggr)\,.
\end{split}
\eqlabel{s8l7order34}
\end{equation}
Eqs. \eqref{s8l7order31}---\eqref{s8l7order34} are solved subject to the asymptotic
expansions,
\nxt in the UV, \ie as $\rho\to 0$,
\begin{equation}
\begin{split}
&\hfl_{g;8;1}=\biggl(\mathcolorbox{orange}{0}+(2\alpha_0+4\beta_1)\ln\rho\biggr) \rho^4+\calo(\rho^5\ln\rho)\,,
\end{split}
\eqlabel{s8luvq332}
\end{equation}
\begin{equation}
\begin{split}
&\hfl_{K;8;1}=\mathcolorbox{green}{0}\rho^4+\mathcolorbox{green}{0}\rho^5+
\left(\frac23-\frac{1}{6}\alpha_0\right) \rho^6
+\calo(\rho^7)\,,
\end{split}
\eqlabel{s8luvq3322}
\end{equation}
\begin{equation}
\begin{split}
&fl_{w;8;1}=\left( \frac{2}{15} \alpha_0+\frac{4}{15} \beta_1\right) \rho^4
+\left(-\frac45 \alpha_0-\frac85 \beta_1\right) \rho^5
+\biggl(\mathcolorbox{red}{0}+\left(
\frac{3}{20} \alpha_0-\frac{1}{10} \beta_1\right) \ln\rho\biggr) \rho^6 \\
&+\calo(\rho^7\ln\rho)\,,
\end{split}
\eqlabel{s8luvq3323}
\end{equation}
\begin{equation}
\begin{split}
&fl_{f;8;1}=\left( \frac{16}{27} \alpha_0+\frac{32}{27} \beta_1\right) \rho^2
+\cdots+\biggl(\mathcolorbox{pink}{0}-\biggl(
\frac{32}{9} \alpha_0 k_{4;0;1}-\frac{32}{15} \beta_1 k_{4;0;1}+\frac{397}{1200}
\alpha_0\\&+\frac{9}{400} \beta_1-2 \gamma_0\biggr)
\ln\rho-\biggl(\frac{1}{12} \alpha_0+\frac{1}{20} \beta_1\biggr) \ln^2\rho
\biggr) \rho^8 
+\calo(\rho^9\ln^2\rho)\,,
\end{split}
\eqlabel{s8luvq3324}
\end{equation}
it is completely specified by $\{\alpha_0$, $\beta_1$, $\gamma_0$, $s_{8;1}\}$;
we further highlighted arbitrary constants, fixed to zero by the overall normalization
\eqref{s8l7zero3}, and the
 extraction of the zero modes  in $fl_{g;8;1}$, $fl_{K;8;1}$  and $fl_{f;8;1}$ \eqref{s8l7deffh3};
\nxt  in the IR, \ie as $y\equiv \frac1\rho\to 0$,
\begin{equation}
\begin{split}
&\hfl_{g;8;1}=\hfl_{g;8;1;0}^h+\calo(y)\,,\qquad
\hfl_{K;8;1}=\hfl_{K;8;1;0}^h+\calo(y)\,,\qquad 
fl_{w;8;1}=fl_{w;8;1;0}^h+\calo(y)\,,\\
&\hfl_{f;8;1}=fl_{f;8;1;0}^h+\calo(y)\,,
\end{split}
\eqlabel{s8lirq332}
\end{equation}
it is completely specified by
\begin{equation}
\{\hfl_{g;8;1;0}^h\,,\, \hfl_{K;8;1;0}^h\,,\,  fl_{w;8;1;0}^h,,\, \hfl_{f;8;1;0}^h\,,\, \alpha_0\,,\,
\beta_{1}\,,\, \gamma_0\,,\,  s_{8;1}\}\,.
\eqlabel{s8lsetq332}
\end{equation}
In total, the UV and IR expansions are completely determined by the parameters
\eqref{s8lsetq332}, which is precisely what is needed to find a unique solution
for four second order ODEs \eqref{s8l7order31}-\eqref{s8l7order34}.
Solving these equations we find
\begin{equation}
\begin{split}
&s_{8;1}=5.0601(3)\,.
\end{split}
\eqlabel{s8lq332resu}
\end{equation}

\subsubsection{Select values of $s_{4\le n\le 8;1}$}
\label{cstable}

Extending the computations of sections \ref{sdetailss3} and \ref{sdetailss7l},
we collect in the table below leading corrections to the conformal spectra on
branches $(A_s)$, $(B_s)$ and $(C_s)$ for $4\le n\le 8$,
\begin{center}
\begin{tabular}{| c| c| c|} 
 \hline
 $n$ & $s_{n;1}^{(A)\&(B)}$ & $s_{n;1}^{(C)}$  \\
 \hline
 4 & $\pm \frac{\sqrt{130}}{5}$ & $-$  \\
 \hline
 5 & $\pm \sqrt{2}$ & $-$ \\
 \hline
 6 & $\pm \frac{\sqrt{1390}}{25}$ & $6.03(2)$ \\
 \hline
 7 & $\pm \frac{\sqrt{1490}}{25}$ & $5.39(7)$ \\
 \hline
 8 & $\pm \frac{2\sqrt{779}}{35}$  & $5.06(0)$  \\
 \hline
 \end{tabular}
\end{center}
These results are used to highlight the features of the spectra
presented in fig.~\ref{figure3s}.

\section{Critical point $H=H_{crit_3}$}
\label{appdd}

The chiral symmetry breaking mode of fluctuations about TypeA$ _s$ background becomes
marginal at $H=H_{crit_3}$, see figure \ref{figure4}. It signals the origin of
TypeA$ _b$ background \cite{Buchel:2019pjb}, which exists only for $H>H_{crit_3}$.
In this section we first construct TypeA$ _b$ background perturbatively in
$A\propto \sqrt{H-H_{crit_3}}$, and then study the $H=H_{crit_3}$ marginal 
mode in this perturbative  TypeA$ _b$ background geometry.
We find that this  mode becomes unstable, \ie
\begin{equation}
\Im[\ww_{\csb}]\bigg|_{{\rm TypeA}_b}\ =\ 0+36.0098(5)\cdot A^2+ \calo(A^4) \,,
\eqlabel{wpert}
\end{equation}
where the precise definition of $A$ is given by \eqref{defa}.

\subsection{TypeA$ _b$ background in the vicinity of $H=H_{crit_3}$}

TypeA$ _s$ background is a special case of TypeA$ _b$ background, constraint by
\eqref{sdfpb}. From  \eqref{ksfa} and  \eqref{ksfb},
\begin{equation}
\begin{split}
&f_a-f_b=\underbrace{\left(2 f_{a,3,0}+\frac12 f_{a,1,0}\right)}_{\equiv 2A} \rho^3
-\frac{f_{a,1,0}}{2} \left(2 f_{a,3,0}+\frac12 f_{a,1,0}\right) \rho^4
\\&+\biggl(
\frac{8 f_{a,1,0}^2+4 K_0-9}{32} \left(2 f_{a,3,0}+\frac12 f_{a,1,0}\right)+\frac14 k_{2,3,0}
+\frac18 \left(2 f_{a,3,0}+\frac12 f_{a,1,0}\right) \ln\rho\biggr) \rho^5
\\&+\calo(\rho^6\ln\rho)\,,
\end{split}
\eqlabel{defa}
\end{equation}
which provides a precise definition of $A$. $A$ vanishes exactly
at\footnote{We use computation SchemeI with $b\equiv 1$, see \cite{Buchel:2019pjb}.} $K_0=K_{0,crit_3}$
\begin{equation}
{K_0}\bigg|_{crit_3}=\ln \frac{H_{crit_3}^2}{\Lambda^2}\,.
\eqlabel{ksdef}
\end{equation}

Perturbatively in $A$, TypeA$ _b$ background can be represented as
\begin{equation}
\begin{split}
&f_a=f_3(\rho)+\sum_{k=1}^\infty A^{2k-1} \cdot \df_{a;2k-1}(\rho)+\sum_{k=1}^\infty A^{2k} \cdot \df_{a;2k}(\rho)\,,
\end{split}
\eqlabel{pert1}
\end{equation}
\begin{equation}
\begin{split}
&f_b=f_3(\rho)-\sum_{k=1}^\infty A^{2k-1} \cdot \df_{a;2k-1}(\rho)+\sum_{k=1}^\infty A^{2k} \cdot \df_{a;2k}(\rho)\,,
\end{split}
\eqlabel{pert2}
\end{equation}
\begin{equation}
\begin{split}
&f_c=f_2(\rho)+\sum_{k=1}^\infty A^{2k} \cdot \df_{c;2k}(\rho)\,,
\end{split}
\eqlabel{pert3}
\end{equation}
\begin{equation}
\begin{split}
&K_1=K(\rho)+\sum_{k=1}^\infty A^{2k-1} \cdot \dk_{1;2k-1}(\rho)+\sum_{k=1}^\infty A^{2k} \cdot \dk_{1;2k}(\rho)\,,
\end{split}
\eqlabel{pert4}
\end{equation}
\begin{equation}
\begin{split}
&K_3=K(\rho)-\sum_{k=1}^\infty A^{2k-1} \cdot \dk_{1;2k-1}(\rho)+\sum_{k=1}^\infty A^{2k} \cdot \dk_{1;2k}(\rho)\,,
\end{split}
\eqlabel{pert5}
\end{equation}
\begin{equation}
\begin{split}
&K_2=1+\sum_{k=1}^\infty A^{2k-1} \cdot \dk_{2;2k-1}(\rho)\,,
\end{split}
\eqlabel{pert6}
\end{equation}
\begin{equation}
\begin{split}
&g\bigg|_{{\rm TypeA}_b}=g(\rho)\bigg|_{{\rm TypeA}_s}+\sum_{k=1}^\infty A^{2k} \cdot \dg_{2k}(\rho)\,,
\end{split}
\eqlabel{pert7}
\end{equation}
\begin{equation}
\begin{split}
&h\bigg|_{{\rm TypeA}_b}=h(\rho)\bigg|_{{\rm TypeA}_s}+\sum_{k=1}^\infty A^{2k} \cdot \dh_{2k}(\rho)\,.
\end{split}
\eqlabel{pert8}
\end{equation}
To compute \eqref{wpert} we need perturbative solution of TypeA$ _b$ background
to order $k=3$ inclusive.
As we now explain, orders $k=\{0,1\}$, and $k=\{2,3\}$ must be solved simultaneously.

\subsubsection{ $k=\{0,1\}$}

At leading $k=0$ order we have TypeA$ _s$ background, labeled by $K_0$; namely,
a coupled system of 4 second-order ODEs for $\{f_3,K,g,h\}$ and a single first-order
ODE for $f_2$.
At order $k=1$, the equations for $\{\df_{a;1},\dk_{1;1},\dk_{2;1}\}$ are just the
equations for the marginal mode --- they are equivalent to \eqref{sfl1}-\eqref{sfl3},
see also \eqref{sdfpf}, 
with the following identification,
\begin{equation}
\df_{a;1}\equiv f_3\cdot F\,,\qquad \dk_{1;1}\equiv \chi_1\,,\qquad \dk_{2;1}\equiv \chi_2\,,
\eqlabel{matchk0}
\end{equation}
and with
\begin{equation}
s=0\,.
\eqlabel{szero}
\end{equation}
They are solved subject to the asymptotics:
\nxt In the UV, \ie as $\rho \to 0$,
\begin{equation}
\begin{split}
&\df_{a;1}= \mathcolorbox{red}{1}\cdot r^3 +\cdots
+\biggl(\df_{a;1;7,0}+\biggl(\frac{275}{384}+\frac{3}{64}\dk_{1;1;3,0}\left(K_0-\frac54\right)
+\frac{7}{256} K_0\\&+3 f_{c,4,0}\biggr) \ln\rho
+\biggl(-\frac{15}{128}-\frac{3}{64} \dk_{1;1;3,0}+\frac{9}{64} K_0\biggr) \ln^2\rho
-\frac18\ln^3\rho\biggr) \rho^7+\calo(\rho^8\ln^3\rho)\,,
\end{split}
\eqlabel{order11}
\end{equation}
\begin{equation}
\begin{split}
\dk_{1;1}=\rho^3 (\dk_{1;1;3,0}+2 \ln\rho)+\rho^4
\biggl(-\frac12 (3 \dk_{1;1;3,0}+2) f_{a,1,0}-3 f_{a,1,0} \ln\rho\biggr)
+\calo(\rho^5\ln^2\rho)\,,
\end{split}
\eqlabel{order12}
\end{equation}
\begin{equation}
\begin{split}
\dk_{2;1}=\rho^3 \left(-1+\frac32 \dk_{1;1;3,0}+3 \ln\rho\right)-\rho^4
\biggl(
\frac94 f_{a,1,0} \dk_{1;1;3,0}+\frac92 f_{a,1,0} \ln\rho
\biggr)+\calo(\rho^5\ln^2\rho)\,,
\end{split}
\eqlabel{order13}
\end{equation}
it is characterized by 2 parameters
\begin{equation}
\{\ \dk_{1;1;3,0}\,,\, \df_{a;1;7,0}\ \}\,.
\eqlabel{uvorder1}
\end{equation}
In \eqref{order11} we highlighted the overall normalization, dictated by our definition
of the amplitude $A$, see \eqref{defa}. Of course,
the asymptotic expansions \eqref{order11}-\eqref{order13} depend on the parameters
of the $k=0$ order background, \ie
\begin{equation}
\{\ K_0\,,\, f_{a,1,0}\,,\ g_{4,0}\,,\, f_{c,4}\,,\, f_{a,6,0}\,,\, f_{a,8,0}\ \}\,.
\eqlabel{uvorder0}
\end{equation}
Comparing with \eqref{uvparks}, because of the constraint \eqref{sdfpb},
we find that $\{f_{a,3,0}$, $k_{2,3,0}$, $f_{a,7,0}\}$
are not independent and instead are determined by  \eqref{uvorder0}:
\begin{equation}
\begin{split}
&f_{a,3,0}=-\frac14  f_{a,1,0}\,,\qquad k_{2,3,0}=0\,,\\
&f_{a,7,0}=\frac{431}{76800} f_{a,1,0} K_0^2+\biggl(-\frac{981}{1024000}
+\frac{1}{40} f_{c,4,0}-\frac{53}{1920} f_{a,1,0}^2\biggr) f_{a,1,0} K_0
+\biggl(-\frac{1362319}{61440000}\\&+\frac{1}{80} f_{a,1,0}^4
+\frac{77}{46080} f_{a,1,0}^2-\frac{1}{320} f_{c,4,0}-2 f_{a,6,0}
-\frac{1}{40} g_{4,0}\biggr) f_{a,1,0}\,.
\end{split}
\eqlabel{order0const}
\end{equation}
\nxt In the IR, \ie as $y\equiv \frac1\rho \to 0$,
\begin{equation}
\df_{a;1}=\frac 1y \biggl(\df_{a;1;0}^h+\calo(y)\biggr)\,,\qquad
\dk_{1;1}=\dk_{1;1;0}^h+\calo(y)\,,\qquad \dk_{2;1}=\dk_{2;1;0}^h+\calo(y)\,,
\eqlabel{order1h}
\end{equation}
it is characterized by 3 parameters
\begin{equation}
\{\ \df_{a;1;0}^h\,,\, \dk_{1;1;0}^h\,,\, \dk_{2;1;0}^h\ \}\,.
\eqlabel{uvorder1h}
\end{equation}
As in UV, the asymptotic expansions \eqref{order1h} depend on the parameters
of the $k=0$ order background, \ie
\begin{equation}
\{\ f_{a,0}^h\,,\,  f_{c,0}^h\,,\, K_{1,0}^h\,,\, g_{0}^h\ \}\,.
\eqlabel{uvorder0h}
\end{equation}
Comparing with \eqref{irph1par}, because of the constraint \eqref{sdfpb},
we find that $\{f_{b,0}^h$, $K_{2,0}^h$, $K_{3,0}^h\}$
are not independent and instead are determined by  \eqref{uvorder0h}:
\begin{equation}
\begin{split}
&f_{b,0}^h=f_{a,0}^h\,,\qquad K_{3,0}^h=K_{1,0}^h\,,\qquad K_{2,0}^h=1\,.
\end{split}
\eqlabel{order0consth}
\end{equation}

In total we have 7 second-order ODEs (4 from $k=0$ order and 3 from $k=1$ order)
and 1 additional first-order ODE from the $k=0$ order. Thus in total, we need
$7\times 2 +1 =15$ adjustable parameters to find a solution. This is precisely what
we have: $6+4=10$ parameters from order $k=0$, see \eqref{uvorder0}
and \eqref{uvorder0h}, and $2+3=5$ parameters from order $k=1$, see \eqref{uvorder1}
and \eqref{uvorder1h}. Note the coupling of
orders $k=0$ and $k=1$ occurs because we traded the parameter
$s$, we set it zero in \eqref{szero}, for a requirement to tune $K_0$
to insure that the $k=1$ order deformation $\{\df_{a;1},\dk_{1;1}\,\dk_{2;1}\}$
is normalizable, \ie the corresponding fluctuations (see \eqref{matchk0})
are marginal.

Solving the order $k=0$ and $k=1$ equations numerically we recover
\begin{equation}
K_0=K_0\bigg|_{crit_3}=-0.1636(3)\,,
\eqlabel{k0crit}
\end{equation}
originally reported in \cite{Buchel:2019pjb}.

\subsubsection{ $k=\{2,3\}$}

We will not present the equations for order $k=\{2,3\}$ perturbative representation of
the background TypeA$ _b$: they can be straightforwardly derived from the
general equations for this background  (see appendix B of \cite{Buchel:2019pjb}) using the ansatz
\eqref{pert1}-\eqref{pert8}. Since the equation for $f_c$ is of the first-order, so will be
the equations for $\df_{c;2k}$. The equations for the other functions are always of the
second-order. We will discuss the asymptotics and count the parameters.

At order $k=2$ we have a coupled system
of 4 second-order ODEs for $\{\df_{a;2}$, $\dk_{1;2}$, $\dg_{2}$, $\dh_2\}$
and the first-order ODE for $\df_{c;2}$. They are solved subject
to the asymptotics:
\nxt In the UV, \ie as $\rho \to 0$,
\begin{equation}
\begin{split}
&\df_{a;2}=\rho\ \df_{a;2;1,0}+\rho^2 \biggl(
\frac12 \df_{a;2;1,0} f_{a,1,0}-\frac14 \dk_{1;2;0,0}\biggr)
+\cdots+
\rho^6\biggl(\df_{a;2;6,0}+\cdots\\&+\frac{3}{640} \dk_{1;2;0,0} \ln^3\rho\biggr)
+\calo(\rho^7\ln^4\rho)\,,
\end{split}
\eqlabel{order21}
\end{equation}
\begin{equation}
\begin{split}
&\df_{c;2}=\rho\ \df_{a;2;1,0}+\rho^2 \biggl(
\frac12 \df_{a;2;1,0} f_{a,1,0}-\frac14 \dk_{1;2;0,0}\biggr)
-\frac14 \rho^3 \df_{a;2;1,0}+\rho^4 \biggl(
\df_{c;2;4,0}\\&+\frac{1}{16} \dk_{1;2;0,0} \ln\rho\biggr)+\calo(\rho^5\ln^2\rho)\,,
\end{split}
\eqlabel{order22}
\end{equation}
\begin{equation}
\begin{split}
&\dk_{1;2}=\dk_{1;2;0,0}+\rho \df_{a;2;1,0}+\rho^2 \biggl(
-\frac12 \df_{a;2;1,0} f_{a,1,0}+\frac18 \dk_{1;2;0,0}\biggr)+\calo(\rho^3\ln\rho)\,,
\end{split}
\eqlabel{order23}
\end{equation}
\begin{equation}
\begin{split}
&\dg_{2}=\frac12 \rho^3 \df_{a;2;1,0}+\rho^4 \biggl(\dg_{2;4,0}
+\biggl(-\frac38 \df_{a;2;1,0} f_{a,1,0}-\frac{5}{64} \dk_{1;2;0,0}+3 \df_{c;2;4,0}
\biggr) \ln\rho\\&+\frac{3}{32} \dk_{1;2;0,0} \ln^2\rho\biggr)+\cdots+
\rho^8\biggl(\dg_{2;8,0}+\cdots-\frac{3}{512} \dk_{1;2;0,0} \ln^5\rho\biggr)
+\calo(\rho^9\ln^6\rho)\,,
\end{split}
\eqlabel{order24}
\end{equation}
\begin{equation}
\begin{split}
&\dh_{2}=\frac14 \dk_{1;2;0,0}+\rho \biggl(
-\frac12 \df_{a;2;1,0} K_0-\frac12 \dk_{1;2;0,0} f_{a,1,0}+\df_{a;2;1,0} \ln\rho
\biggr)+\calo(\rho^2\ln\rho)\,,
\end{split}
\eqlabel{order25}
\end{equation}
it is characterized by 6 parameters
\begin{equation}
\{\ \dk_{1;2;0,0}\,,\,
\df_{a;2;1,0}\,,\, \df_{c;2;4,0}\,,\, \dg_{2;4,0}\,,\, \df_{a;2;6,0}\,,\, \dg_{2;8,0}\ \}\,.
\eqlabel{uvorder2}
\end{equation}
\nxt In the IR, \ie as $y\equiv \frac 1\rho\to 0$,
\begin{equation}
\begin{split}
&\df_{a;2}=\frac 1y \biggl(\df_{a;2;0}^h+\calo(y)\biggr)\,,\qquad
\df_{c;2}=\frac 1y \biggl(\df_{c;2;0}^h+\calo(y)\biggr)\,,\ \
\dk_{1;2}=\dk_{1;2;0}^h+\calo(y)\,,\\
&\dg_{2}=\dg_{2;0}^h+\calo(y)\,,\qquad
\dh_{2}=y^3\cdot \biggl(\frac{(K_{1,0}^h)^2}{(f_{a,0}^h)^4 f_{c,0}^h}
\left(-\frac{2 (\df_{a;1,0}^h)^2}{(f_{a,0}^h)^2}+\frac{4 \df_{a;2,0}^h}{f_{a,0}^h}
+\frac{\df_{c;2,0}^h}{f_{c,0}^h}\right)
\\&+\frac{2 K_{1,0}^h}{(f_{a,0}^h)^4 f_{c,0}^h}
( \dk_{1;1;0}^h \dk_{2;1;0}^h-\dk_{1;2;0}^h)
+\biggl(
-\frac{9}{40 (f_{a,0}^h)^2 f_{c,0}^h}+\frac{3}{5 (f_{a,0}^h)^3}-\frac{f_{c,0}^h}{10(f_{a,0}^h)^4}
\\&-\frac{9g_0^h}{10 (f_{a,0}^h)^4 f_{c,0}^h}\biggr) (\df_{a;1,0}^h)^2
+\biggl(
-\frac{3}{5 (f_{a,0}^h)^2}+\frac{f_{c,0}^h}{5(f_{a,0}^h)^3}+\frac{3g_0^h}{5(f_{a,0}^h)^3 f_{c,0}^h}
\biggr) \df_{a;2,0}^h
\\&-\frac{3g_0^h}{10(f_{a,0}^h)^2 f_{c,0}^h} (\dk_{2;1;0}^h)^2
+\biggl(-\frac{1}{10 (f_{a,0}^h)^2}+\frac{3g_0^h}{10(f_{a,0}^h)^2 (f_{c,0}^h)^2}\biggr) \df_{c;2,0}^h
-\frac{3}{10 (f_{a,0}^h)^2 f_{c,0}^h} \dg_{2;0}^h\\
&-\frac{6g_0^h \df_{a;1,0}^h \dk_{2;1;0}^h}{5(f_{a,0}^h)^3 f_{c,0}^h}
-\frac{27(\dk_{1;1;0}^h)^2}{40(f_{a,0}^h)^2 g_0^h f_{c,0}^h}\biggr)+\calo(y^4)\,,
\end{split}
\eqlabel{order2h}
\end{equation}
it is characterized by 4 parameters
\begin{equation}
\{\ \df_{a;2;0}^h\,,\, \df_{c;2;0}^h\,,\,  \dk_{1;2;0}^h\,,\, \dg_{2;0}^h\ \}\,.
\eqlabel{irorder2}
\end{equation}
At order $k=3$ we have a coupled system of 3 second-order ODEs for $\{\df_{a;3}$,
$\dk_{1;3}$, $\dk_{2;3}\}$. They are solved subject to the asymptotics:
\nxt In the UV, \ie as $\rho \to 0$,
\begin{equation}
\begin{split}
&\df_{a;3}= \mathcolorbox{red}{0}\cdot\rho^3-\frac 12 \rho^4 \df_{a;2;1,0}+\rho^5
\biggl(
\frac12 \df_{a;2;1,0} f_{a,1,0}+\frac18 \dk_{1;2;0,0}+\frac{3}{16} \dk_{1;3;3,0}\biggr)
\\&+\rho^7\biggl(
\df_{a;3;7,0}+\cdots+\left(\frac{9}{64} \dk_{1;2;0,0}-\frac{3}{64} \dk_{1;3;3,0}\right) \ln^2\rho
\biggr)
+\calo(\rho^8\ln^3\rho)\,,
\end{split}
\eqlabel{order31u}
\end{equation}
\begin{equation}
\begin{split}
&\dk_{1;3}=\rho^3 \dk_{1;3;3,0}+\rho^4 \biggl(
-\frac32 \df_{a;2;1,0} \dk_{1;1;3,0}-\frac32 \dk_{1;3;3,0} f_{a,1,0}-\df_{a;2;1,0}
-3 \df_{a;2;1,0} \ln\rho\biggr)
\\&+\calo(\rho^5\ln\rho)\,,
\end{split}
\eqlabel{order32u}
\end{equation}
\begin{equation}
\begin{split}
&\dk_{2;3}=\frac32 \rho^3 \dk_{1;3;3,0}-\frac 94\rho^4
\biggl(
\df_{a;2;1,0} \dk_{1;1;3,0}+\dk_{1;3;3,0} f_{a,1,0}+2 \df_{a;2;1,0} \ln\rho
\biggr)
+\calo(\rho^5\ln\rho)\,,
\end{split}
\eqlabel{order33u}
\end{equation}
it is characterized by 2 parameters
\begin{equation}
\{\ 
\df_{a;3;7,0}\,,\, \dk_{1;3;3,0}\ \}\,.
\eqlabel{uvorder23}
\end{equation}
In \eqref{order31u} we highlighted the parameter fixed to zero,
as dictated by our definition of the amplitude $A$, see \eqref{defa}.
\nxt In the IR, \ie as $y\equiv \frac1\rho \to 0$,
\begin{equation}
\df_{a;3}=\frac 1y \biggl(\df_{a;3;0}^h+\calo(y)\biggr)\,,\qquad
\dk_{1;3}=\dk_{1;3;0}^h+\calo(y)\,,\qquad \dk_{2;3}=\dk_{2;3;0}^h+\calo(y)\,,
\eqlabel{order3h}
\end{equation}
it is characterized by 3 parameters
\begin{equation}
\{\ \df_{a;3;0}^h\,,\, \dk_{1;3;0}^h\,,\, \dk_{2;3;0}^h\ \}\,.
\eqlabel{irorder3h}
\end{equation}

In total we have 7 second-order ODEs (4 from $k=2$ order and 3 from $k=3$ order)
and 1 additional first-order ODE from the $k=2$ order. Thus in total, we need
$7\times 2 +1 =15$ adjustable parameters to find a solution. This is precisely what
we have: $6+4=10$ parameters from order $k=2$, see \eqref{uvorder2}
and \eqref{irorder2}, and $2+3=5$ parameters from order $k=3$, see \eqref{uvorder23}
and \eqref{irorder3h}. Here the coupling of
orders $k=2$ and $k=3$ occurs because we have an {\it additional}  parameter
at order $k=2$,  and we are {\it lacking} one parameter at order $k=3$. Specifically,
$\dk_{1;2;0,0}$, see  \eqref{uvorder2}, is needed to parameterize background solutions
TypeA$ _b$, away from $K_0=K_{0,crit_3}$:
\begin{equation}
K_0-K_{0,crit_3} = \ln \frac{H^2}{H_{crit_3}^2}=\dk_{1;2;0,0}\cdot A^2+\calo(A^4)\,.
\eqlabel{defk0}
\end{equation}
On the other had, at order $k=3$ we have 3 second-order ODEs for $\{\df_{a;3}$, $\dk_{1;3}$,
$\dk_{2;3}\}$, however we have only $2+3=5$ adjustable parameter (see \eqref{uvorder23}
and \eqref{irorder3h}) --- the {\it missing} parameter is the highlighted one
in \eqref{order31u}, that we are forced to set to zero as part of the
definition of the amplitude $A$ \eqref{defa}.

Solving the order $k=2$ and $k=3$ equations numerically we find
\begin{equation}
\dk_{1;2;0,0}=6.4889(0)\,.
\eqlabel{ks2res}
\end{equation}

\subsubsection{$K_0(A)$ and its perturbative approximation}

\begin{figure}[t]
\begin{center}
\psfrag{a}{{$A$}}
\psfrag{k}{{$K_0$}}
\includegraphics[width=4in]{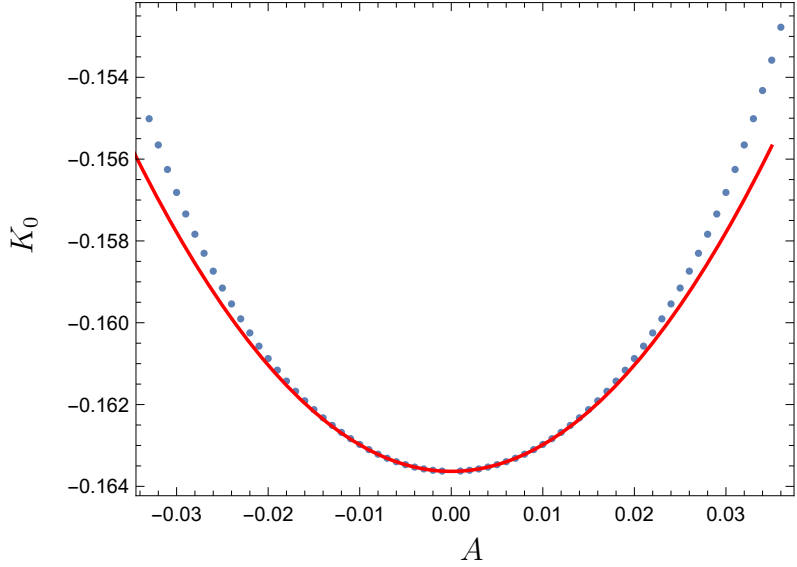}
\end{center}
  \caption{We construct TypeA$ _b$ background geometry parameterizing it with
  $A$, defining the deviation from the critical point $H_{crit_3}$, see \eqref{defa}.
  The same geometry was parameterized with $K_0$ in \cite{Buchel:2019pjb}.
  Dots represent $K_0(A)$ for select values of $A$. The solid red curve
  is the perturbative approximation \eqref{defk0}. 
} \label{K0a}
\end{figure}

To identify TypeA$ _b$ DFP instability it is most convenient to
construct  numerically the corresponding TypeA$ _b$ background
by parameterizing it with $A$, as defined in \eqref{defa},
rather than using $K_0$, as it is done in  \cite{Buchel:2019pjb}.
This allows us to use the near-critical analysis of the
marginal mode of section \ref{qnmmaginal} as an approximation for the
spectral analysis of this mode at finite $A$, see fig.~\ref{figure1b}.
In fig.~\ref{K0a} we compare $K_0(A)$ with its perturbative in $A$ approximation
given by \eqref{defk0}. We find from numerical interpolation an excellent
agreement, 
\begin{equation}
\biggl[\frac{1}{2\dk_{1;2;0,0}} \frac{d^2 K_{0}(A)}{dA^2}-1\biggr]\bigg|_{A=0} =
1.33\cdot 10^{-15}\,.
\eqlabel{compa}
\end{equation}

\subsection{TypeA$ _b$ background instability in the vicinity of $H=H_{crit_3}$}
\label{qnmmaginal}

Perturbatively in $A$, the chiral symmetry breaking, marginal at $H=H_{crit_3}$, mode
can be represented as, see \eqref{sdfpf}, 
\begin{equation}
fl_a=f_3(\rho)\cdot F(\rho)+\sum_{k=1}^\infty A^{2k-1}\cdot fl_{a;2k-1}(\rho)
+\sum_{k=1}^\infty A^{2k}\cdot fl_{a;2k}(\rho)\,,
\eqlabel{flaa}
\end{equation}
\begin{equation}
fl_b=-f_3(\rho)\cdot F(\rho)+\sum_{k=1}^\infty A^{2k-1}\cdot fl_{a;2k-1}(\rho)-
\sum_{k=1}^\infty A^{2k}\cdot fl_{a;2k}(\rho)\,,
\eqlabel{flba}
\end{equation}
\begin{equation}
fl_c=\sum_{k=1}^\infty A^{2k-1}\cdot fl_{c;2k-1}(\rho)\,,
\eqlabel{flca}
\end{equation}
\begin{equation}
fl_{K_1}=\chi_1(\rho)+\sum_{k=1}^\infty A^{2k-1}\cdot fl_{K_1;2k-1}(\rho)
+\sum_{k=1}^\infty A^{2k}\cdot fl_{K_1;2k}(\rho)\,,
\eqlabel{flk1a}
\end{equation}
\begin{equation}
fl_{K_2}=\chi_2(\rho)+\sum_{k=1}^\infty A^{2k}\cdot fl_{K_2;2k}(\rho)\,,
\eqlabel{flk2a}
\end{equation}
\begin{equation}
fl_{K_3}=-\chi_1(\rho)+\sum_{k=1}^\infty A^{2k-1}\cdot fl_{K_1;2k-1}(\rho)
-\sum_{k=1}^\infty A^{2k}\cdot fl_{K_1;2k}(\rho)\,,
\eqlabel{flk3a}
\end{equation}
\begin{equation}
fl_{g}=\sum_{k=1}^\infty A^{2k-1}\cdot fl_{g;2k-1}(\rho)\,,
\eqlabel{flga}
\end{equation}
\begin{equation}
fl_{h}=\sum_{k=1}^\infty A^{2k-1}\cdot fl_{h;2k-1}(\rho)\,,
\eqlabel{flha}
\end{equation}
with
\begin{equation}
-\Im[\ww_{\csb}]\bigg|_{{\rm TypeA}_b}\equiv s=0+\sum_{k=1}^\infty A^{2k}\cdot s_{2k}\,.
\eqlabel{wmagpert}
\end{equation}
The equations of motion for the terms of the perturbative
expansion of the fluctuations can  be derived from the 
general equations of appendix  \ref{fgframe}, using
the perturbative TypeA$ _b$ background ansatz \eqref{pert1}-\eqref{pert8},
and \eqref{flaa}-\eqref{wmagpert}.
Since $fl_h$ can always be algebraically determined
from the remaining modes, see \eqref{flh}, we find that
the same is true for its perturbative terms $fl_{h,2k-1}$.

\begin{figure}[t]
\begin{center}
\psfrag{a}{{$A$}}
\psfrag{b}{{$\beta$}}
\includegraphics[width=4in]{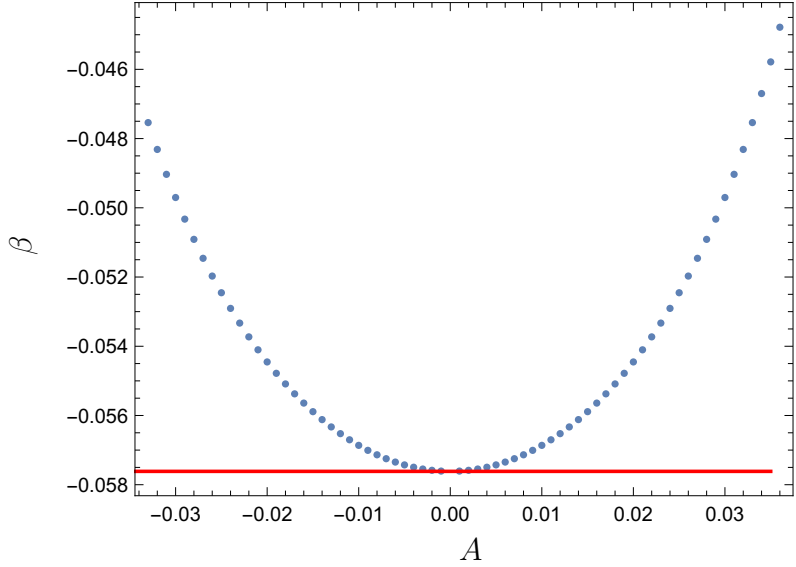}
\end{center}
  \caption{Dots represent the amplitude of the zero mode
  $\beta(A)$ as given by \eqref{defbetaa}. The solid
  red line is the perturbative approximation \eqref{beta0}.
} \label{betaa}
\end{figure}

We summarize below the salient features of the numerical analysis.
\begin{itemize}
\item Order $k=0$. Here, the fluctuations are represented by the marginal
chiral symmetry breaking mode, see \eqref{sdfpf}, \eqref{sfl1}-\eqref{sfl3}
with $s=0$.
\item At any even order in $k$ there is a zero mode:
if $\{fl_{a;2k},fl_{K_1;2k},fl_{K_2;2k}\}$ is a solution of the equations of motion,
so is 
\begin{equation}
\{\ fl_{a;2k}+f_3 F\,,\, fl_{K_1;2k}+\chi_1\,,\, fl_{K_2;2k}+\chi_2\ \}\,,
\eqlabel{zmarg}
\end{equation}
with an arbitrary amplitude $f_{3,0}\equiv \beta_{2k}$, see \eqref{collsdfp}.
These arbitrary at order $2k$ parameters are fixed at order $2k+1$.
For example, we find in this manner
\begin{equation}
\beta_0=-0.05761(4)\,.
\eqlabel{beta0}
\end{equation}
\item Order $k=1$. At this order the fluctuations are
$\{fl_{a;1}$, $fl_{c;1}$, $fl_{K_1;1}$, $fl_{g_1}\}$. Since there is
no contribution to $s$ at this order, \ie $s_1=0$ in \eqref{wmagpert},
the zero mode amplitude at the previous order, $\beta_0$, is needed
to find a unique solution.
\item Order $k=2$. At this order the fluctuations are
$\{fl_{a;2}$, $fl_{K_1;2}$, $fl_{K_2;2}\}$; additionally, the equations
explicitly depend on $s_2$ parameter in \eqref{wmagpert}.
The equations for the fluctuations
also require the input of the background TypeA$ _b$
up to order $k=2$ inclusive.
\item In fig.~\ref{betaa} we compare the zero mode amplitude $\beta(A)$, extracted
in computing numerically $s(A)$ in TypeA$ _b$ DFP at finite $A$,   
\begin{equation}
\beta(A)=\lim_{\rho\to 0}\frac{fl_a(\rho)-fl_b(\rho)}{2\rho^3} \equiv \sum_{k=0}^\infty A^{2k}\cdot \beta_{2k}\,,
\eqlabel{defbetaa}
\end{equation}
with its perturbative approximation at $A=0$, see \eqref{beta0}.
Numerically interpolating the finite $A$ results we find a good agreement,
\begin{equation}
\biggl[\frac{\beta(A)}{\beta_0}-1\biggr]\bigg|_{A=0}=-2.0(7)\cdot 10^{-5}\,.
\eqlabel{compare}
\end{equation}
\item Numerical analysis at order $k=2$ provide the value of $s_2$,
\begin{equation}
s_2=-36.0098(5)\,,
\eqlabel{s2pert}
\end{equation}
which implies that  marginal at $H=H_{crit_3}$ fluctuations become unstable
in TypeA$ _b$ for $H>H_{crit_3}$. The frequency of this mode at finite $A$
is presented in fig.~\ref{figure1b}.
Numerically interpolating the finite $A$ results for $s(A)$ we find a good agreement
with the leading nontrivial order perturbative approximation, \eqref{wmagpert},
\begin{equation}
\biggl[\frac{1}{2s_2}\frac{d^2 s(A)}{dA^2}-1\biggr]\bigg|_{A=0}=1.1(7)\cdot 10^{-3}\,.
\eqlabel{comparesa}
\end{equation}
\end{itemize}

\bibliographystyle{JHEP}
\bibliography{cascadingdfp}

\end{document}